\documentclass[twocolumn,twocolappendix]{aastex701}    

\usepackage{newtxtext,newtxmath} 
\usepackage{amsmath}

\newcommand{\hunit}{km s$^{-1}$ Mpc$^{-1}$}

\newcommand{\kmps}{km s$^{-1}$}
\newcommand{\remaj}{\textit{$R_{\rm e}^{\rm maj}$}}
\newcommand{\as}{$^{\prime \prime}$}

\newcommand{\newadd}[1]{\textcolor{black}{{#1}}}

\newcommand{\jamcyl}{$\textsc{JAM}_{\mathrm{cyl}}$}
\newcommand{\jamsph}{$\textsc{JAM}_{\mathrm{sph}}$}
\newcommand{\tslope}{$\gamma_{\rm T}$}

\hypersetup{linkcolor=red,citecolor=blue}     

\NewDocumentCommand\michele{O{Rewritten}m}{\textsf{\color{green}[Michele: #1]\(\rightarrow\)[``\emph{#2}'']}}

\begin{document}

\title{\Large MAGNUS. III. Mild Evolution of the Total Density Slope in Massive Early-type Galaxies Since z$\sim$1 from Dynamical Modeling of MUSE Integral-field Stellar Kinematics}

\author[0000-0002-8593-7243]{Pritom Mozumdar}\thanks{corresponding author:pmozumdar@astro.ucla.edu} 
\email{pmozumdar@astro.ucla.edu}
\affiliation{Department of Physics and Astronomy, University of California, Los Angeles, CA 90095, USA}

\author[0000-0002-1283-8420]{Michele Cappellari}
\email{michele.cappellari@physics.ox.ac.uk}
\affiliation{Sub-Dep. of Astrophysics, Dep. of Physics, University of Oxford, Denys Wilkinson Building, Keble Road, Oxford, OX1 3RH, UK}

\author[0000-0002-4030-5461]{Christopher D. Fassnacht}
\email{fassnacht@ucdavis.edu}
\affiliation{Department of Physics and Astronomy, University of California, Davis, CA 95616, USA}

\author[0000-0002-8460-0390]{Tommaso Treu}
\email{tommaso.treu@physics.ucla.edu}
\affiliation{Department of Physics and Astronomy, University of California, Los Angeles, CA 90095, USA}

\begin{abstract}
We investigate the total mass density slope evolution in massive early-type galaxies (ETGs) over the last 6.5 billion years ($0 < z < 0.75$). We perform a detailed dynamical analysis of approximately 200 ETGs spanning the redshift range $0.24 < z < 0.75$, utilizing spatially resolved stellar kinematics derived from high signal-to-noise ratio (S/N) MUSE-DEEP spectroscopy and surface brightness models from high-resolution HST imaging. We constrain mass distributions using the Jeans Anisotropic Modeling (JAM) technique coupled with Multi-Gaussian Expansion (MGE) method. To rigorously constrain evolutionary trends, we combine this intermediate-redshift dataset with a local ETG sample ($z \sim 0.05$) from the MaNGA survey. We adopt dynamical constraints for the local sample derived using an identical homogeneous methodology, ensuring a strictly consistent comparison. \newadd{We found that the total density profiles of the intermediate-redshift ETG sample are approximately isothermal and exhibit a median mass-weighted total density slope, $<\gamma_{\rm T}>=2.19 \pm 0.01$ at $<z>=0.44$, which is shallower than the local baseline of $<\gamma_{\rm T}> = 2.26 \pm 0.01$ at $<z>=0.04$.} This structural shift corresponds to a redshift gradient of \newadd{$\mathrm{d} \gamma_{\rm T}/\mathrm{d} z \approx -0.20 \pm 0.03$}, detected at $\sim$6$\sigma$ significance. We demonstrate that this trend is robust against model assumptions and persists even when restricting the analysis to high-velocity dispersion systems ($\sigma_e > 150$ \kmps). Our findings are consistent with previous lensing-based studies and in tension with cosmological simulations. The observed steepening suggests that dissipative processes, such as gas-rich accretion and mergers, must play a non-negligible role in the late-stage assembly of massive ETGs.
\end{abstract}

\keywords{Galaxy evolution --- total density slope --- resolved kinematics --- Jeans' anisotropic modeling}


\section{Introduction} \label{sec:intro}

Galaxies form within dark matter (DM) halos as baryonic matter cools and condenses at their centers \citep[e.g.,][]{White_Frenk_1991}. The combined distribution of baryonic and dark matter determines the total mass profile, which is effectively described by a power-law radial density distribution, $\rho(r) \propto r^{-\gamma_{\rm T}}$ (in this work we adopt the convention \tslope\ $> 0$) \citep[e.g.][]{Treu&Koopmans_2002a, Treu_Koopmans_2004}. The interplay between these components dictates the internal structure of galaxies, which evolves over cosmic time driven by physical processes such as mergers, gas accretion, and environmental interactions \citep[e.g.,][]{Naab_Ostriker_2017}. Consequently, the total density slope serves as a fundamental diagnostic for constraining the mechanisms of galaxy formation and evolution.

The assembly of the mass distribution within massive galaxies is theorized to proceed broadly in two phases \citep[e.g.,][]{Oser_2010, van_Dokkum_2015}. During the first phase ($z \gtrsim 2$), cold gas collapses toward the center and triggers intense \textit{in situ} star formation, causing the central region to become denser and the total density slope to steepen with decreasing redshift. Subsequently, as \textit{in situ} star formation is quenched, the mass distribution evolves primarily through mergers and accretion \citep[e.g.,][]{Naab_2009, van_Dokkum_2010}. In this second phase, two scenarios generally diverge based on the gas content of the interactions. Gas-poor (dry) minor mergers tend to deposit mass in the galaxy outskirts, causing the total density slope to become shallower over cosmic time \citep{Naab_2009, Hilz_Naab_Ostriker_2013}. Conversely, gas-rich (wet) mergers or accretion events can channel mass toward the central regions, maintaining or even steepening the density profile \citep{Barnes_Hernquist_1991, Hopkins_2009}. Given that galaxies may experience a combination of these processes, the resulting mass assembly history is often complex. Therefore, investigating the evolution of the total density slope since $z \sim 1$ is crucial for disentangling these mechanisms. In this context, massive early-type galaxies (ETGs) are of particular interest, as they represent the end-products of these evolutionary tracks and encapsulate the cumulative history of galaxy formation.


Cosmological simulations generally predict that the total mass density slope of massive ETGs evolves toward isothermality, converging to $\gamma_{\rm T} \approx 2$ by the present epoch ($z=0$). However, the predicted evolutionary pathways to this state differ significantly depending on the simulation suite. For instance, the IllustrisTNG simulations \citep{Pillepich_2018} suggest that the average slope decreases from $\gamma_{\rm T} \approx 2.2$ at $z=2$ to $\gamma_{\rm T} \approx 2.0$ at $z=0$, with mild evolution (i.e., $d\gamma_{\rm T} / dz \sim 0$) below $z < 1$ \citep{Wang_2019}. Similarly, the Magneticum simulations \citep{Dolag_2015} exhibit a flattening trajectory, with the slope decreasing from $\gamma_{\rm T} \approx 2.3$ to $\gamma_{\rm T} \approx 2.0$ between $z=1$ and $z=0$ (i.e., $d\gamma_{\rm T} / dz > 0$) \citep{Remus_2017}. In contrast, the Horizon-AGN simulations \citep{Peirani_2019} predict a continuous steepening of the total density slope from $z=2$ to $z=0$, albeit still converging to an isothermal profile ($\gamma_{\rm T} \approx 2.0$) at the present day (i.e., $d\gamma_{\rm T} / dz < 0$). This diversity in predicted trends highlights the profound sensitivity of the total density slope to the specific implementation of sub-grid baryonic physics, such as gas cooling, stellar winds, and active galactic nucleus (AGN) feedback \citep{Wang_2020, Mukherjee_2021}.

Observational studies in the local Universe ($z \sim 0$) consistently indicate that massive ETGs possess total density profiles that are, on average, slightly steeper than the isothermal prediction. \citet{Cappellari_2015} analyzed a sample of 14 fast rotator ETGs from the ATLAS\textsuperscript{3D} survey \citep{ATLAS_I_Cappellari_2011} and found that their density profiles are consistent with a single power-law from $R_{\rm e}/10$ to $4R_{\rm e}$. They measured an average slope of $\gamma_{\rm T} = 2.19 \pm 0.03$ with very small intrinsic scatter, which is significantly steeper than the isothermal value. Remarkably, this characteristic value of $\gamma_{\rm T} \approx 2.2$ has been confirmed with high precision by subsequent studies using various samples and data qualities, demonstrating the robustness and reproducibility of this kind of measurements. For example, analyzing a sample of 142 ETGs from the ATLAS\textsuperscript{3D} survey with stellar velocity dispersion $\lg(\sigma_e/$\kmps$) > 2.1$, \citet{Poci_2017} found a median slope of $\gamma_{\rm T} = 2.19 \pm 0.02$ with an intrinsic scatter of 0.17. Importantly, starting with \citet{Poci_2017}, subsequent studies \citep[e.g.][]{Li_2019, Dynpop_III_Zhu_2024} also revealed a clear break in $\lg(\sigma_e/$\kmps$) \approx 2.1$ below which the total slope becomes less steep. Further supporting these results, \citet{Bellstedt_2018} extended the analysis of \citet{Cappellari_2015} using similar data for 21 fast rotators out to 4$R_e$, measuring $\gamma_{\rm T}=2.25\pm0.05$. Similarly, \citet{Li_2019} utilized a sample of 2000 galaxies from the MaNGA survey \citep{MaNGA_survey_Bundy_2015} to derive a median $\gamma_{\rm T} = 2.24$ within 1$R_e$ for ETGs with $\lg(\sigma_e/$\kmps$) > 2$. More recently, using the final MaNGA sample of 10,000 galaxies, \citet[fig.~8]{Dynpop_III_Zhu_2024} measured $\gamma_{\rm T} \approx 2.18$ within 1$R_e$ for galaxies above a characteristic velocity dispersion of $\lg(\sigma_e/$\kmps$)\gtrapprox2.2$. All of these studies consistently relied on the Jeans Anisotropic Modeling (JAM) approach \citep{Jampy_Cappellari_2008, Jampy_spherical_Cappellari_2020} to constrain the total density slope by fitting observed two-dimensional stellar kinematic maps. \\


Observational studies at intermediate redshifts also present a conflicting picture compared to theoretical predictions, with different methodologies yielding divergent evolutionary trends. Most studies utilizing strong gravitational lensing indicate that the total density slopes of massive ETGs were shallower in the past and have steepened over cosmic time (i.e., $d\gamma_{\rm T} / dz < 0$). For instance, measurements of lens galaxies from the Lenses Structure and Dynamics (LSD) survey, Sloan Lens ACS Survey (SLACS), Strong Lenses in the Legacy Survey (SL2S), and BOSS Emission-Line Lens Survey (BELLS) consistently find that $\gamma_{\rm T}$ is systematically steeper at lower redshifts \citep[e.g.,][]{Koopmans_2006, Auger_2010, Ruff_2011, Bolton_2012, Sonnenfeld_2013, Li_2018}. 

Distinct from both the lensing results and the simulation predictions, recent dynamical modeling studies do not detect evolution. \citet{Derkenne_2021} and \citet{Derkenne_2023}, analyzing spatially resolved kinematics of quiescent galaxies at $z \sim 0.35$, suggest that the total density slope has remained constant relative to local baselines (i.e., $d\gamma_{\rm T} / dz \approx 0$).  In conclusion, there is potentially a three-way disagreement regarding the evolution of the total density slope in massive ETGs since $z \sim 1$: broadly, simulations predict a decrease, lensing suggests an increase, and stellar dynamics are consistent with stability. However, it is important to note that for both observational approaches, the available statistical samples of ETGs are relatively small beyond $z > 0.4$. \newadd{Therefore, their uncertainties are relatively large and insufficient to detect mild evolution if present.} \\

In this third paper of the MAGNUS (`MUSE-deep Analysis of Galaxy kinematics and dyNamical evolution across redShift') series, we aim to reach a conclusive observational answer. To meet this goal, we measure the total density slope for a substantial sample of more than 200 ETGs (hereafter the MAGNUS sample) at intermediate redshifts ($0.24 < z < 0.75$) and constrain the $\gamma_{\rm T}-z$ relation over the range of $0 < z < 1$ by comparing these data with a statistically matched local ETG sample from the MaNGA survey \citep{MaNGA_survey_Bundy_2015}. We utilize the Jeans Anisotropic Modeling (JAM) approach, a technique extensively validated in both local and intermediate-redshift studies \citep[e.g.,][]{Cappellari_2015, Poci_2017, LEGAC_van_Houdt_2021, Derkenne_2021, Derkenne_2023, Dynpop_I_Zhu_2023}. To robustly characterize the redshift evolution of $\gamma_{\rm T}$, it is essential to utilize a statistically significant sample with comparable data quality distributed evenly across the redshift baseline, analyzed via a homogeneous methodology. This work satisfies these requirements, providing a complementary perspective to previous lensing and dynamical studies. Furthermore, to quantify potential systematics, we employ six distinct dynamical model variations, exploring different assumptions regarding the mass distribution and the orientation of the velocity ellipsoid (spherical versus cylindrical alignment). Finally, we contextualize our findings by comparing our measurements with other samples available in the literature.


The paper is organized as follows. In \autoref{sec:data}, we describe the sample selection and data extraction processes for both the MAGNUS and MaNGA datasets. \autoref{sec:analysis_method} presents the theoretical framework of the Jeans Anisotropic Modeling (JAM) technique and the Multi-Gaussian Expansion (MGE) parameterization of surface brightness. This section also details the adopted dynamical models, the fitting methodology, and the measurement of the total density slope. In \autoref{sec:results}, we present the results and an analysis of systematic uncertainties for the MAGNUS and MaNGA samples. We compare our measurements with literature samples and discuss their physical implications in \autoref{sec:discussion}. Finally, we summarize our key conclusions in \autoref{sec:conclusion}. Throughout this study, we assume a flat $\Lambda$CDM cosmology with $H_0 = 70.0$ \hunit\ and $\Omega_{\rm m} = 0.3$.

\section{Data}
\label{sec:data}
 In the first paper of the MAGNUS series \citep[][hereafter MAGNUS I]{Mozumdar_MAGNUS_I}, we introduced the sample and detailed the extraction of spatially resolved stellar kinematics and morphological properties from the MUSE DEEP data cubes. In this work, we utilize this sample along with the kinematic maps and photometric data products derived in MAGNUS I. Below, we provide a brief summary of the sample selection and the relevant data products utilized for the dynamical modeling.
 
\subsection{MAGNUS sample }
MUSE is an integral-field spectrograph mounted on the Very Large Telescope (VLT) of the European Southern Observatory (ESO). In its wide-field mode, the instrument provides a spatial sampling of 0.2\as\ over a $1^\prime \times 1^\prime$ field-of-view (FOV), often assisted by an adaptive optics (AO) system \citep{MUSE_Bacon_2010}. It covers a wavelength range of $4800-9300$ \AA\ with a spectral sampling of 1.25 \AA, yielding an average spectral resolution of $R \sim 2000$. This corresponds to an instrumental dispersion of $\sigma_{\text{inst}} \sim 65$ \kmps\ at $\sim 5000$ \AA. The MUSE-DEEP data release \citep{ESO2017} combines observations from single or multiple MUSE programs targeting specific sky locations into deep IFU data cubes, achieving significant effective exposure times. Leveraging these high quality data, MAGNUS I compiled a sample of 212 ETGs from 35 data cubes within the redshift range $0.24 < z < 0.75$, featuring spatially resolved stellar kinematics extending to several effective radii. For a detailed description of the selection process, we refer the reader to MAGNUS I.

The selected data cubes cover a diverse set of fields, including five of the six Frontier Fields galaxy clusters: Abell S1063, Abell 2744, Abell 370, MACS J0416, and MACS J1149 \citep{Lotz_2017}. The sample also encompasses parts of the Hubble Ultra Deep Field \citep[HUDF;][]{Bacon_2017} and the COSMOS field \citep{Scoville_2007}. Additional well-known clusters and fields covered by the data set include Abell 1835, Abell 2895, the Bullet Cluster, MACS J0257, MACS J0940, MACS J1206, MACS J2129, MACS J2214, MS 0451-03, RX J1347–1145, RX J2129+0005, SDSS J1029+2623, and SMACS 2131. The complete list of target fields, corresponding ESO archive IDs, original MUSE program IDs, effective exposure times, number of selected galaxies per cube, and catalog references are reported in Table 1 of MAGNUS I. We note that several data cubes utilized in this study were previously analyzed by \citet{Derkenne_2021} to measure the total density slopes of intermediate-redshift ETGs. However, due to distinct selection criteria, there is only a partial overlap ($\sim 20\%$) between the two ETG samples.

The MAGNUS sample consists primarily of massive, luminous ETGs. The integrated velocity dispersion, $\sigma_\mathrm{e}$, spans a range of approximately $100-300$ \kmps\, with a median of 180 \kmps. The median surface brightness in the HST F814W band is 18.27 mag arcsec$^{-2}$, ranging from $16.62-20.34$ mag arcsec$^{-2}$. In terms of physical size, the semi-major axis, \remaj\, ranges from 0.6 kpc to $\sim 10$ kpc; the majority of the sample ($\sim 80\%$) lies between 1 and 4 kpc, with a median of 2.2 kpc. Similarly, the stellar mass, $\log_{10}(M_{\ast}/M_{\odot})$, ranges from 10.5 to 12.0, with a median value of 11.15. The coordinates, redshifts, and other derived morphological and stellar population properties for individual galaxies are available in MAGNUS I.

\subsection{Stellar kinematics and photometry}
While MAGNUS I provides a detailed description of the derivation of the 2D kinematic maps and photometry, we summarize the key data reduction steps here. The MUSE-DEEP collection consists of reduced 3D data and variance cubes processed via the standard MUSE data reduction pipeline \citep{MUSE_pipeline_Weilbacher_2020}. This pipeline\footnote{For more details, see \url{https://www.eso.org/rm/api/v1/public/releaseDescriptions/102}} corrects for instrumental signatures and sky background, then resamples and combines observations into a final data cube. Subsequently, individual data cubes for each target galaxy were extracted from the parent MUSE-DEEP cubes.

To measure spatially resolved kinematics, we first isolated the target galaxy signal from the background. Spaxels exceeding a S/N threshold were selected, where the S/N for each spaxel was calculated as the mean value within the rest-frame wavelength range of $3990-4190$ \AA. The selection threshold was determined for each galaxy, typically corresponding to a median S/N of $\sim 1$ per pixel (ranging from 0.5 to 5.5 depending on the depth of the observations). We then performed Voronoi binning on the selected spaxels using the \textsc{VorBin} Python package \citep{Voronoi_Cappellari2003}. Given the varying data quality across the sample, we adopted target S/N values empirically for each galaxy, ranging from 5 to 25 per \AA. This ensured that key absorption features remained measurable in the binned spectra while minimizing spatial over-smoothing. In general, we enforced a minimum target S/N $\geq 5$ for low-S/N galaxies to ensure the resulting Voronoi map contained at least $\sim 10$ spatial bins. For high-S/N galaxies, we utilized higher target thresholds to maximize spectral quality while limiting the total number of bins.

Using the derived Voronoi tessellation, spectra were co-added to produce a single high-S/N spectrum per bin. Stellar kinematics were extracted using the penalized pixel-fitting code \textsc{pPXF} \citep{Cappellari_2017_ppxf, LEGAC_ppxf_Cappellari_2023} within the rest-frame wavelength range of $3780-4600$ \AA. The blue-side cutoff of this range was constrained by the instrument coverage and galaxy redshift, while the red-side cutoff was imposed to avoid contamination from strong sky lines beyond 8000 \AA\ in the observed frame.

Using a high-S/N subset of the MAGNUS sample, \citet[][hereafter KM25]{Knabel_Mozumdar_2025} demonstrated that defective stellar templates—such as those suffering from flux calibration errors or inaccurate telluric corrections—can introduce significant bias into kinematic measurements. Furthermore, independent analyses by both KM25 and \citet{Mozumdar_Knabel_2025} identified template mismatch as the dominant source of systematic uncertainty in the stellar kinematics of ETGs. To mitigate these effects, we employed three distinct stellar template libraries: Indo-US \citep{Indo_US_lib}, MILES \citep{MILES_library_2011}, and the X-shooter Spectral Library \citep[XSL;][]{Verro_2022}. For each library, we restricted our template selection to the `clean' subsets identified by KM25, resulting in final libraries containing 989, 789, and 462 templates for Indo-US, MILES, and XSL, respectively.

To minimize bin-to-bin scatter, we did not fit the full `clean' library to every spatial bin. Instead, for each galaxy and each library, we constructed an optimal template subset. This subset was generated by fitting the high-S/N central spectrum of the galaxy using the full clean library and retaining only those templates assigned non-zero weights in the fit. During the kinematic fitting process, we included additive and multiplicative Legendre polynomials of degree 4 and 2, respectively, to correct for the continuum shape. Following the methodology of KM25, the final mean velocity, $\overline{v}$, and velocity dispersion, $\overline{\sigma_{\rm{v}}}$, along with the associated systematic ($\Delta \overline{\sigma_{\rm{v}}}$) and statistical ($\delta \overline{\sigma_{\rm{v}}}$) uncertainties, were defined as the equally weighted average of the measurements obtained from the three independent libraries. In this work, we employ both the spatially resolved kinematic maps and the integrated velocity dispersions ($\sigma_e$) extracted via this process. Visualizations of these kinematic maps and the corresponding HST image cutouts describe below are available in MAGNUS I.

Individual galaxy cutouts, typically spanning dimensions of $10-15$\as, were extracted from HST imaging covering the same fields as the MUSE-DEEP data cubes. We primarily utilized images taken with the F814W filter; in cases where F814W data were unavailable, we substituted observations in the F555W or F606W bands. A detailed summary of the archival HST data, including instrument setups, filters, and exposure times, is provided in Table 2 of MAGNUS I. For each cutout, the background level was estimated and subtracted. These sky-subtracted images served as inputs to model the surface brightness distribution as a series of 2D Gaussian profiles (see \autoref{sec:mge_model}), implemented using the \textsc{MgeFit} Python package \citep{MGE_Cappellari_2002}. \newadd{For each galaxy, we utilized the local FWHM PSF estimates provided in the datacube headers rather than a single global value to account for the complex construction of the MUSE-DEEP mosaic. Since the final datacube is a co-addition of observations spanning multiple programs, epochs, and atmospheric conditions, the effective seeing varies significantly across the field of view. The list of the FWHM PSF used for the MAGNUS sample are available in \citet[][hereafter MAGNUS II]{Mozumdar_2025_MAGNUS_II}. } 

Previous studies have established that the total density slope strongly correlates with the integrated velocity dispersion ($\sigma_e$) below a specific threshold while the correlation is mild above this threshold \citep[e.g.,][]{Poci_2017, Li_2019, DynPop_VI_Li_2024}. The observed trends suggest the \tslope\ flattens as $\sigma_e$ decreases. \citet{Poci_2017} and \citet{Li_2019} identified this transition at approximately 100 \kmps\ using the ATLAS\textsuperscript{3D} and MaNGA samples, respectively. However, using the latest MaNGA data release, \citet{DynPop_VI_Li_2024} recently suggested that for ETGs, this threshold may be closer to 150 \kmps. To minimize $\sigma_e$ as a confounding factor in the evolutionary analysis while maximizing sample statistics, we adopted a conservative lower limit of $\sigma_e \geq 100$ \kmps. This selection criterion reduces the final MAGNUS sample to 198 ETGs. We also explored the implications of a more stringent constraint ($\sigma_e > 150$ \kmps); while this stricter cut reduces the MAGNUS sample size by approximately 20\%, the qualitative results regarding the density slope evolution remain unchanged (see \autoref{sec:high_vd_cut}).

\subsection{Comparison sample: MaNGA}
\label{sec:manga_sample}
The MaNGA survey provides spatially resolved spectral measurements for approximately 10,000 nearby galaxies in the redshift range $0.01 < z < 0.15$ \citep{MaNGA_survey_Bundy_2015}. The observations cover a spatial extent of up to 2.5 effective radii ($R_e$) \citep{MaNGA_observation_Wake_2017}. The spectroscopic data span a wavelength range of $\lambda = 3600-10000$ \AA\ with a spectral resolution of $R \approx 2000$, corresponding to an instrumental dispersion $\sigma_{\text{inst}} \approx 72$ \kmps\ \citep{MaNGA_DRP_Law_2016}. The final data cubes were generated from spectrophotometrically calibrated raw data using the MaNGA Data Reduction Pipeline \citep[DRP;][]{MaNGA_DRP_Law_2016}. To derive stellar kinematic maps, the data cubes were Voronoi-binned to a target S/N of 10 and fitted using \textsc{pPXF}. These steps were performed using the official MaNGA Data Analysis Pipeline \citep[DAP;][]{MaNGA_DAP_Westfall_2019}.

\citet[][hereafter DynPop I]{Dynpop_I_Zhu_2023} conducted detailed dynamical modeling of these high-quality resolved kinematics by solving the axisymmetric Jeans equations using \textsc{JamPy} Python package \citep{Jampy_Cappellari_2008, Jampy_spherical_Cappellari_2020}. Their analysis employed four distinct mass models and two orientations of the velocity ellipsoid (cylindrically and spherically aligned), assuming a constant anisotropy in the ratio of radial-to-tangential velocity dispersion. Model parameters were constrained by comparing the observed second velocity moments, $V_{\rm rms}$, to model predictions within a Bayesian framework. The JAM approach requires the surface brightness distribution of the galaxy to serve as the tracer density $\zeta$ and  DynPop I utilized SDSS $r$-band images to model this distribution via the MGE formalism \citep{MGE_Emsellem_1994, MGE_Cappellari_2002}. For each model configuration, the mass-weighted total density slope was derived from the best-fit parameters.

In this work, we adopt an identical methodology for the MAGNUS sample, employing the MGE formalism to model the surface brightness (see \autoref{sec:mge_model}) and \textsc{JamPy} for modeling the observed $V_{\rm rms}$ maps (see \autoref{sec:jam_method}). We utilize the same mass model parameterizations and velocity ellipsoid orientations,  and measure \tslope\, as the mass-weighted average within one effective radius ($1 R_e$), identical to the definition used in DynPop I. Given that the data reduction, analysis pipelines, and derivation techniques are consistent between the two datasets, we adopt the dynamical quantities reported by DynPop I and utilize the MaNGA dataset as our primary comparison sample from the local Universe. We retrieved the relevant galaxy properties from the DynPop I catalog\footnote{\href{https://github.com/manga-dynpop/manga-dynpop.github.io/tree/main/catalogs/JAM}{https://github.com/manga-dynpop/manga-dynpop.github.io/tree/main/catalogs/JAM}}, including redshifts, JAM model quality flags, integrated velocity dispersions ($\sigma_e$), and the mass-weighted total density slopes (\tslope) for the four model variants utilized in this study. Additionally, we obtained stellar mass measurements ($M_{\ast}$) for the MaNGA sample from \citet{DynPop_II_Lu_2023}. These masses were derived by fitting FSPS \citep{FSPS_Conroy_2009, FSPS_Conroy_2010} stellar population templates, generated with a Salpeter Initial Mass Function \citep[IMF;][]{Salpeter_1955}, to the integrated spectra within one effective radius using \textsc{pPXF}. The stellar masses of the MAGNUS sample were also measured in a similar fashion (for details check MAGNUS I).

The full MaNGA survey includes both early-type (e.g., E and S0) and late-type (e.g., spiral) galaxies. Since the MAGNUS sample is comprised exclusively of massive ETGs, we restricted the MaNGA comparison sample to ETGs only. The morphological selection criteria for these ETGs are detailed in MAGNUS II. To ensure a fair comparison, we required that the ETGs of the comparison sample have $\log(M_{\ast}/M_{\odot}) \geq 10.5$ and $\sigma_e \geq 100$ \kmps. These criteria correspond to the lower bounds of the MAGNUS sample for the respective properties. Furthermore, following the recommendations of DynPop I, we excluded galaxies with a quality flag (Qual) $< 1$, as their derived dynamical quantities are considered unreliable due to poor data quality or model fits. These criteria yielded a final sample of 1,794 ETGs from the original 3,294 ETGs in the full MaNGA catalog. The median redshift of this MaNGA sample is $z \sim 0.06$. This MaNGA ETG sample effectively populates the epoch $z < 0.1$, serving as a complementary anchor to the MAGNUS data. Together, they provide an excellent opportunity to precisely constrain the evolution of \tslope\ since $z < 1$, leveraging a homogeneously analyzed dataset across cosmic time.

We note that our comparison is done at fixed stellar and mass and stellar velocity dispersion. This is not the same as measuring changes of the slope along the evolutionary path of an individual massive galaxy, which is expected to grow slowly in stellar mass over time, and very slowly in velocity dispersion  \citep[e.g.][]{Nipoti_2012}. Following the evolutionary path of an individual galaxy requires modeling physical phenomena and additional data \citep{Sonnenfeld_2013,Sonnenfeld_2015}. This is beyond the scope of this paper, and left for future work.

\begin{figure*}
    \centering
    {\includegraphics[width=0.9\textwidth]{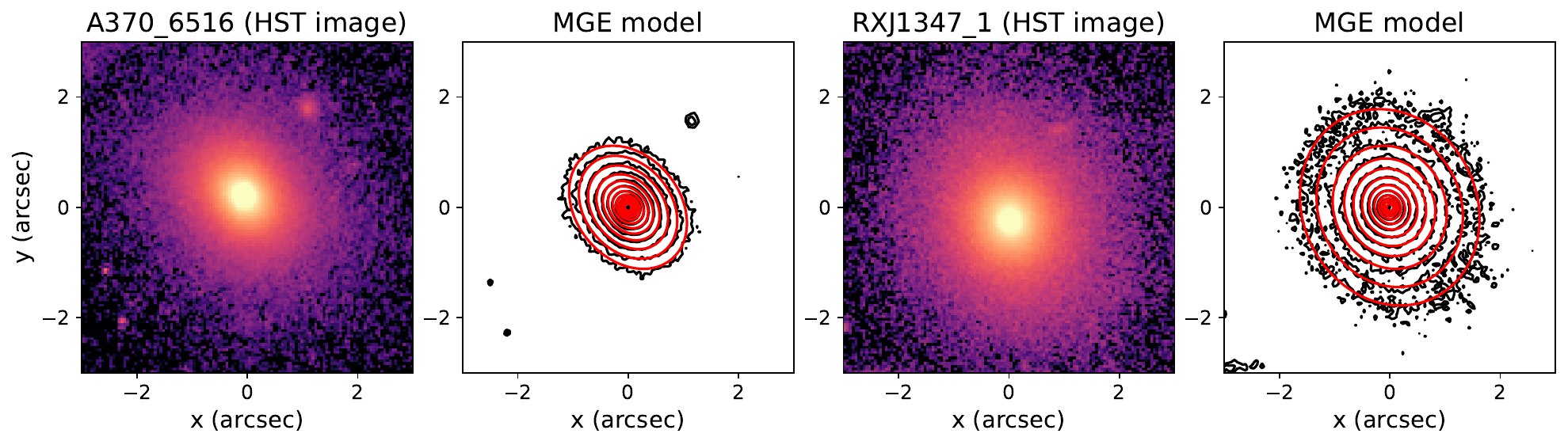}}
    \caption{Representative MGE surface brightness models for the MAGNUS sample. The first and third panel display the sky-subtracted HST images. The second and fourth panel show the observed surface brightness isophotes (black contours) overlaid with the best-fit MGE model (red contours). Galaxy IDs are indicated above the corresponding HST images. \newadd{The MGE models for the remaining galaxies in the sample are provided in Appendix \ref{sec:all_galaxy_model}}. 
    }
\label{fig:mge_example}
\end{figure*}

\section{Analysis Method}
\label{sec:analysis_method}
In this section, we present the dynamical modeling framework used to constrain the total mass distributions of the MAGNUS sample. We base our analysis on the JAM technique, which predicts the galaxy's spatially resolved second velocity moments ($V_{\rm rms}$) by solving the axisymmetric Jeans equations. We first outline the mathematical formalism of these equations in \autoref{sec:jam_method}. Since this approach requires an analytical parameterization of the stellar tracer density, we subsequently describe the MGE modeling of the observed surface brightness in \autoref{sec:mge_model}. Finally, we detail the adopted mass models and anisotropy profiles (\autoref{sec:mass_models}), the Bayesian parameter estimation strategy (\autoref{sec:fitting}), and the methodology for deriving the mass-weighted total density slope (\autoref{sec:measure_dyn_mass_slope}).

\subsection{Jeans anisotropic modelling}
\label{sec:jam_method}
The orbital dynamics of a steady-state, collisionless stellar system are governed by the collisionless Boltzmann equation \citep[CBE;][Eq. 4-13b]{Binney_Tremaine_1987}. This fundamental equation describes the evolution of the phase-space distribution function, $f(\boldsymbol{x}, \boldsymbol{v})$, within a total gravitational potential, $\Phi$, and is expressed as:

\begin{equation}
    \sum_{i=1}^3\left(v_i \frac{\partial f}{\partial x_i}-\frac{\partial \Phi}{\partial x_i} \frac{\partial f}{\partial v_i}\right)=0
\end{equation}

However, without further assumptions and simplifications, we can not practically use this equation. Assuming an axisymmetric stellar system (i.e., $\partial \Phi / \partial \phi=\partial f / \partial \phi=0$), we obtain the following Jeans equations in cylindrical coordinates, $(R, z, \phi)$ \citep[][eq. 4-29 a,c]{Binney_Tremaine_1987} as
\begin{equation}
    \begin{aligned}
        \frac{\zeta (\overline{v_R^2}- \overline{v_\phi^2})}{R}+\frac{\partial\left(\zeta \overline{v_R^2}\right)}{\partial R}+\frac{\partial\left(\zeta\ \overline{v_R v_z}\right)}{\partial z} &=-\zeta \frac{\partial \Phi}{\partial R}, \\
\frac{\zeta\ \overline{v_R v_z}}{R}+\frac{\partial\left(\zeta\ \overline{v_z^2}\right)}{\partial z}+\frac{\partial\left(\zeta\ \overline{v_R v_z}\right)}{\partial R} &=-\zeta \frac{\partial \Phi}{\partial z},
    \end{aligned}
\end{equation}

where $\zeta$ is the number density 
of the tracer population from which one measures the kinematics such that 

\begin{equation}
    \zeta\ \overline{v_k v_j} \equiv \int v_k v_j f \mathrm{~d}^3 v 
\end{equation}

This set of equations is general, relying only on the assumptions of steady-state equilibrium and axial symmetry. However, because the system is not closed—containing more unknowns than independent equations—additional constraints on the velocity ellipsoid are required to obtain a unique solution. Following \citet{Jampy_Cappellari_2008}, we assume that the velocity ellipsoid is aligned with the cylindrical coordinate system $(R, z, \phi)$, which implies that the off-diagonal moment vanishes (i.e., $\overline{v_R v_z} = 0$). Furthermore, we assume that the velocity anisotropy in the meridional plane, $\beta_z$, is constant, defined as:

\begin{equation}
    \beta_z \equiv 1-\frac{\overline{v_z^2}}{\overline{v_R^2}}=1-\frac{\sigma_z^2}{\sigma_R^2}=1-\frac{1}{b}
\end{equation}
then Equation 2 reduces to \citep[][eq. 8 and 9]{Jampy_Cappellari_2008}

\begin{equation} 
\label{equ:cyl}
    \begin{aligned}
        \frac{ {\zeta\ (b \overline{v_z^2}}-\overline{v_\phi^2})}{R}+\frac{\partial\left(b \zeta\ \overline{v_z^2}\right)}{\partial R} &=-\zeta\ \frac{\partial \Phi}{\partial R} \\
\frac{\partial\left(\zeta\ \overline{v_z^2}\right)}{\partial z} &=-\zeta\ \frac{\partial \Phi}{\partial z},
    \end{aligned}
\end{equation}

This equation set can be solved to yield unique solutions for $\overline{v_z^2}$ and $\overline{v_\phi^2}$ under the boundary condition $\zeta\ \overline{v_z^2}=0$ as $z \rightarrow \infty$. \\

Similarly, if one assumes that the velocity ellipsoid is aligned with the spherical coordinates $(r, \theta, \phi)$ and the velocity anisotropy is defined as
\begin{equation}
    \beta_r \equiv 1-\frac{\overline{v_\theta^2}}{\overline{v_r^2}}=1-\frac{\sigma_\theta^2}{\sigma_r^2}
\end{equation}

then Jeans equations transform to the following equations \citep[][eq. 7]{Jampy_spherical_Cappellari_2020}

\begin{equation}
\label{equ:sph}
    \begin{aligned}
        \frac{\partial\left(\zeta\ \overline{v_r^2}\right)}{\partial r}+\frac{\left(1+\beta_r\right) \zeta ( \overline{v_r^2}- \overline{v_\phi^2})}{r}=-\zeta\ \frac{\partial \Phi}{\partial r} \\
\left(1-\beta_r\right) \frac{\partial\left(\zeta\ \overline{v_r^2}\right)}{\partial \theta}+\frac{\left(1-\beta_r\right) \zeta\ (\overline{v_r^2}- \overline{v_\phi^2})}{\tan \theta}=-\zeta\ \frac{\partial \Phi}{\partial \theta}
    \end{aligned}
\end{equation}
As before these equations could be solved for unique solutions of $\overline{v_r^2}$ and $\overline{v_{\phi}^2}$ under the boundary condition $\zeta \overline{v_r^2} = 0$ as r $\to \infty$.

Consequently, for a given gravitational potential $\Phi$, tracer density distribution $\zeta$, and velocity anisotropy ($\beta_z$ or $\beta_r$ for cylindrically or spherically aligned ellipsoids, respectively), the intrinsic second velocity moments---either $(\overline{v_z^2}, \overline{v_{\phi}^2})$ or $(\overline{v_r^2}, \overline{v_{\phi}^2})$---can be derived by solving respective equations. In practice, we adopt a specific mass density profile and anisotropy model corresponding to the chosen velocity ellipsoid orientation, while the tracer density is directly constrained by the galaxy's observed surface brightness. The projected line-of-sight (LOS) second velocity moment is then calculated as:
\begin{equation}
    \left\langle v_{\text {los }}^2\right\rangle S(x, y)=\int_{-\infty}^{\infty} \mathrm{d} z \zeta\left\langle v_z^2\right\rangle.
\end{equation}
where $S(x, y)$ is the surface density of the luminous dynamical tracer. The luminosity-weighted observed LOS velocity dispersion is given by 
\begin{equation}
    \left[\left\langle v_{\text {los }}^2\right\rangle\right]_{\text {obs }}=\frac{\int_{\text {ap }} \mathrm{d} x \mathrm{~d} y I\left\langle v_{\text {los }}^2\right\rangle \otimes \mathrm{PSF}}{\int_{\text {ap }} \mathrm{d} x \mathrm{~d} y I \otimes \mathrm{PSF}},
\end{equation}

where the symbol ``$\otimes$ PSF" denotes a convolution with the PSF. The surface brightness profile, $I(x, y)$, of the dynamical tracer works as a substitute for the surface density $S(x, y)$,  since they are related by a constant mass-to-light ratio that cancels out in this expression.

The model-predicted $\langle v_{\text{los}}^2 \rangle$ is directly comparable to the observed root-mean-square velocity, defined as $V_{\text{rms}} \equiv \sqrt{V^2 + \sigma_{\rm v}^2}$, where $V$ and $\sigma_{\rm v}$ represent the observed stellar velocity and velocity dispersion, respectively. By optimizing the agreement between the predicted and observed $V_{\text{rms}}$ maps, we constrain the model parameters and derive the total density slope, \tslope. To perform this analysis, we utilize the \textsc{JamPy} package, applying three distinct mass density models and assuming a spatially constant velocity anisotropy. To ensure the robustness of our derived quantities against structural assumptions, we adopt both cylindrically and spherically aligned velocity ellipsoid configurations, as the \textsc{JamPy} framework accommodates both geometries. A fundamental prerequisite for this technique is an analytical description of the stellar tracer density, which serves as the input for the Jeans equations. To achieve this, we model the galaxy surface brightness using the MGE formalism, which we detail in the following section.

\begin{deluxetable*}{l c c c c c c}
    \tablecaption{\label{table:dyn_model}Summary of the dynamical model parameterizations employed in this study. For each model configuration, we list the free parameters and the corresponding prior distributions or boundary constraints applied during the Bayesian inference. The notations \jamcyl\ and \jamsph\ denote models assuming cylindrically and spherically aligned velocity ellipsoids, respectively.}
    \tablehead{\colhead{Model} & \multicolumn{6}{c}{ Fitted parameters with prior} \\ 
               \colhead{} & \colhead{$q_{\text {min }}$} & \colhead{$\frac{\sigma_z}{\sigma_R} $ or $\frac{\sigma_\theta}{\sigma_r}$} & \colhead{log $\left[\frac{M}{L}\right] \left[\frac{\textrm{M}_{\odot}} {\textrm{L}_{\odot}}\right]$} & \colhead{$\rho_s$} & \colhead{$f_{\textrm{DM}}$} & \colhead{$\gamma$ }}

    \startdata
\jamcyl: power-law  & {$\left[0.05, q_{\min }^{\prime}\right]$} & $\mathcal{N}(1.0, 0.07), {\frac{\sigma_z}{\sigma_R} \le 1}$ & - & [0, 4] & - & [1.5, 2.5] \\
\jamsph: power-law & {$\left[0.05, q_{\min }^{\prime}\right]$} & $\mathcal{N}(1.0, 0.07)$ & - & [0, 4] & - & [1.5, 2.5]  \\
\jamcyl: Stars + NFW & {$\left[0.05, q_{\min }^{\prime}\right]$} & $\mathcal{N}(1.0, 0.07), {\frac{\sigma_z}{\sigma_R} \le 1}$ & $[-2,2]$ & - & $[0, 1]$ & - \\
\jamsph: Stars + NFW & {$\left[0.05, q_{\min }^{\prime}\right]$} & $\mathcal{N}(1.0, 0.07)$ & $[-2,2]$ & - & $[0,1]$ &  - \\
\jamcyl: Stars + gNFW & {$\left[0.05, q_{\min }^{\prime}\right]$} & $\mathcal{N}(1.0, 0.07), {\frac{\sigma_z}{\sigma_R} \le 1}$ & $[-2,2]$ & - & $[0,1]$ & $[0, 1.6]$\\
\jamsph: Stars + gNFW & {$\left[0.05, q_{\min }^{\prime}\right]$} & $\mathcal{N}(1.0, 0.07)$ & $[-2,2]$ & - & $[0,1]$ & $[0, 1.6]$\\
\hline
\enddata
\end{deluxetable*}

\subsection{Multi-Gaussian expansion}
\label{sec:mge_model}
We model the observed surface brightness distribution of each galaxy by performing a MGE fit to the high-resolution HST imaging. In this formalism, the surface brightness $\Sigma(x', y')$ is parameterized as a sum of $N$ two-dimensional Gaussian components:
\begin{equation}
    \Sigma\left(x^{\prime}, y^{\prime}\right)=\sum_{k=1}^N \frac{L_k}{2 \pi \sigma_k^2 q_k^{\prime}} \exp \left[-\frac{1}{2 \sigma_k^2}\left(x^{\prime 2}+\frac{y^{\prime 2}}{q_k^{\prime 2}}\right)\right]
\end{equation}
where $L_k$, $\sigma_k$, and $q_k^{\prime}$ represent the total luminosity, dispersion, and observed axial ratio of the $k$-th Gaussian component, respectively, and $(x', y')$ define the coordinates on the sky plane. A primary advantage of this parameterization is its analytical tractability: operations such as convolution with a PSF and deprojection can be performed independently on each Gaussian component and linearly combined. While the deprojection of a 2D surface brightness profile into a 3D intrinsic density is mathematically non-unique, a unique solution can be obtained by imposing geometric constraints. Under the standard assumption of oblate axisymmetry, the 2D MGE models can be analytically deprojected. In cylindrical coordinates $(R, z)$, the intrinsic luminosity density is given by \citep[][Eq. 8]{Jampy_Cappellari_2008}:

\begin{equation}
    v(R, z)=\sum_{k=1}^N \frac{L_k}{\left(\sqrt{2 \pi} \sigma_k\right)^3 q_k} \exp \left[-\frac{1}{2 \sigma_k^2}\left(R^2+\frac{z^2}{q_k^2}\right)\right]
\end{equation}

where the luminosity $L_k$ and dispersion $\sigma_k$ are conserved from the 2D MGE, while $q_k$ represents the intrinsic 3D axial ratio of the $k$-th Gaussian component, defined as:
\begin{equation}
    q_k=\frac{\sqrt{q_k^{\prime 2}-\cos ^2 i}}{\sin i}.
\end{equation}

This deprojection depends on the galaxy inclination $i$ (where $i=90^{\circ}$ corresponds to an edge-on orientation). For the intrinsic model to be physically valid, the argument of the square root must be non-negative for all components. Consequently, the inclination is constrained by the flattest observed component, $q_{\min }^{\prime}=\min \left(q_k^{\prime}\right)$, and the assumed intrinsic minimum axial ratio, $q_{\min }=\min \left(q_k\right)$. The inclination is related to these extrema via:
\begin{equation}
    \tan^2 i=\frac{1-q^{\prime 2}_{\min }}{q^{\prime 2}_{\min }-q_{\min }^2}.
\end{equation}

We generated the MGE models using the \textsc{MgeFit} Python package. First, we determined the galaxy centroid, position angle (PA), and ellipticity using the \textsc{find\_galaxy} routine. We then adapted our modeling approach based on the galaxy's morphology. For galaxies without isophotal twists, we sampled the surface brightness using \textsc{sectors\_photometry} and performed a regularized fit using the \textsc{mge\_fit\_sectors\_regularized} routine. This sector-based sampling effectively mitigates local irregularities, such as dust lanes, while preserving the global isophotal structure and the regularization minimizes the number of Gaussian components and suppresses unphysical oscillations in the axial ratio distribution. Conversely, for galaxies exhibiting isophotal twists, we utilized \textsc{sectors\_photometry\_twist} followed by the \textsc{mge\_fit\_sectors\_twist} routine to explicitly model the position angle variations. In all cases, we restricted the measurements to regions where the flux exceeded $2\sigma$ above the background. The intrinsic MGE models were analytically convolved with the PSF, derived from a non-saturated star in the same field, and optimized to match the observed surface brightness. \newadd{Figure \ref{fig:mge_example} shows the HST images for two representative galaxies from the sample, comparing the observed surface brightness contours with the best-fitting MGE model reconstructions. The MGE models for the remaining galaxies in the sample are provided in Appendix \ref{sec:all_galaxy_model}. }\\

While the absolute scaling of the luminous tracer is not strictly required to measure the mass-weighted slope of the total density profile, it is essential for constraining dynamical mass-to-light ratios ($M/L$). We normalized the total counts ($C_k$) of each Gaussian component to a peak surface brightness ($I_k$) via:
\begin{equation*}
    I_k = \frac{C_k}{2\pi \sigma_k^2 q_k}.
\end{equation*}
Simultaneously, the dispersion $\sigma_k$ was converted from pixels to arcseconds using the instrument pixel scale. To recover the intrinsic luminosity density, we converted the MGE surface brightness profiles to AB magnitudes and corrected for Galactic extinction. We then applied a scaling factor of $(1 + z)^3$ to account for cosmological surface brightness dimming and the bandwidth effects inherent to the AB system. \newadd{Finally, to express the luminosity in solar units ($L_{\odot}$), we performed a K-correction \citep{K-correction_Hogg_2002}. We determined the reference solar magnitude for the specific HST bandpass used to observe the galaxy (typically F814W) using the \textsc{sps\_util.solar\_mag} routine from the \textsc{pPXF} package. This function computes the magnitude of the Sun as observed through a given filter at the specific redshift of the galaxy.}

\begin{figure}[t]
\centering
\includegraphics[width=\columnwidth]{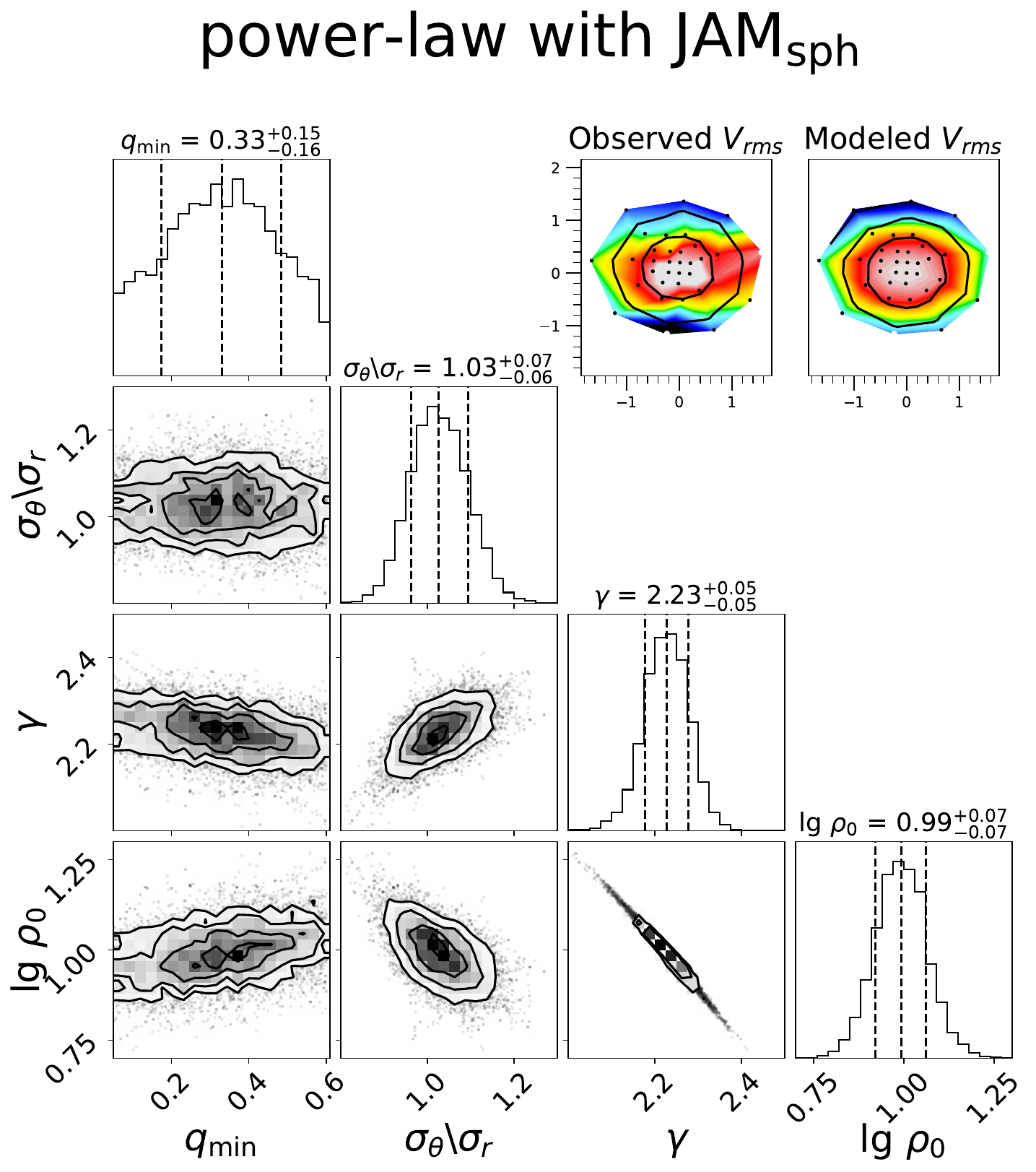}
\caption{Posterior probability distributions for a representative galaxy (ID =`A2744\_4423') derived using the Power-Law mass model with
spherical velocity ellipsoid alignment (\jamsph). The diagonal subplots display the marginalized 1D distributions, where vertical dashed lines indicate the median and the 16th/84th percentiles. The off-diagonal subplots show the covariances between model parameters. In the top-right corner of the plot we display the observed root-mean-square velocity field ($V_{\rm rms}$) compared with the prediction from the respective best-fit model. The stars+NFW and stars+gNFW parameterizations for the same galaxy are shown in Figure~\ref{fig:dymd_mcmc_example_2}. The observed and best-fit $V_{\rm rms}$ maps for the entire sample are presented in Appendix~\ref{sec:all_galaxy_model}.}
\label{fig:dymd_mcmc_example_1}
\end{figure}

\begin{figure}
\centering
\includegraphics[width=\columnwidth]{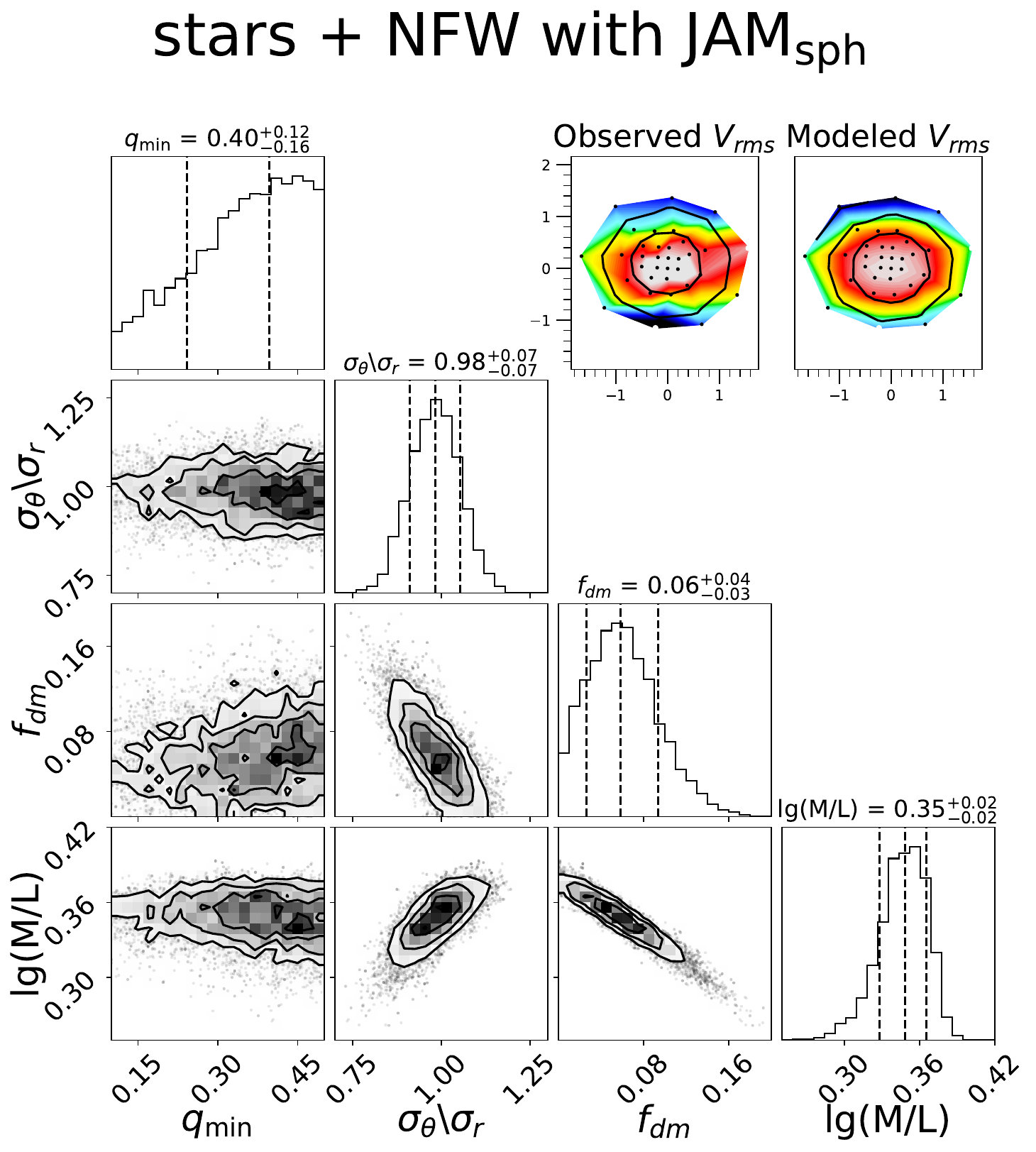}
\\[1em]
\includegraphics[width=\columnwidth]{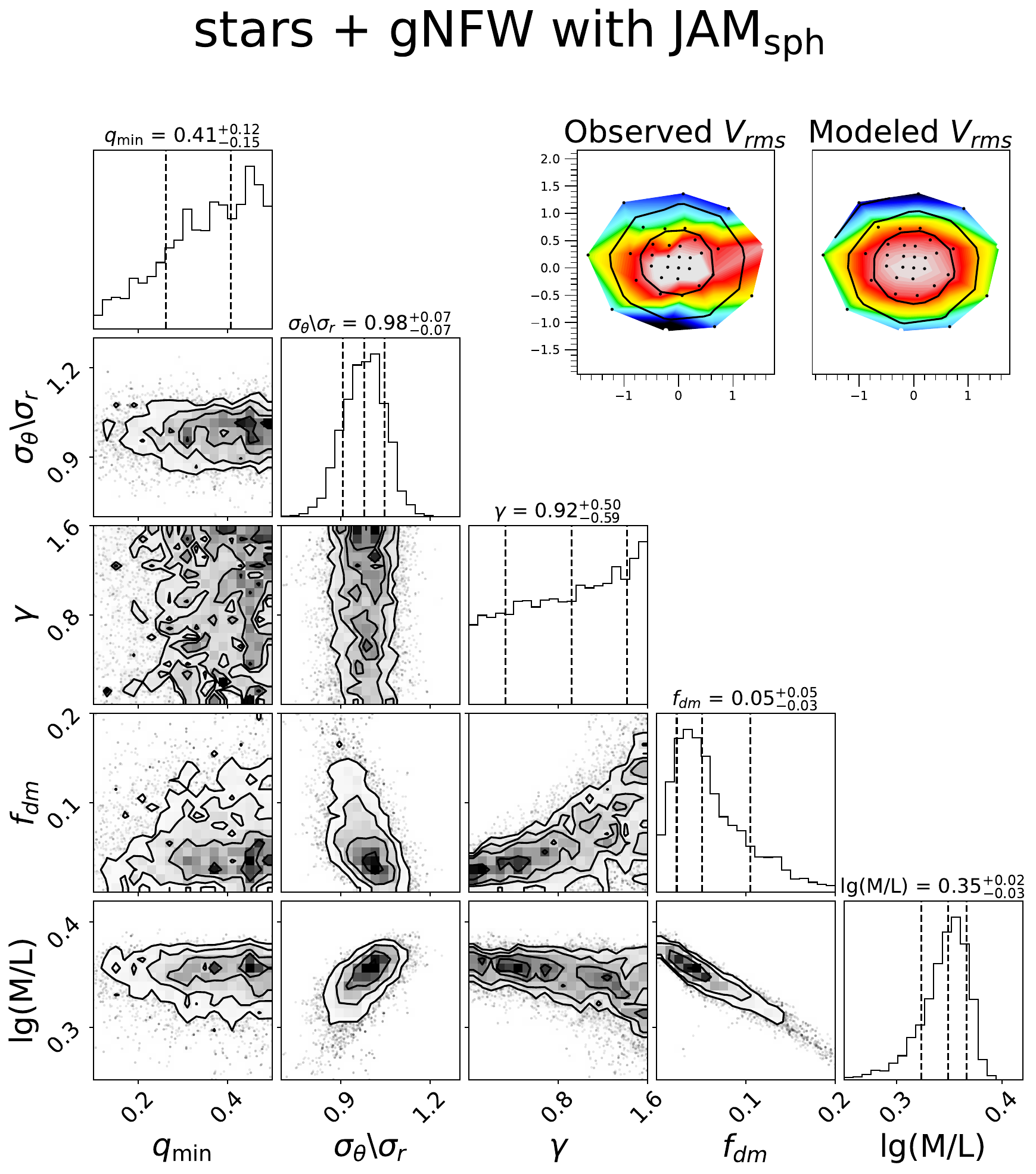}
\caption{Same as Figure~\ref{fig:dymd_mcmc_example_1}, but for the stars+NFW (top panel) and stars+gNFW (bottom panel) mass models, both with
spherical velocity ellipsoid alignment (\jamsph), for the same representative galaxy (ID = `A2744\_4423').}
\label{fig:dymd_mcmc_example_2}
\end{figure}

\subsection{Mass models and kinematic anisotropy}
\label{sec:mass_models} 
\subsubsection{Constant anisotropy}
\label{sec:constant_anisotropy}
We adopted a spatially constant velocity anisotropy for both the cylindrically and spherically aligned velocity ellipsoid configurations, hereafter denoted as \jamcyl\ and \jamsph, respectively. The anisotropy parameter, $\beta$ is parameterized by the ratio of the principal velocity dispersion components: $ \sigma_z/\sigma_R$ for the \jamcyl\ case and $\sigma_{\theta}/\sigma_r$ for the \jamsph\ case.

In our primary analysis, we imposed a Gaussian prior on these ratios, centered at isotropy: $\mathcal{N}(1.0, 0.07)$. For the \jamcyl\ configuration, we additionally enforced an upper bound of unity ($\sigma_z/\sigma_R \leq 1$), as standard practice to avoid the inclination-anisotropy degeneracy of these models \citep[sec.~4.3]{Jampy_Cappellari_2008}, effectively adopting a half-Gaussian prior. This choice is motivated by two key factors. 
First, the intrinsic degeneracy between mass and anisotropy can lead to unphysical solutions when using uninformative (flat) priors, often resulting in models with negligible dark matter content compensated by extreme tangential anisotropy. Since the kinematic data alone are insufficient to break this degeneracy, an informed prior is required to exclude these non-physical regions of parameter space. Second, this prior is empirically grounded in high-resolution dynamical studies of local galaxies. Using Schwarzschild orbit-superposition models for a diverse sample of nearby fast- and slow-rotator ETGs, \citet{Cappellari_review_2026} demonstrated that the radial anisotropy profiles are nearly isotropic ($\sigma_r/\sigma_{\rm tang} \approx 1$). Accordingly, we centered our prior on isotropy (mean $=$ 1.0). The adopted standard deviation of $0.07$ corresponds to the scatter of the mean anisotropy $\sigma_r/\sigma_{\rm tang}$ of the 13 galaxies in \citet[fig.~10]{Cappellari_review_2026}, each first independently averaged over the range $R_e/30 < r < R_e$.

For comparison, we also tested a flat prior configuration following the approach of DynPop I, where the anisotropy ratios were bounded by $\mathcal{R}(q) < \sigma_z / \sigma_R < 1$ and $\mathcal{R}(q) < \sigma_\theta / \sigma_r < 2$, with the lower bound defined as $\mathcal{R}(q)=\sqrt{0.3+0.7 q}$ (where $q$ is the intrinsic axial ratio). We found that while the choice of prior influences the specific best-fit anisotropy and mass parameters, it has a negligible impact on the derived total density slope. In Appendix \ref{sec:anisotropy_prior_slp}, we demonstrate that the \tslope\ values measured using these two different anisotropy priors are statistically consistent.


\subsubsection{Power-law Total-density model}
In this model, the total mass density distribution, $\rho(r)$, is parameterized as a single power-law profile of the form:
\begin{equation*}
    \rho (r)=\rho_s\left(\frac{r}{r_s}\right)^{-\gamma},
\end{equation*}
where $r$ is the spherical radius, $\gamma$ is the constant logarithmic slope of the density profile, \newadd{and $r_s$ serves as a fixed reference radius determining the normalization density $\rho_s$}. This parameterization effectively captures the combined gravitational potential of the baryon-dominated center and the dark matter-dominated halo. Following previous dynamical studies \citep[e.g.,][]{ATLAS_XV_Michele_2013, Bellstedt_2018, Derkenne_2021}, we fixed the \newadd{reference} radius to $r_s = 20$ kpc, as this parameter has a negligible impact on the dynamical analysis. Consequently, this model comprises four free parameters: (1) $\frac{\sigma_z}{\sigma_R}$ for \jamcyl\ case (or $\frac{\sigma_\theta}{\sigma_r}$ for \jamsph\ ), (2) the intrinsic axial ratio ($q_{\min}$)  of the flattest gaussian in the MGE, (3) the total density slope ($\gamma$), and (4) the density normalization ($\rho_s$).

For this and all subsequent models, the analytic 1D mass density profiles were converted into gravitational potentials by fitting with an MGE using the \textsc{mge\_fit\_1d} procedure in the \textsc{MgeFit} package, enabling computation of the model-predicted second velocity moments via the JAM framework.

\subsubsection{Composite model: Stars with NFW dark halo}

This composite model offers greater flexibility than the single power-law profile by explicitly decomposing the total mass distribution into stellar and dark matter components. \newadd{The intrinsic stellar mass density was derived by scaling the MGE luminosity density with a constant dynamical mass-to-light ratio, $\rm {M/L}$, thereby assuming that the stellar mass distribution strictly traces the light with no radial gradients in the stellar population.} The dark matter component is parameterized as a spherical halo following the Navarro-Frenk-White (NFW) profile \citep{NFW_profile_1996}, defined as:
\begin{equation*}
    \rho_{\mathrm{DM}}(r)=\rho_s \left(\frac{r}{r_s}\right)^{-1}\left(\frac{1}{2}+\frac{1}{2} \frac{r}{r_s}\right)^{-2},
\end{equation*}
where $r$ denotes the spherical radius. In this specific parameterization, $\rho_s$ represents the characteristic density at the break radius, $r_s$ (i.e., $\rho_{\mathrm{DM}}(r_s)=\rho_s$). \newadd{Following previous dynamical modeling studies of massive ETGs, we fixed the break radius to $r_s = 20$ kpc \citep[e.g.][]{Cappellari_2015, Poci_2017, Bellstedt_2018, Derkenne_2021, Derkenne_2023}. Systematic tests by \citet{Bellstedt_2018} demonstrated that leaving the break radius as a free parameter yields total density slopes consistent with those derived using a fixed 20 kpc value. Furthermore, \citet{Derkenne_2023} showed that while adopting a significantly smaller break radius ($\sim 10$ kpc) can artificially bias the results toward steeper slopes, increasing $r_s$ beyond 20 kpc has no significant effect on the derived parameters.} 


While the halo normalization is formally set by $\rho_s$, this parameter is degenerate with the stellar mass. To facilitate robust parameter estimation, we re-parameterize the normalization using the dark matter fraction, $f_{\rm DM}$ (calculated within the effective radius). This approach allows us to impose precise physical bounds ($0 < f_{\rm DM} < 1$) and sample the dark matter content more intuitively. The reduced covariance also allow for a more efficient sampling of the model posterior during the MCMC. Consequently, this model comprises four free parameters: (1) $\frac{\sigma_z}{\sigma_R}$ for \jamcyl\ case (or $\frac{\sigma_\theta}{\sigma_r}$ for \jamsph\ ), (2) the minimum intrinsic axial ratio ($q_{\min}$), (3) the stellar mass-to-light ratio ($M/L$), and (4) the dark matter fraction ($f_{\rm DM}$).

\subsubsection{Composite model: Stars with gNFW dark halo}
Finally, to allow for a more complex interplay between dark and baryonic matter, we adopt a composite model utilizing a generalized NFW \citep[gNFW;][]{gnFW_Wyithe_2001} halo profile. We still assume no $M/L$ gradient for the stars, identical to the NFW model described above. However, unlike the standard NFW halo which enforces a fixed inner slope ($\gamma=1$), the gNFW profile treats the inner logarithmic density slope ($\gamma$) as a free parameter. This added flexibility enables the model to account for the dynamical response of the dark matter halo to baryonic assembly processes, such as adiabatic contraction (which steepens the profile; e.g. \citealt{Blumenthal1986, Gnedin2004}) or feedback-driven expansion (which flattens it; e.g. \citealt{Governato2010}). The dark matter density distribution is described as:
\begin{equation*}
    \rho_{\text {DM}}(r)=\rho_s\left(\frac{r}{r_s}\right)^{-\gamma} \left(\frac{1}{2}+\frac{1}{2} \frac{r}{r_s}\right)^{\gamma-3},
\end{equation*}
where the standard NFW profile is recovered for $\gamma=1$. Similar to the NFW model, we fixed the break radius to $r_s = 20$ kpc and constrained the halo density normalization, $\rho_s$, via the dark matter fraction ($f_{\rm DM}$). Consequently, this model depends on five free parameters: (1)  $\frac{\sigma_z}{\sigma_R}$ for \jamcyl\ case (or $\frac{\sigma_\theta}{\sigma_r}$ for \jamsph\ ), (2) the minimum intrinsic axial ratio ($q_{\min}$), (3) the stellar mass-to-light ratio ($M/L$), (4) the dark matter fraction ($f_{\rm DM}$), and (5) the inner dark matter slope ($\gamma$).

For all dynamical models employed in this study, we assumed a central supermassive black hole mass of zero. Given the redshift range and spatial resolution of our data, the sphere of influence of the central black hole is unresolved and thus has a negligible effect on the global kinematic fit. A summary of the adopted models, including the parameter priors and boundaries, is presented in \autoref{table:dyn_model}.

\begin{figure*}
  \centering

  \newcommand{\panelwidth}{0.32\textwidth} 
  \newcommand{\rowsep}{3pt} 
  \setlength{\tabcolsep}{3pt} 
  \renewcommand{\arraystretch}{1.1}  

  \newcommand{\panellabelsize}{\footnotesize}

  \begin{tabular}{ccc}
    \includegraphics[width=\panelwidth,keepaspectratio]{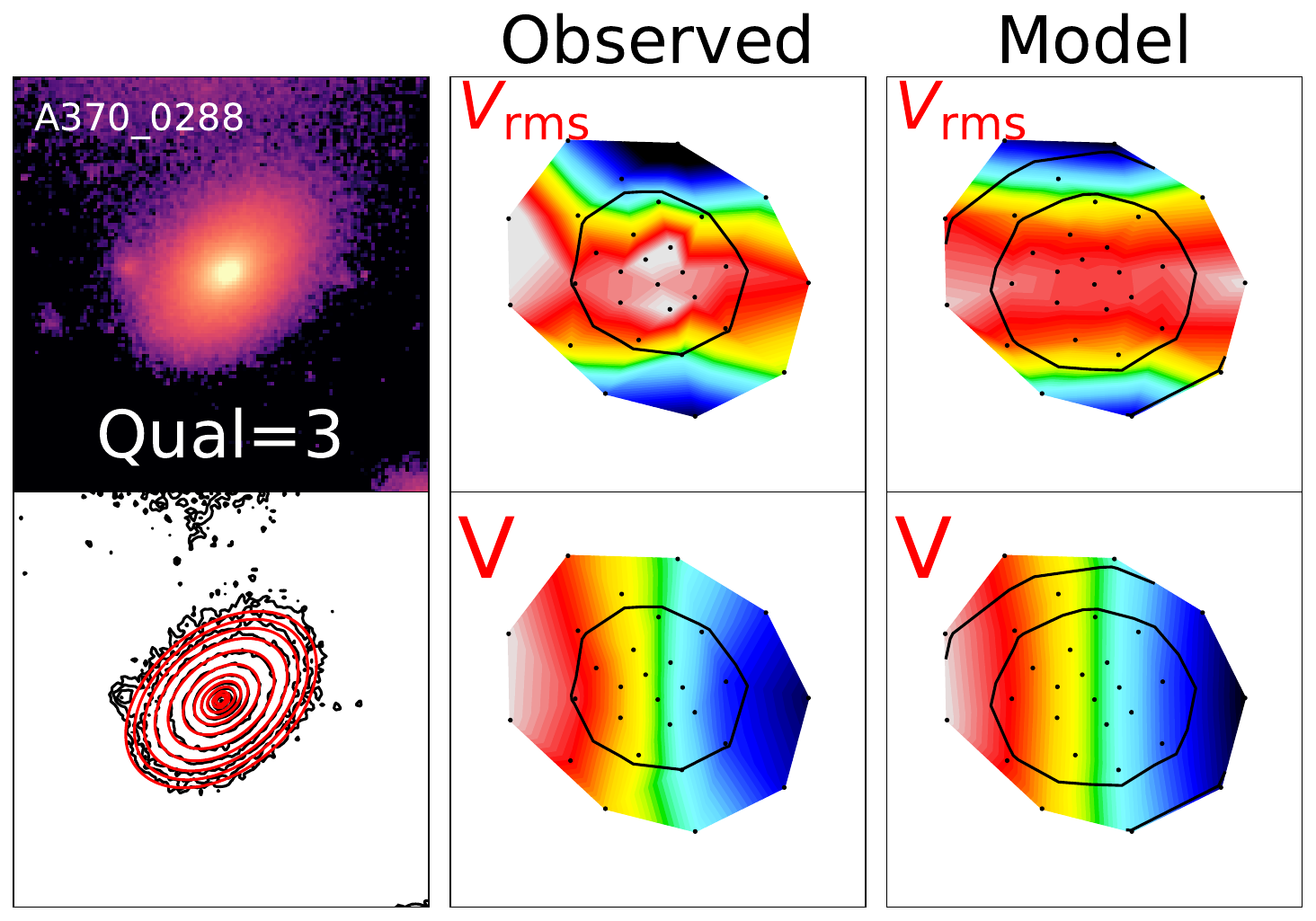} &
    \includegraphics[width=\panelwidth,keepaspectratio]{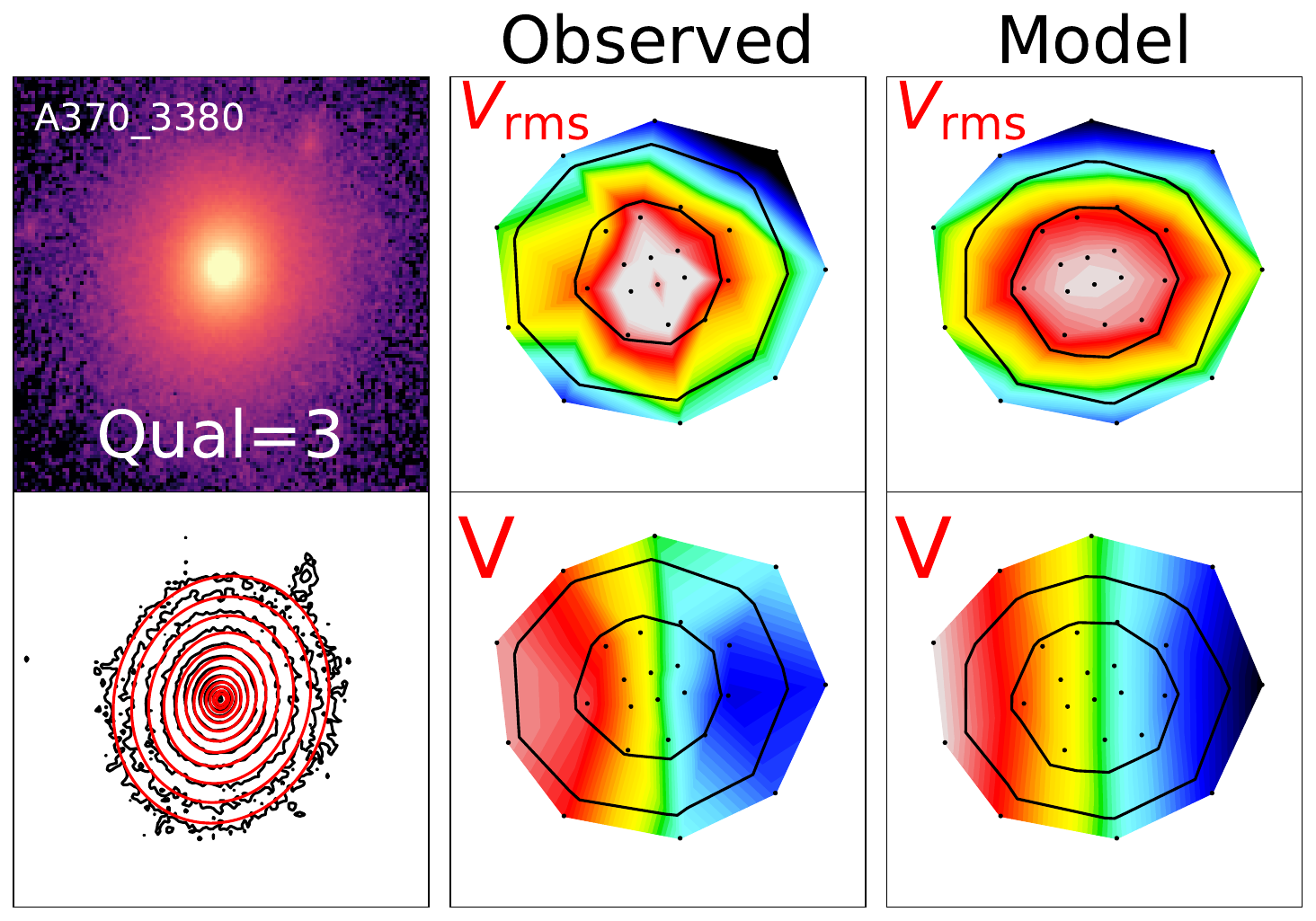} &
    \includegraphics[width=\panelwidth,keepaspectratio]{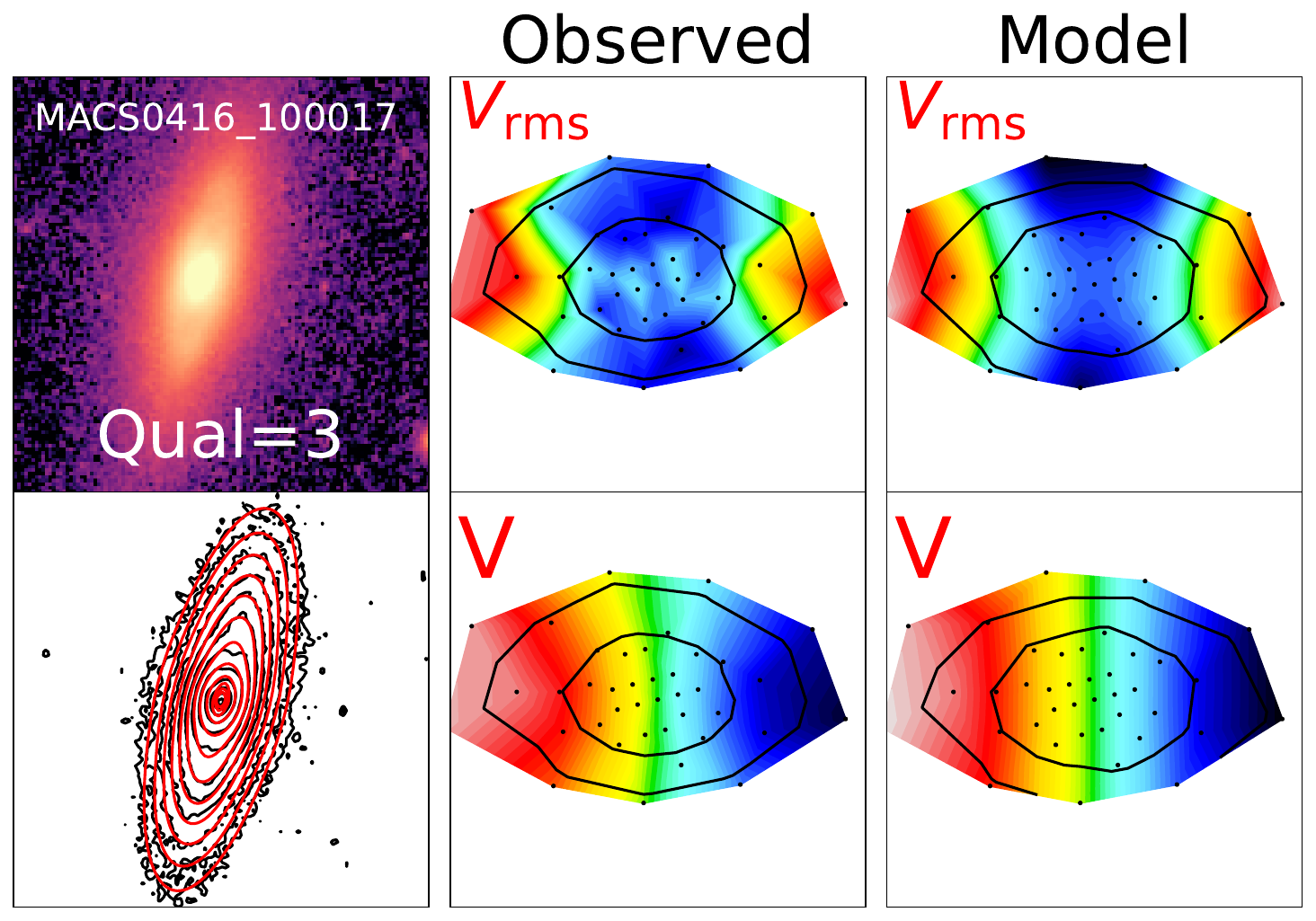} \\[\rowsep]

    \includegraphics[width=\panelwidth,keepaspectratio]{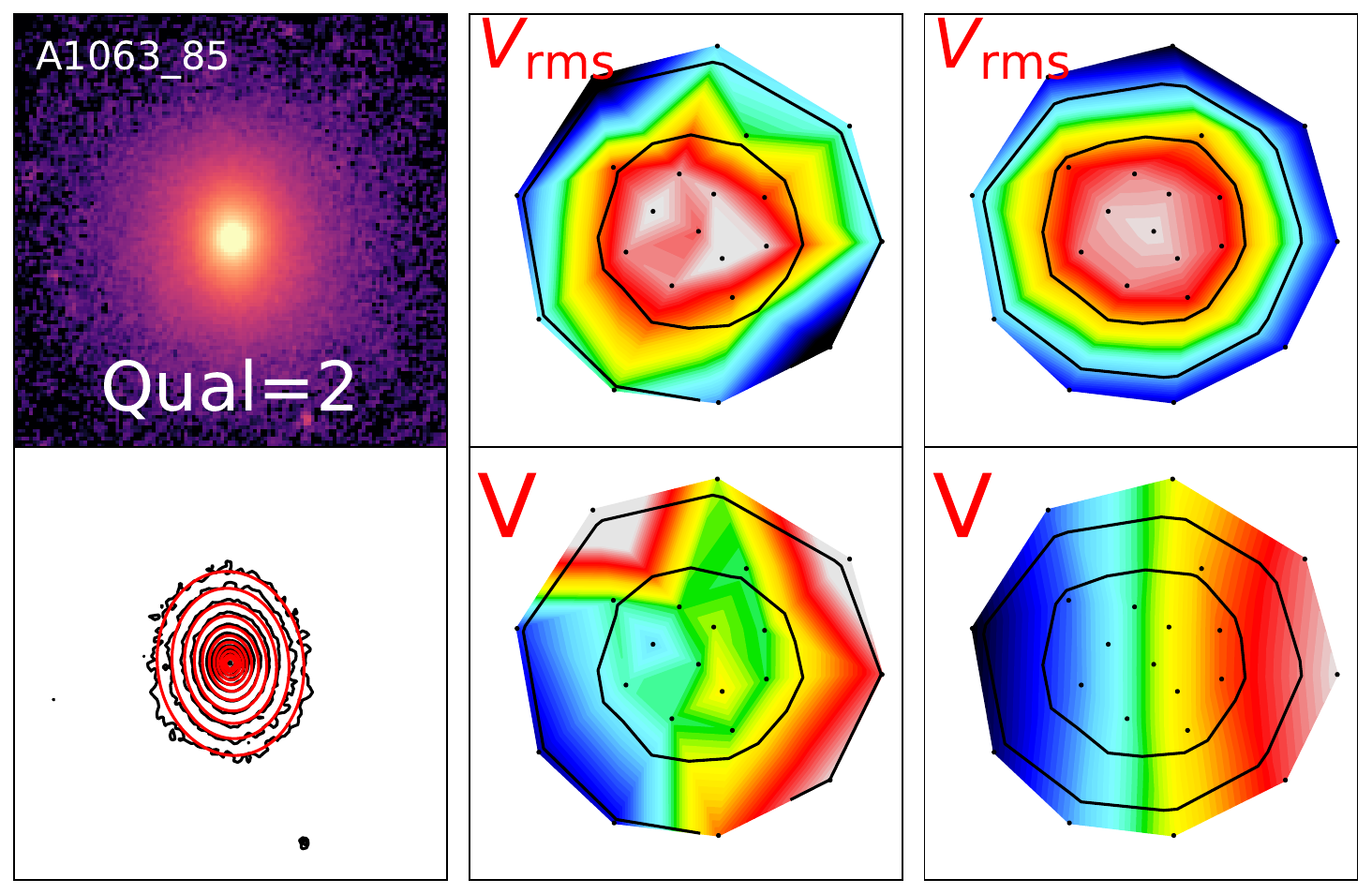} &
    \includegraphics[width=\panelwidth,keepaspectratio]{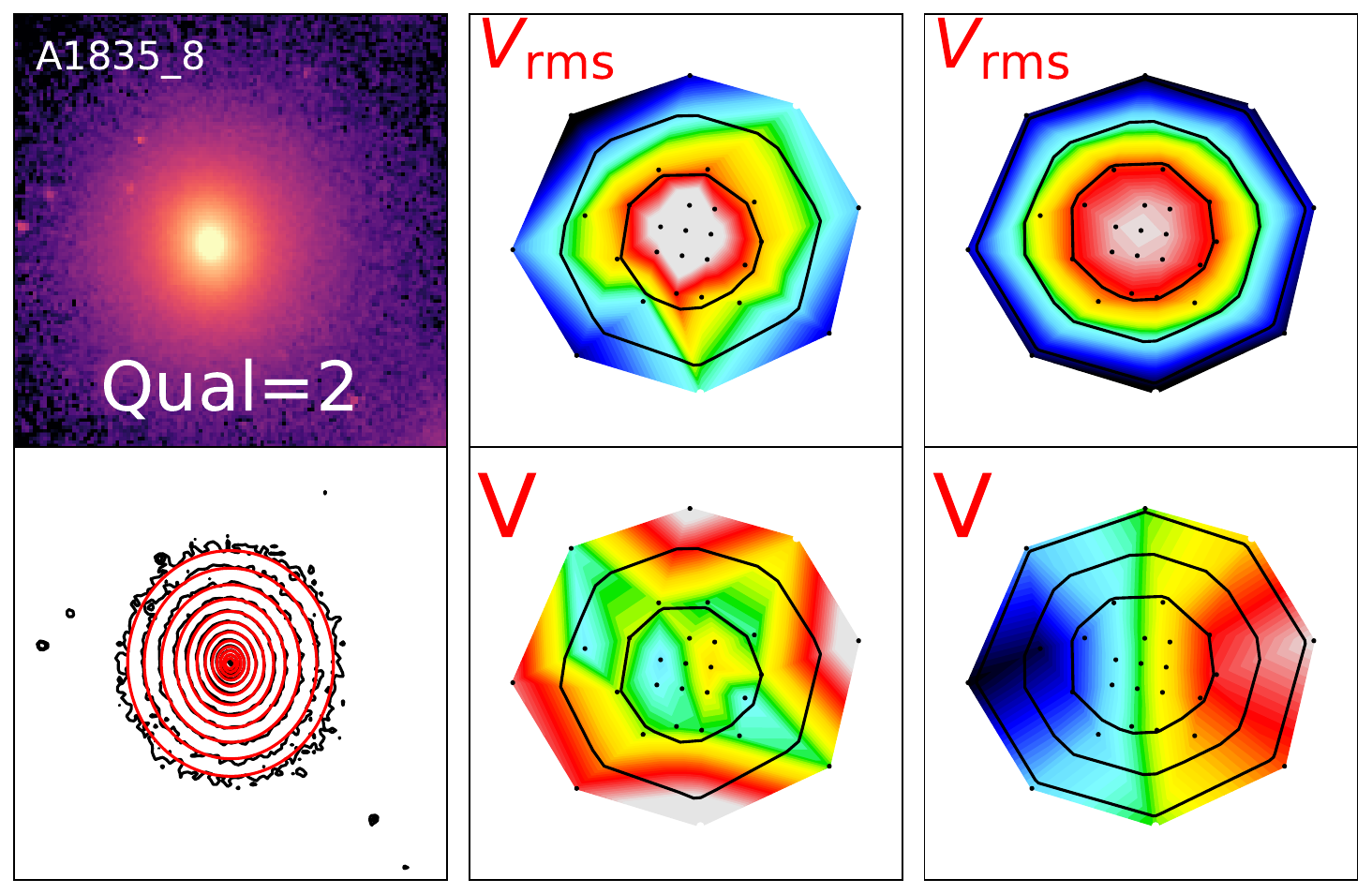} &
    \includegraphics[width=\panelwidth,keepaspectratio]{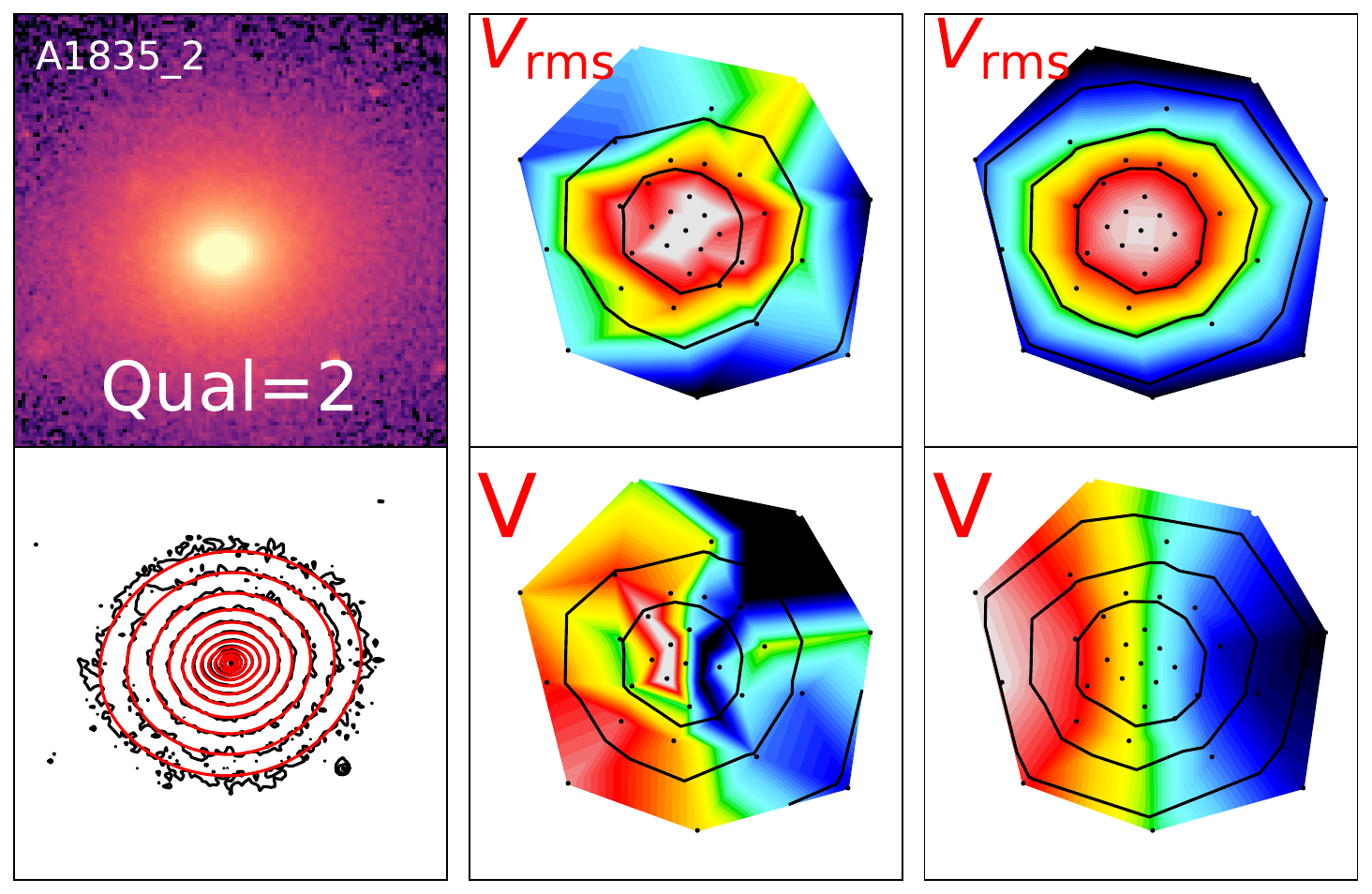} \\[\rowsep]

    \includegraphics[width=\panelwidth,keepaspectratio]{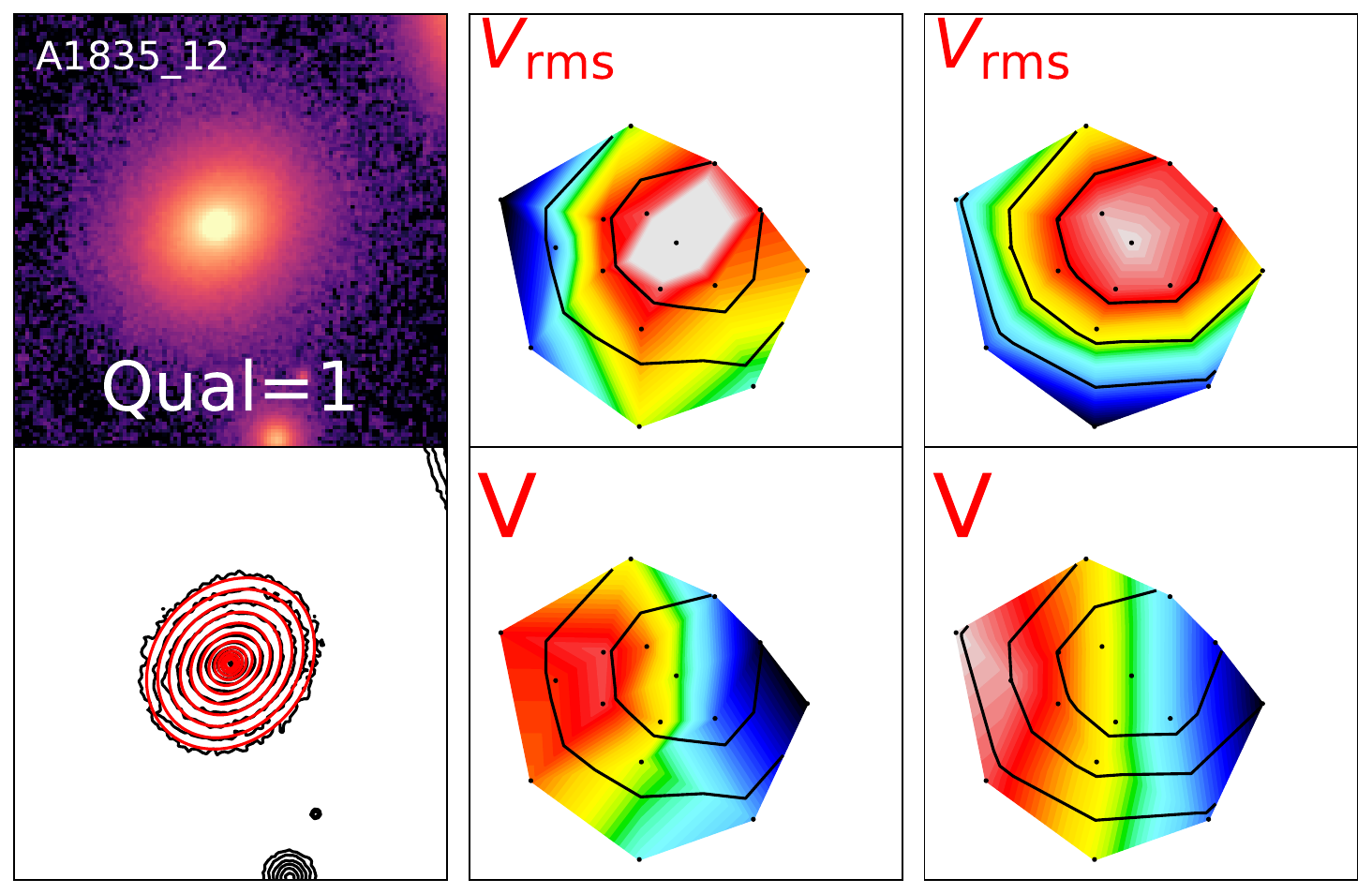} &
    \includegraphics[width=\panelwidth,keepaspectratio]{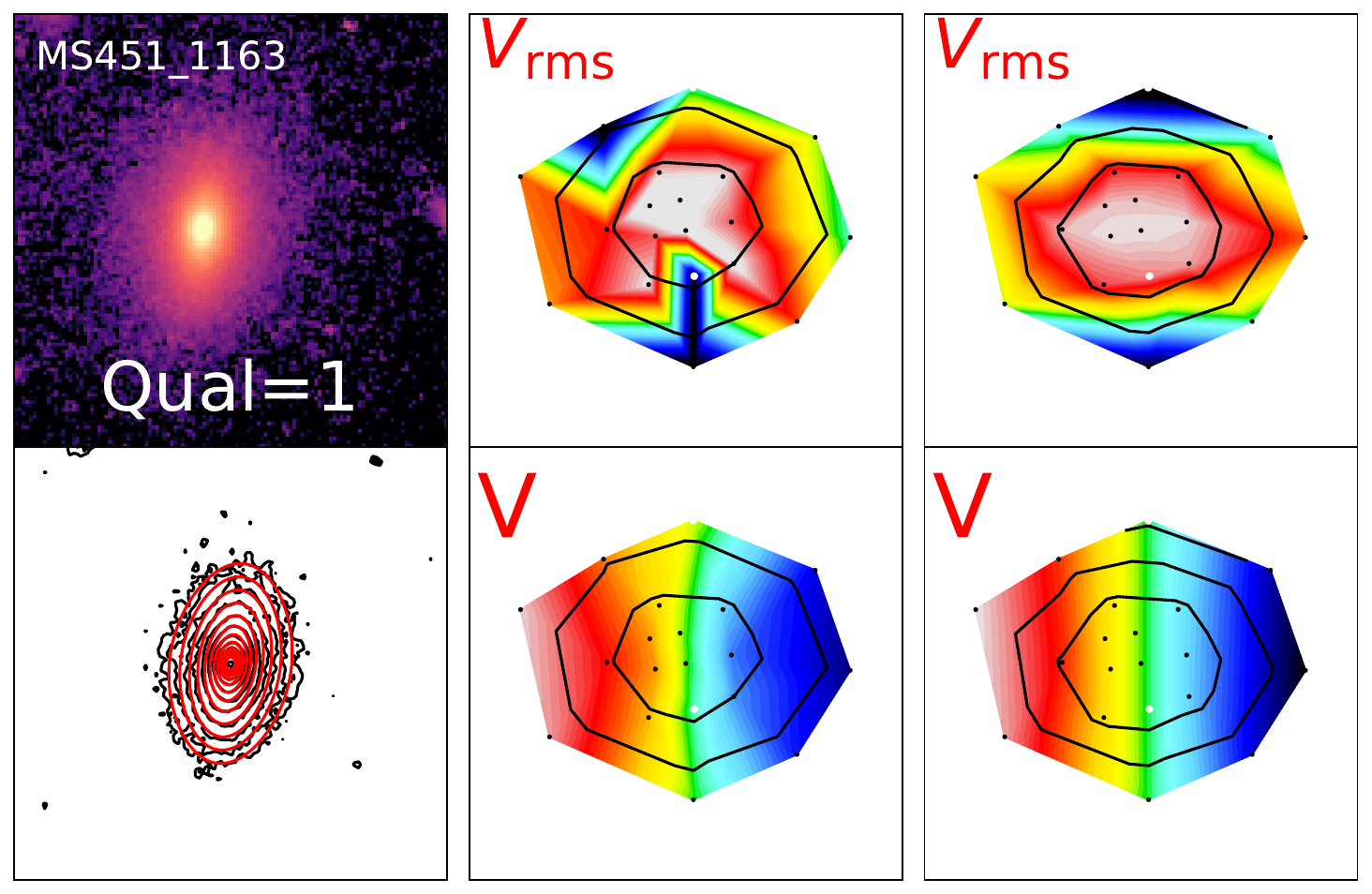} &
    \includegraphics[width=\panelwidth,keepaspectratio]{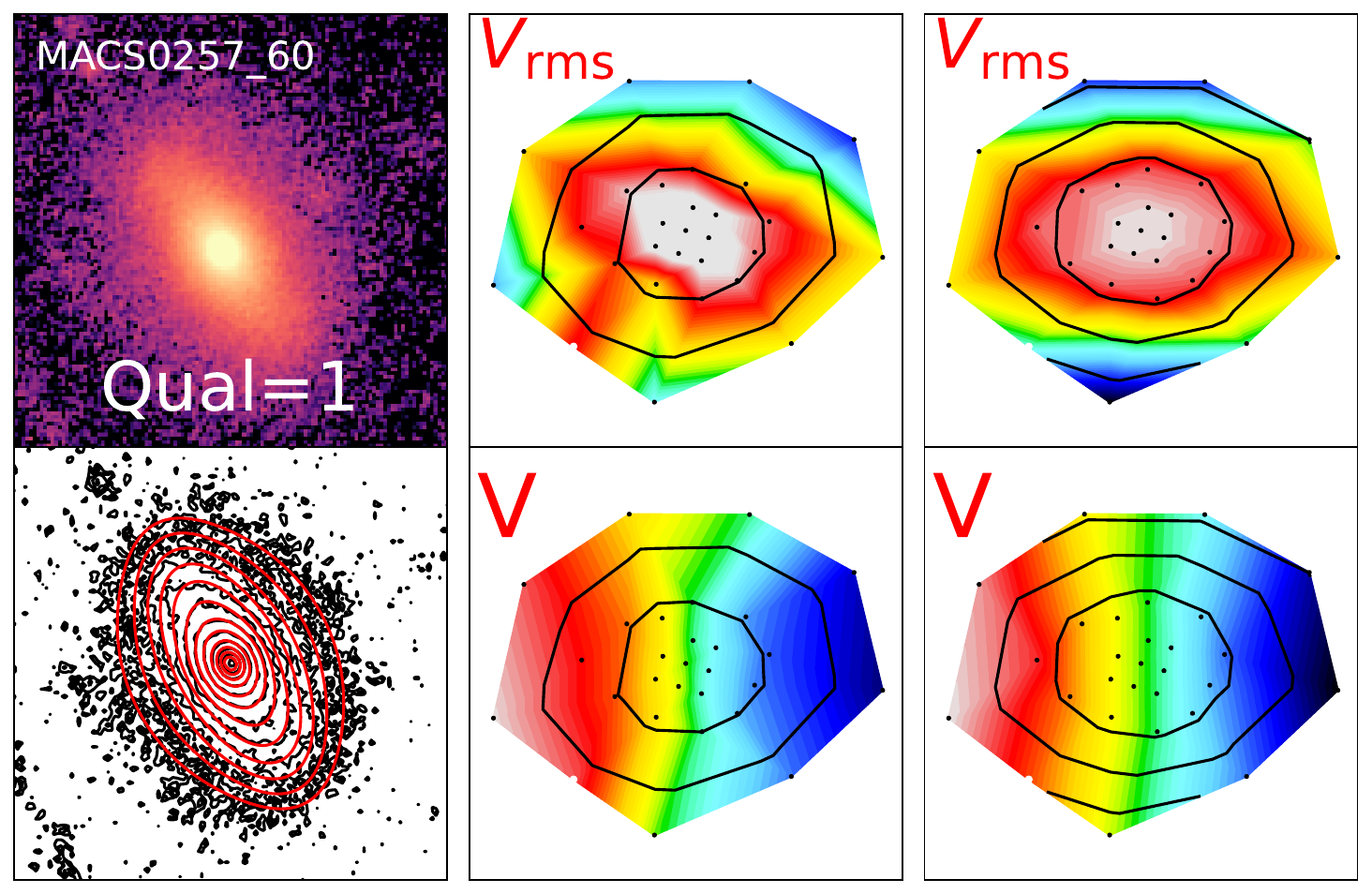} \\[\rowsep]

    \includegraphics[width=\panelwidth,keepaspectratio]{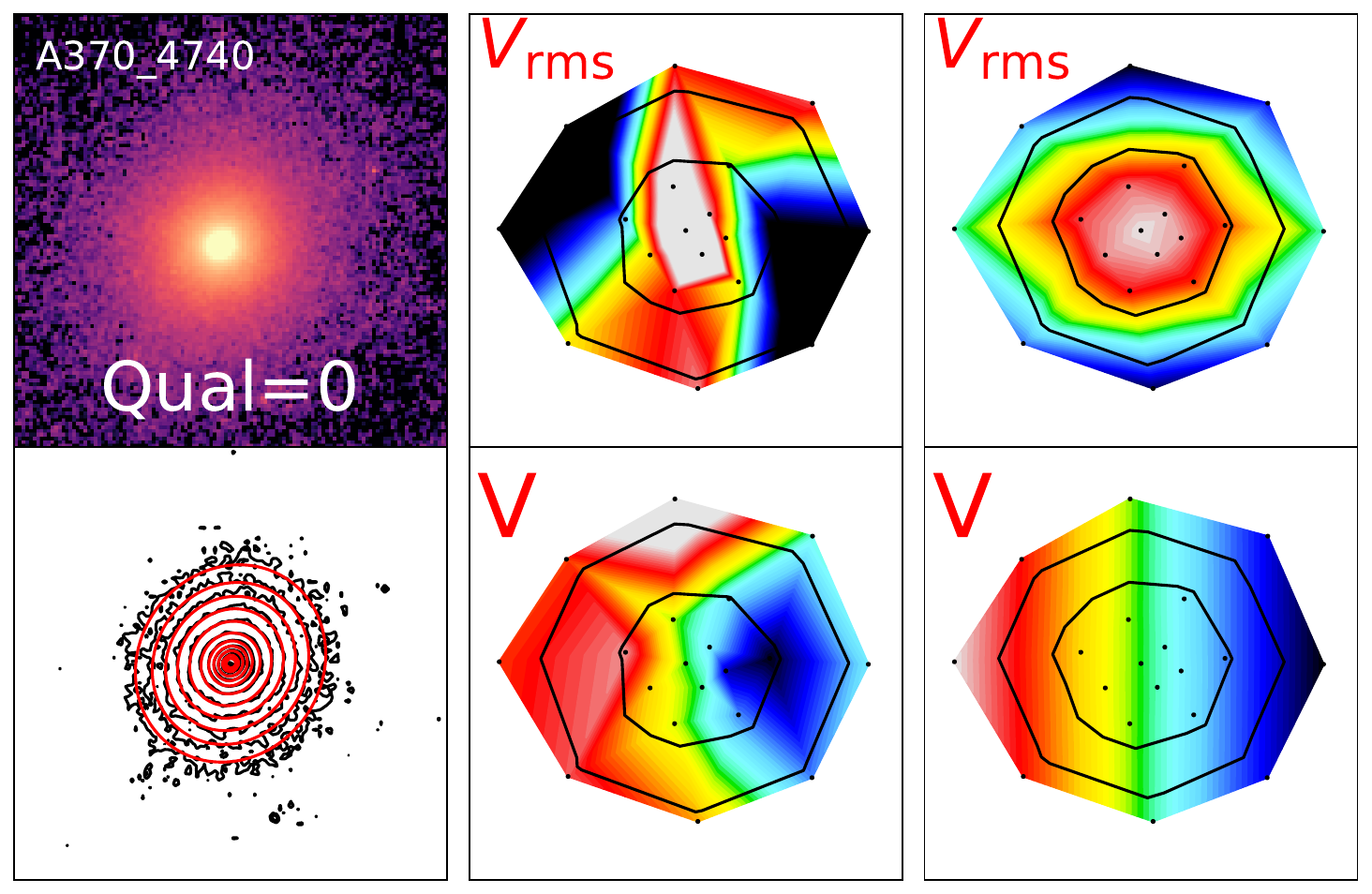} &
    \includegraphics[width=\panelwidth,keepaspectratio]{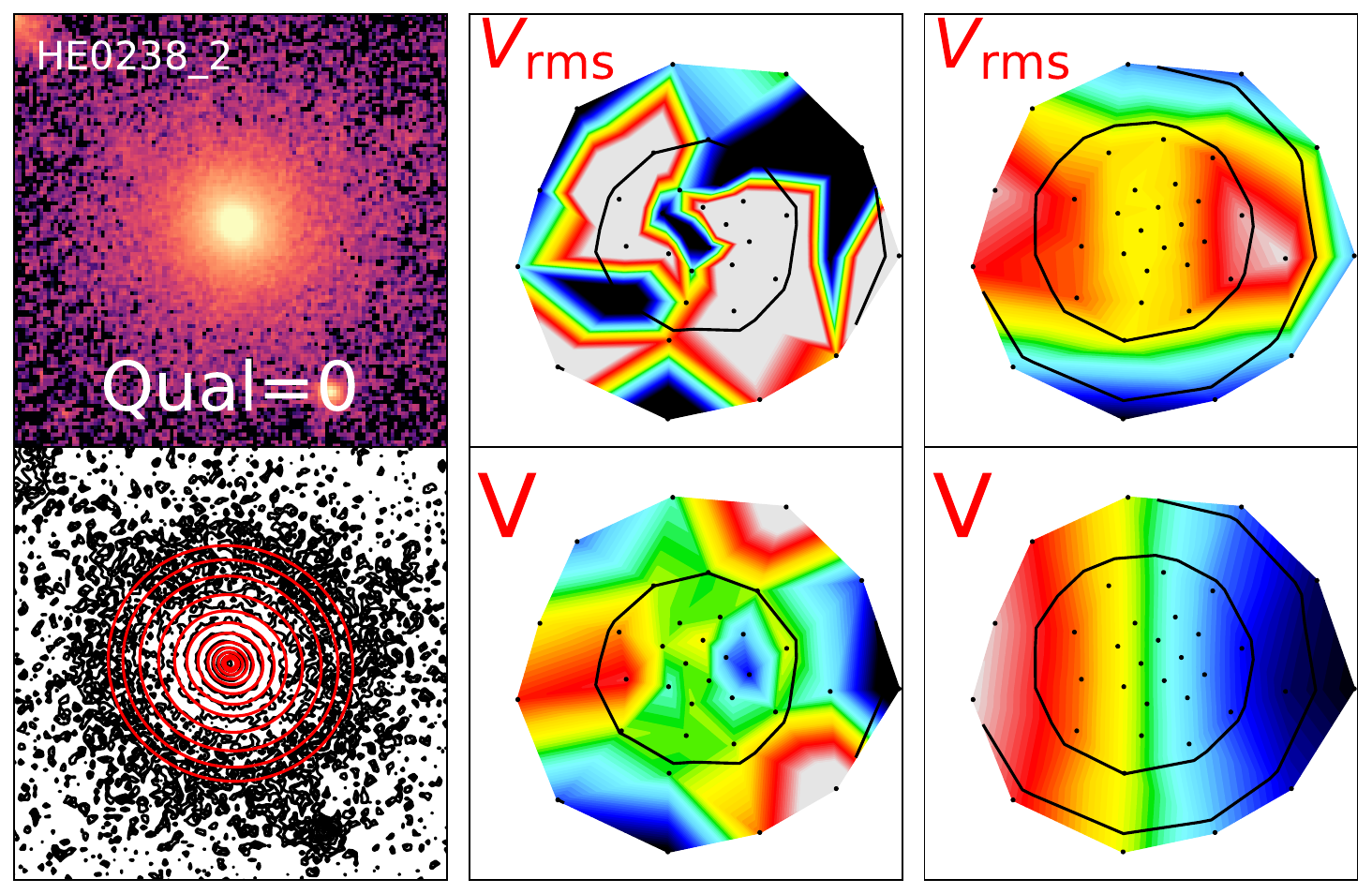} &
    \includegraphics[width=\panelwidth,keepaspectratio]{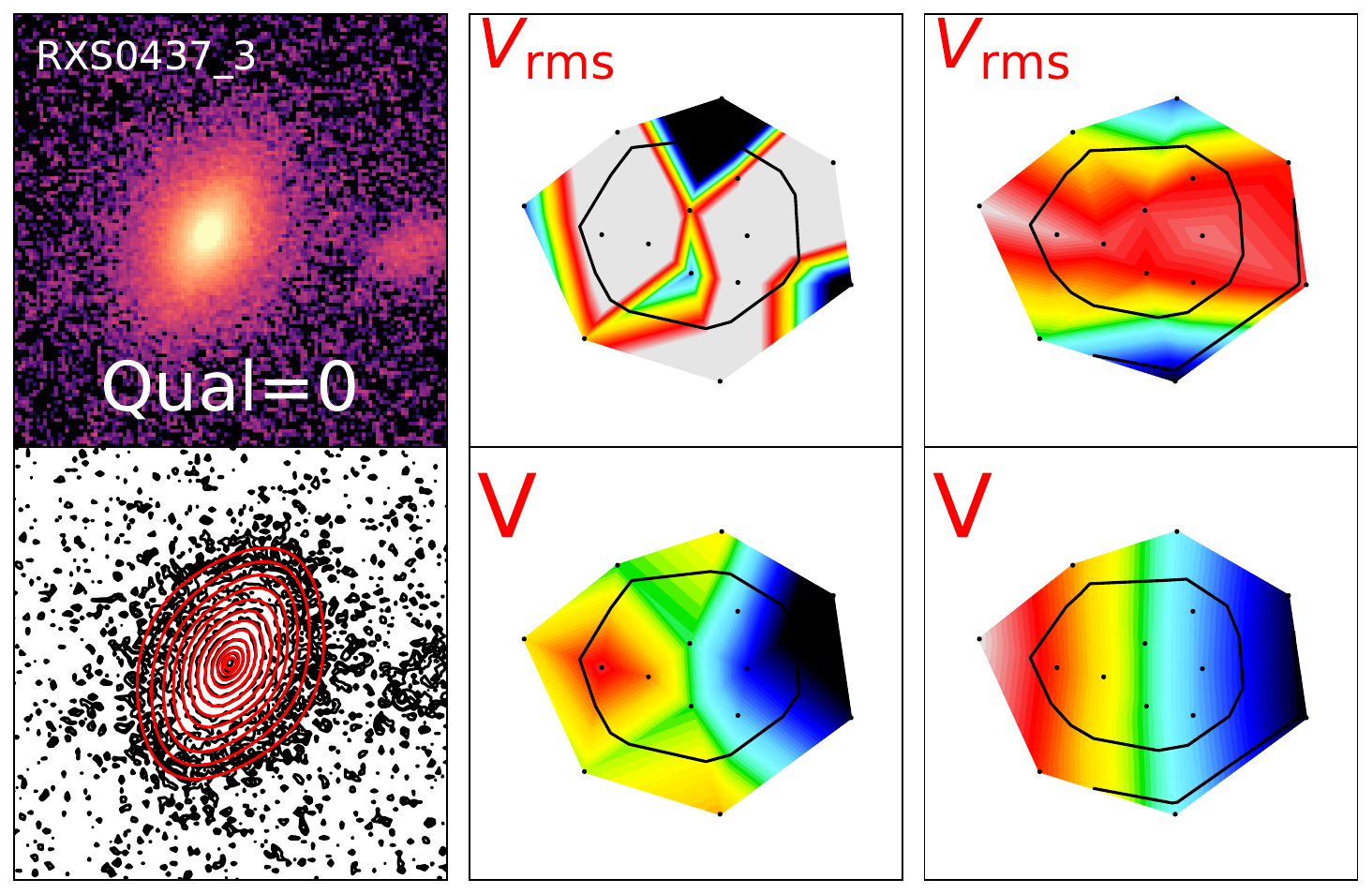} \\[\rowsep]

    \includegraphics[width=\panelwidth,keepaspectratio]{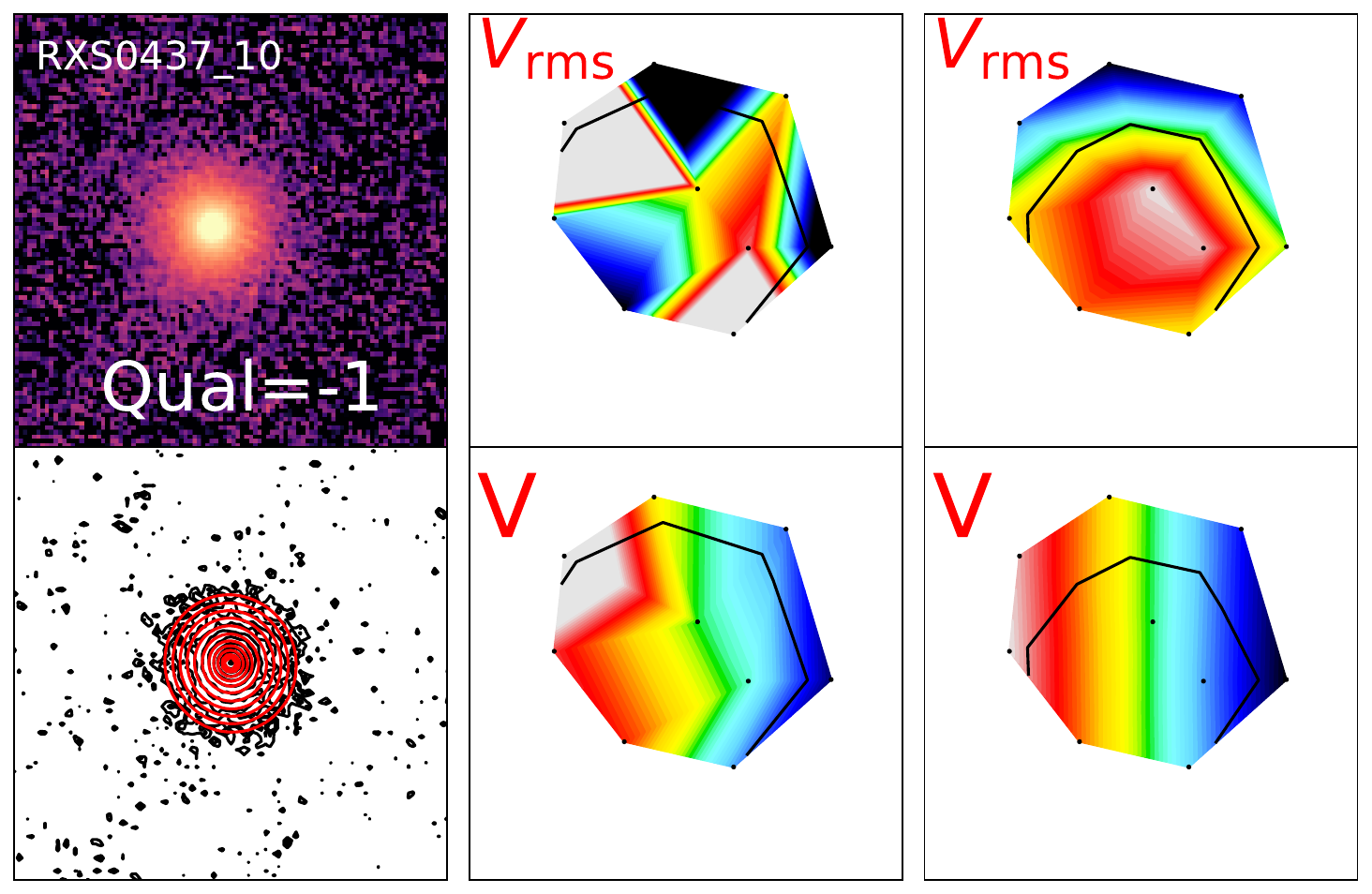} &
    \includegraphics[width=\panelwidth,keepaspectratio]{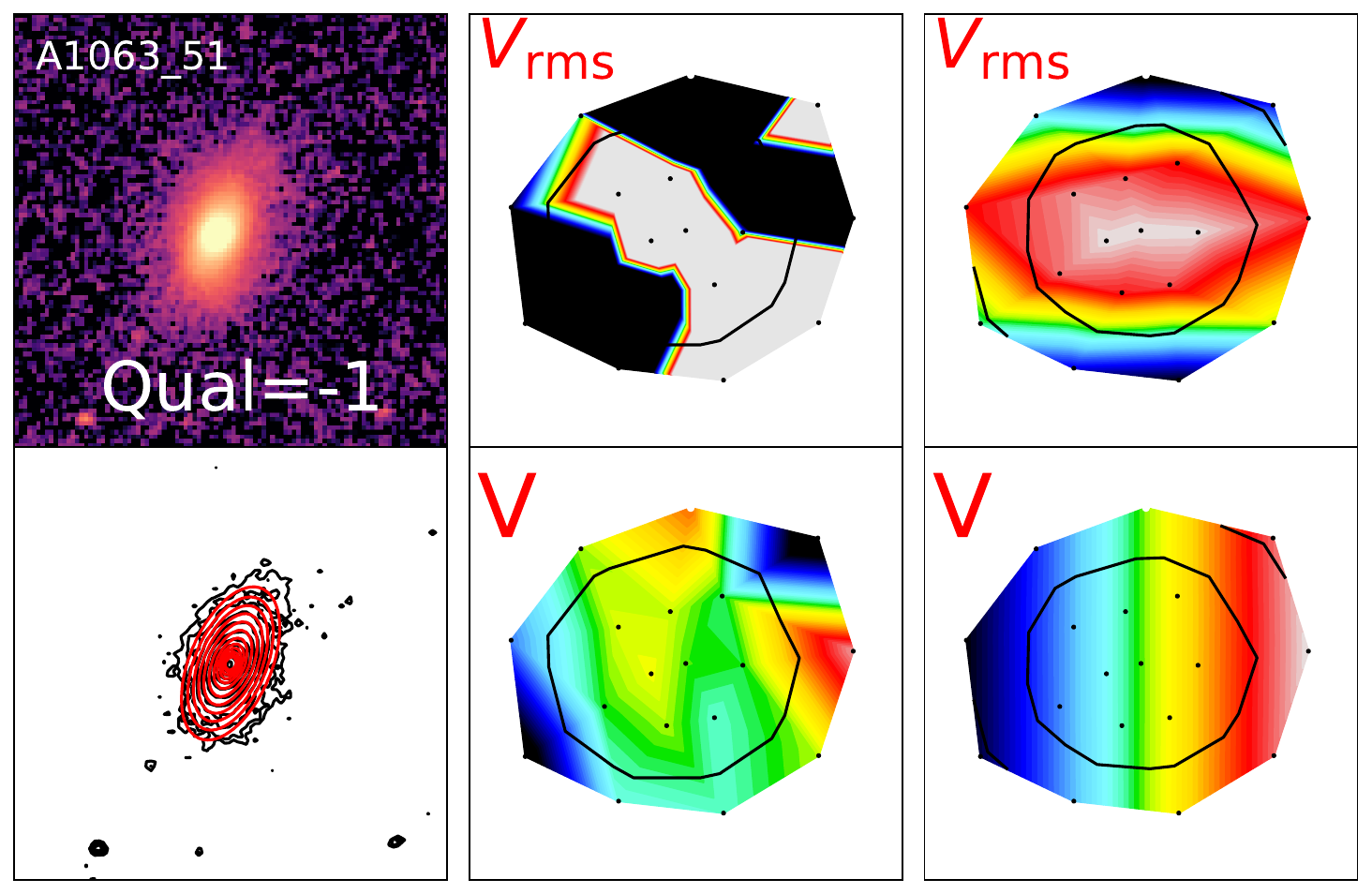} &
    \includegraphics[width=\panelwidth,keepaspectratio]{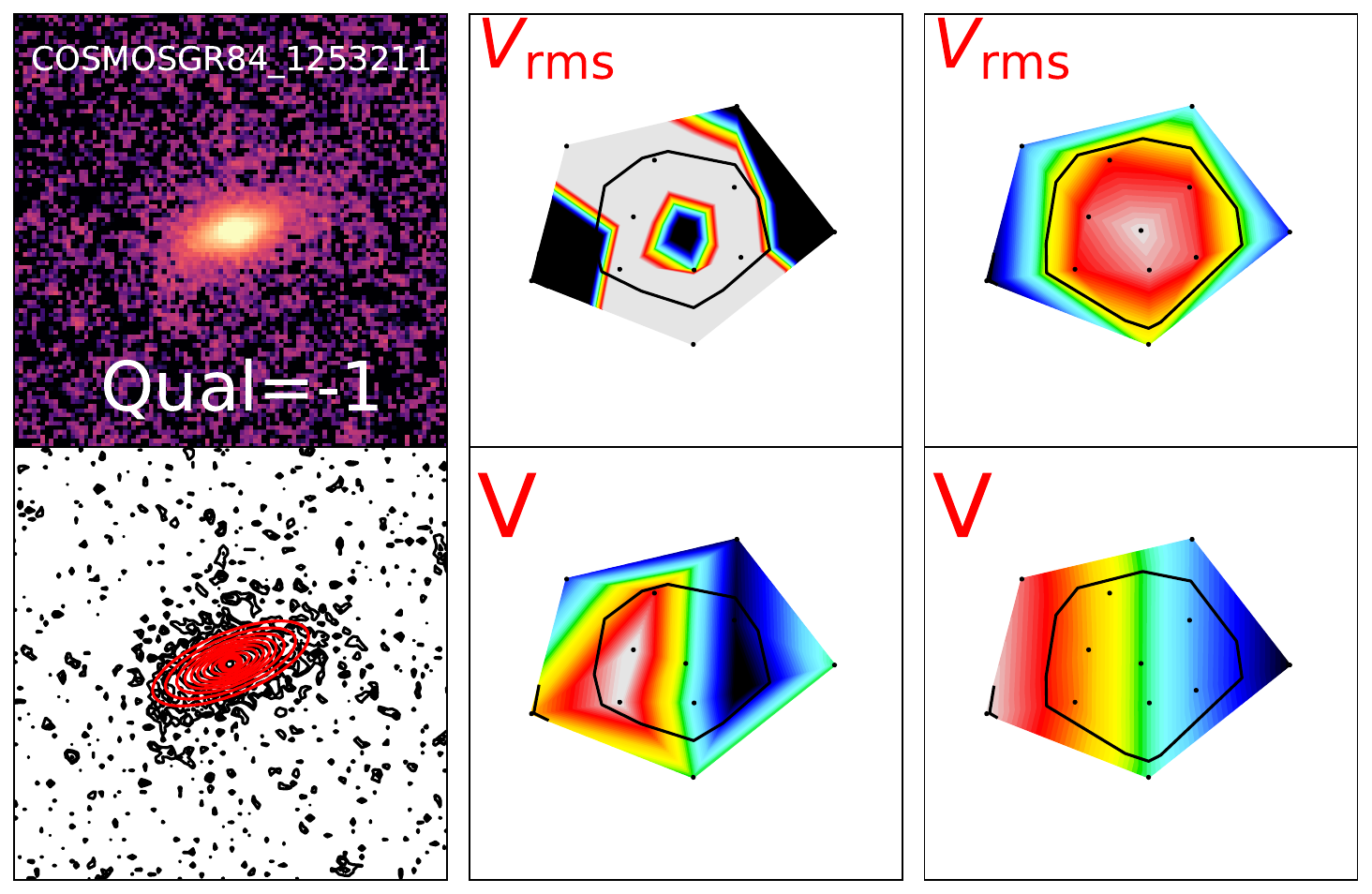}
  \end{tabular}

  \caption{Example of visual assessment of the dynamical model fit quality. We present three galaxies for each quality class (defined in \autoref{sec:fit_quality}). Each block of two rows corresponds to a specific quality class, displaying three galaxies arranged side-by-side. For an individual galaxy, the top row shows the HST image (labeled with the Galaxy ID and Quality flag), followed by the observed and modeled root-mean-square velocity ($V_{\rm rms}$) maps. The bottom row displays the MGE fit, where black and red contours indicate the observed and modeled isophotes, respectively, followed by the observed and predicted mean velocity ($V$) fields. In the observed and fitted kinematic maps, the overlaid black contours trace the observed and modeled surface brightness distribution. Color bars are not provided, as the quality classification is based on visual rather than quantitative assessment.}
  \label{fig:visual_quality}
\end{figure*}

\subsection{Fitting procedure}
\label{sec:fitting}
The parameter estimation is performed in two stages. First, we employed an iterative outlier-rejection scheme to identify and exclude spurious Voronoi bins. We fitted a preliminary Mass-Follows-Light (MFL) model with cylindrical alignment (\jamcyl) to the observed kinematics, \newadd{utilizing the non-linear least-squares routine \textsc{scipy.optimize.least\_squares}.} The model has three free parameters: the stellar mass-to-light ratio ($M/L$), $\frac{\sigma_z}{\sigma_R}$, and the minimum intrinsic axial ratio ($q_{\min}$). At each iteration, we computed the logarithmic residuals between the data and model, and quantified the scatter of these residuals ($\sigma_{\rm scatter}$) using a robust bi-weight estimator \citep{Hoaglin_1983}. Any Voronoi bin with an absolute residual exceeding $3\sigma_{\rm scatter}$ was flagged as an outlier and excluded from the next iteration. This process was repeated until convergence, defined as the point where no new bins were rejected. To ensure consistency, the final mask of valid bins was applied uniformly across all subsequent dynamical model fits.

Next, we utilized a Bayesian inference framework to constrain the parameters of the six dynamical models and quantify their uncertainties. We sampled the posterior probability distributions using the Affine Invariant Markov chain Monte Carlo (MCMC) ensemble sampler implemented in the \textsc{emcee} Python package \citep{emcee_Foreman_2013}. The posterior probability of the model parameters $\boldsymbol{\theta}$, conditioned on the observed data $\boldsymbol{D}$, is given by Bayes' theorem:
\begin{equation*}
    P(\boldsymbol{\theta} \mid \boldsymbol{D}) \propto P(\boldsymbol{D} \mid \boldsymbol{\theta}) P(\boldsymbol{\theta})
\end{equation*}
where $P(\boldsymbol{\theta})$ represents the prior probability and $P(\boldsymbol{D} \mid \boldsymbol{\theta})$ is the likelihood function. Assuming Gaussian uncertainties for the kinematic measurements, the likelihood is proportional to $\exp(-\chi^2/2)$, with the $\chi^2$ statistic defined as:
\begin{equation*}
    \chi^2=\sum_j\left(\frac{\sqrt{\overline{v_{\mathrm{los}, j}^2}}-V_{\mathrm{rms}, j}}{\varepsilon_{V_{\mathrm{rms}, j}}}\right)^2
\end{equation*}
Here, $\sqrt{\overline{v_{\mathrm{los}, j}^2}}$ denotes the model-predicted line-of-sight RMS velocity, while $V_{\mathrm{rms}, j}$ and $\varepsilon_{V_{\mathrm{rms}, j}}$ represent the observed RMS velocity and its associated measurement uncertainty in the $j$-th Voronoi bin, respectively. The summation extends over all spatial bins retained after the outlier-rejection phase.

To sample the parameter space, we deployed 20 independent walkers evolving for 1,000 steps, generating a total of 20,000 posterior samples for each model. As discussed in \autoref{sec:mass_models}, we applied the informative Gaussian priors to the anisotropy parameters, while adopting flat (uniform) priors for all other parameters within the physical boundaries specified in \autoref{table:dyn_model}. Figs.~\ref{fig:dymd_mcmc_example_1} and \ref{fig:dymd_mcmc_example_2} displays representative corner plots showing the posterior distributions for three distinct JAM models fitted to the same galaxy.

\subsection{Measuring dynamical quantities} \label{sec:measure_dyn_mass_slope}
From the posterior distributions of the model parameters, we derived the total enclosed mass, $M_{\rm T}(<R_e)$, and the mass-weighted total density slope, \tslope, within the effective radius. The total mass was calculated analytically using the \textsc{mge\_radial\_mass} routine within the \textsc{JamPy} package. Following \citet{Dutton_2014}, we defined \tslope\ as the mass-weighted average of the local logarithmic slope within $\rm{R_e}$. This definition is robust against local fluctuations in the density profile and is expressed as:
\begin{equation*}
\resizebox{\linewidth}{!}{
   $ \gamma_{\rm T} = \frac{1}{M_{\mathrm{T}}(<R_{\mathrm{e}})} \int_0^{R_{\mathrm{e}}} -\frac{\mathrm{d} \log \rho_{\mathrm{T}}}{\mathrm{d} \log r} 4 \pi r^2 \rho_{\mathrm{T}}(r) \mathrm{d} r = 3 - \frac{4 \pi R_{\mathrm{e}}^3 \rho_{\mathrm{T}}(R_{\mathrm{e}})}{M_{\mathrm{T}}(<R_{\mathrm{e}})},
    $}
\end{equation*}
where $\rho_{\mathrm{T}}(r)$ represents the spherically averaged total density profile. We computed this quantity analytically, on spherical shells, from the axisymmetric MGE densities utilizing the \textsc{mge\_weighted\_slope} function in \textsc{JamPy}.

To quantify the statistical uncertainties, we generated posterior distributions for \tslope\ by propagating the MCMC samples from each JAM model. We first discarded the initial 100 steps (2,000 samples) as a burn-in phase to ensure chain convergence. The remaining chains were thinned by a factor of 2 to reduce autocorrelation, yielding a final set of 9,000 independent samples per model. We adopted the median of these derived distributions as our fiducial measurement, with the 16th and 84th percentiles defining the $1\sigma$ confidence intervals. This approach is superior to relying on a single best-fit solution or the medians of marginalized parameters, as it explicitly propagates the intrinsic covariances between model parameters into the final physical quantities.


\begin{figure*}
    \centering
    {\includegraphics[width=0.95\textwidth]{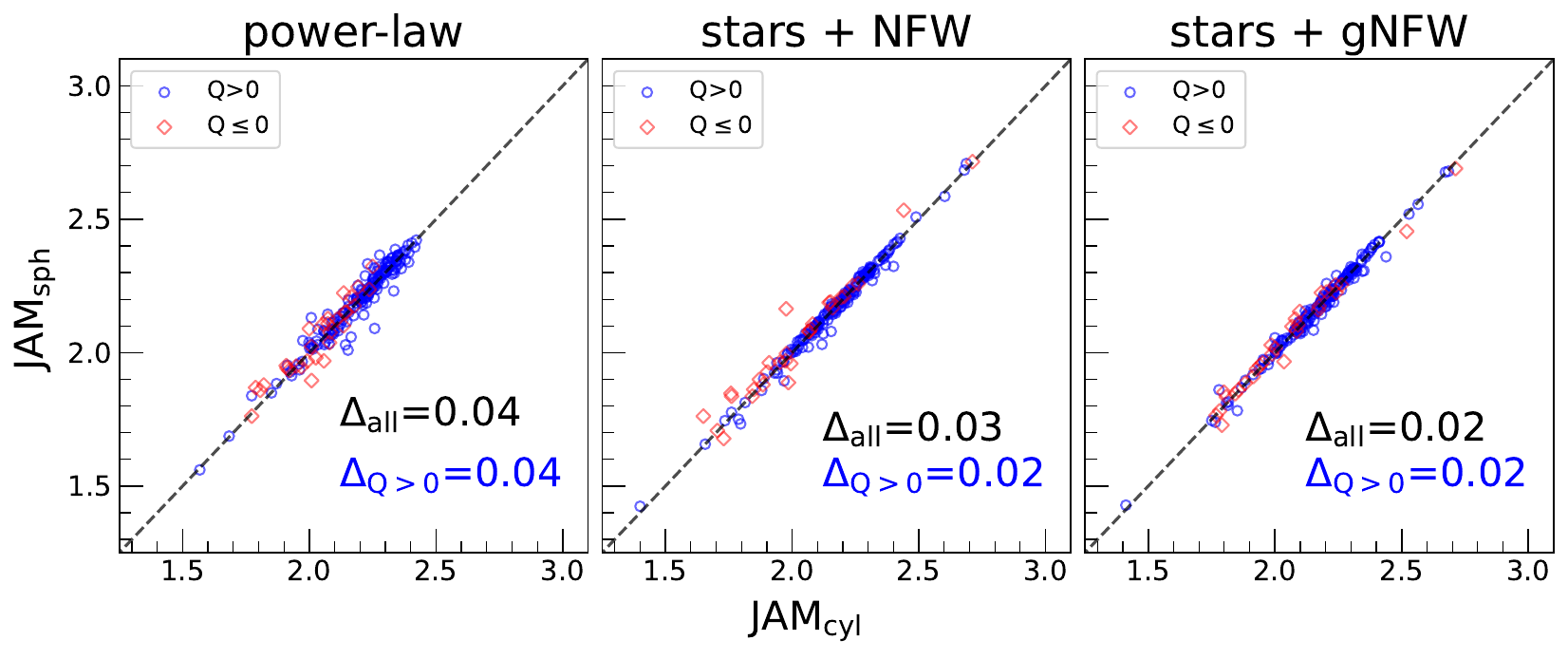}}
    \caption{Comparison of the total density slope (\tslope) measured using cylindrically (\jamcyl) versus spherically (\jamsph) aligned velocity ellipsoids for same mass model. The three panels correspond to the power-Law (left), stars + NFW (middle), and stars + gNFW (right) mass models. Galaxies with reliable kinematic fits (Qual $> 0$) are plotted as blue circles, while those with poor or unreliable fits (Qual $\le 0$) are marked with \newadd{red diamonds}. The black dashed line indicates the one-to-one relation. The intrinsic scatter ($\Delta$) between the two orientations, quantified using the \textsc{LtsFit} package, is reported within each panel for both the full sample and the high-quality subset.
    }
    \label{fig:slope_same_mass_model}
\end{figure*}

\section{Results}
\label{sec:results}
In this section, we present the measurements of the total density slope derived from our dynamical analysis. We begin by assessing the reliability of the dynamical model fits and defining our quality control criteria in \autoref{sec:fit_quality}. We then evaluate the internal consistency of our measurements across different model assumptions (mass profiles and velocity ellipsoid alignments) in \autoref{sec:slope_consistency} and explore potential correlations with global galaxy properties in \autoref{sec:slope_correlations}. The core evolutionary trends are presented in \autoref{sec:z_evolution_magnus} for the MAGNUS sample alone, and in \autoref{sec:combined_analysis} for the combined dataset incorporating the local MaNGA baseline.

\subsection{Quality of dynamical modeling fits}\label{sec:fit_quality}
While the JAM formalism provides a robust method for recovering galaxy mass profiles, the reliability of the derived parameters depends critically on the model's ability to reproduce the observed kinematics. We found that in cases of low S/N or complex kinematic substructures, the models occasionally failed to converge to a physically meaningful solution. To quantify this, we performed a visual quality-of-fit assessment by comparing the observed velocity ($V$) and second moment ($V_{\rm rms}$) maps with the model predictions. We observed a high degree of consistency across the different mass models; specifically, if the data quality precluded a good fit for the most flexible model configuration (stars + gNFW with \jamsph), the simpler model variants invariably failed as well. Consequently, we adopted the stars + gNFW with \jamsph\ fit as a representative proxy to classify the entire sample. We used a classification scheme where a quality flag (Qual) ranging from $-1$ to $3$ was assigned to each galaxy based on the following criteria \citep[e.g.][]{ATLAS_XV_Michele_2013, Poci_2017, Dynpop_I_Zhu_2023}:

\begin{itemize}
    \item \textbf{Qual = 3 (Excellent):} The model accurately reproduces both the observed $V$ and $V_{\rm rms}$ maps. These galaxies are typically fast rotators with best quality data.
    \item \textbf{Qual = 2 (Good):} The predicted $V_{\rm rms}$ maps are of high quality, comparable to Qual = 3, but the fit to the mean velocity field ($V$) shows minor discrepancies. This category is dominated by slow rotators, where the velocity maps lack clear rotation signal and thus  intrinsically harder to constrain, yet the global mass distribution remains well-recovered via $V_{\rm rms}$.
    \item \textbf{Qual = 1 (Marginal):} The modeled $V_{\rm rms}$ maps are less accurate than the higher quality tiers but still capture the essential radial and azimuthal features of the observed data. In these cases, the $V$ maps are generally well-recovered, except for slow rotators.
    \item \textbf{Qual = 0 (Poor):} The model fails to capture the structure of the observed $V_{\rm rms}$ map, even if the $V$ field is reasonably fitted.
    \item \textbf{Qual = -1 (Unreliable):} The model fails to reproduce both the observed $V$ or $V_{\rm rms}$ maps significantly.
\end{itemize}
\autoref{fig:visual_quality} presents representative examples of galaxies from each quality group. Of the 198 galaxies in the MAGNUS sample, 168 ($\sim 85\%$) were assigned a fit quality of Qual $\geq 1$. This visual quality-of-fit assessment was performed by the corresponding  author and reviewed and agreed upon by all co-authors. Although we did not follow a strictly formal blinding procedure, the total density slope  did not influence the visual classification process. As the dynamical constraints for Qual $< 1$ galaxies are deemed unreliable, we excluded these 30 objects from all subsequent analyses. This exclusion is justified not only visually but also statistically. The excluded galaxies are characterised by systematically lower S/N per \AA\, measured from their integrated spectra (median  S/N per \AA\ 25, 30 and 48 for Qual $< 1$, Qual\,$=$\,1 and Qual\,$\geq$\,2, respectively). We also found that the median relative uncertainty on the derived total density slope (\tslope) for the excluded group (Qual $< 1$) is approximately double that of the retained sample. For instance, using the stars+gNFW with \jamsph\ model, the median uncertainty for Qual $\geq 1$ galaxies is $2.68\%$ (with a standard deviation of $3.27\%$), compared to $4.85\%$ (standard deviation $4.11\%$) for the low-quality fits. Furthermore, the retained sample typically exhibits reduced $\chi^2$ values in the range of $1-2$, confirming that the models provide a statistically satisfactory description of the data. The fitted $V$, $V_{\rm rms}$ maps and $\gamma_{\rm T}$ posterior distributions for all 198 galaxies, including those with Qual\,$<$\,1, are provided in the figure set (Figure~\ref{fig:galaxy_atlas}) and Table~\ref{table:slope_table}, available in the online journal.

\subsection{Consistency of Total Density Slope Measurements} \label{sec:slope_consistency}
To assess the robustness of our results, we first examined the sensitivity of the measured \tslope\ to the assumed geometry of the velocity ellipsoid. In \autoref{fig:slope_same_mass_model}, we compare the slopes derived using the same mass model but with different alignments (\jamcyl\ vs. \jamsph). We found that, for a fixed mass profile, the derived \tslope\ values are highly consistent regardless of the geometric assumption. This robustness is expected, as the imposition of a nearly isotropic prior on the velocity anisotropy effectively minimizes the impact of the velocity ellipsoid orientation. Using the \textsc{LtsFit} Python package \citep{ATLAS_XV_Michele_2013} with a $5\sigma$ outlier clipping, we measured the intrinsic scatter ($\Delta$) between the two orientations. For the power-law model, we found a negligible scatter of $\Delta = 0.04$, which remained constant even after excluding unreliable fits (Qual $\leq 0$). Similarly, the stars + NFW and stars + gNFW models exhibited very tight correlations with scatters of $\Delta = 0.03$ and $\Delta = 0.02$, respectively (improving to $\Delta = 0.02$ for the NFW case when excluding low-quality fits). These comparisons demonstrate that our measurements of \tslope\ are largely independent of the assumed velocity ellipsoid orientation.

\begin{figure*}
    \centering
    {\includegraphics[width=0.95\textwidth]{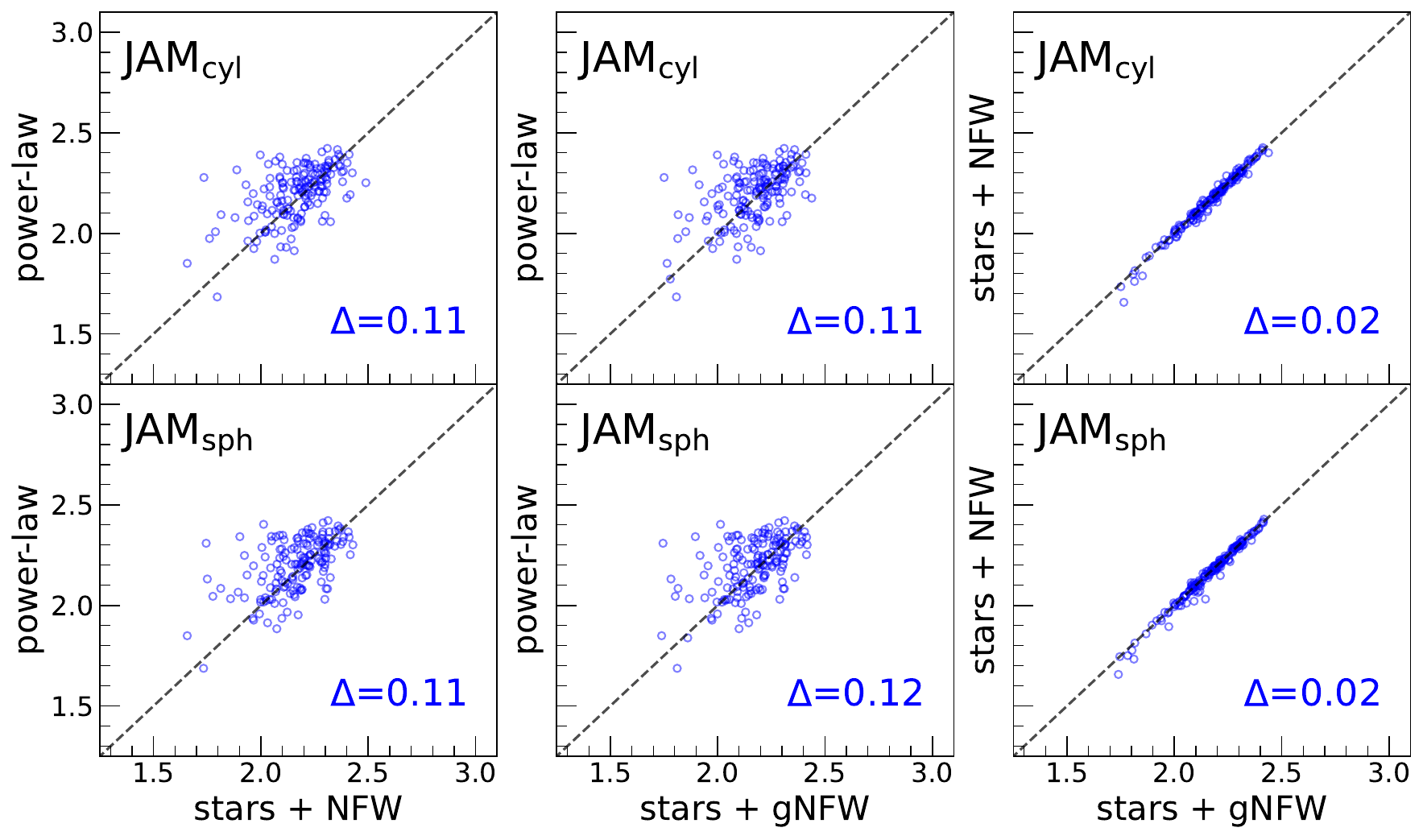}}
    \caption{Comparison of the total density slopes ($\gamma_{\rm T}$) derived using same velocity ellipsoid orientation but different mass models. This comparison is restricted to the subsample with reliable dynamical models (Qual $\ge 1$). The top and bottom rows display results for the cylindrically (\jamcyl) and spherically (\jamsph) aligned velocity ellipsoids, respectively. The black dashed line marks the one-to-one relation, and the intrinsic scatter ($\Delta$) between the models is indicated in each panel.}
    \label{fig:slope_diff_mass_model}
\end{figure*}

Next, we evaluated the consistency of \tslope\ across the different mass density parameterizations (power-Law, stars+ NFW, and stars+gNFW), considering only galaxies with reliable fits (Qual $\geq 1$). As shown in \autoref{fig:slope_diff_mass_model}, the NFW and gNFW models are in excellent agreement with each other, exhibiting a minimal intrinsic scatter of $\Delta = 0.02$ for both velocity ellipsoid configurations. This tight correlation is unsurprising given that the gNFW profile is a generalization of the NFW form; however, it confirms that the additional freedom in the inner halo slope of the gNFW model does not introduce significant instability in the recovered total density slope.

\newadd{In contrast, the single power-law models exhibit larger scatter relative to the composite models, though with no systematic bias. Approximately 50\% of the sample yields \tslope\ values agreeing within 1\%, while deviations in the remainder are bidirectional. This increased dispersion most probably stems from the structural rigidity of the power-law parameterization, which enforces self-similarity unlike the radially varying slopes of the composite models. Nevertheless, the measurements remain statistically compatible, with an intrinsic scatter of $\approx 0.11$, consistent with typical measurement uncertainties.}

\subsection{Correlations with Galaxy Properties} \label{sec:slope_correlations}
Previous studies have suggested that the total density slope of early-type galaxies (ETGs) correlates with structural properties such as mass, size, and velocity dispersion. \citet{Auger_2010} reported weak correlations between the slope and the effective radius ($R_{\rm e}$) as well as the central mass density, while \citet{Sonnenfeld_2013} found that the total slope correlates most strongly with the surface mass density. Early lensing-based results showed only weak trends, largely because lens galaxies preferentially inhabit the high-$\sigma_e$ regime, where galaxies tend to have similar density slopes. However, when lower-mass systems from dynamical studies were included, significantly stronger correlations emerged. In particular, the strongest trend was found between the total density slope $\gamma_{\rm T}$ and the stellar velocity dispersion $\sigma_e$, or equivalently, a near constancy of the slope along lines of constant $M/R_{\rm e}$ (\citealt{Cappellari_review_2016}, fig.~22c; \citealt{Poci_2017}). Extensions to dynamical samples from the MaNGA survey, spanning all morphological types and a wide range of stellar ages, further demonstrated that although $\sigma_e$ remains the best predictor of the total density slope, at fixed $\sigma_e$ the slope decreases systematically for younger galaxies (\citealt{Li_2019}; \citealt{Dynpop_III_Zhu_2024}, fig.~8). More recently, \citet{Derkenne_2023} reported an additional dependence of $\gamma_{\rm T}$ on environment.

\begin{figure*}
    \centering
    {\includegraphics[width=0.95\textwidth]{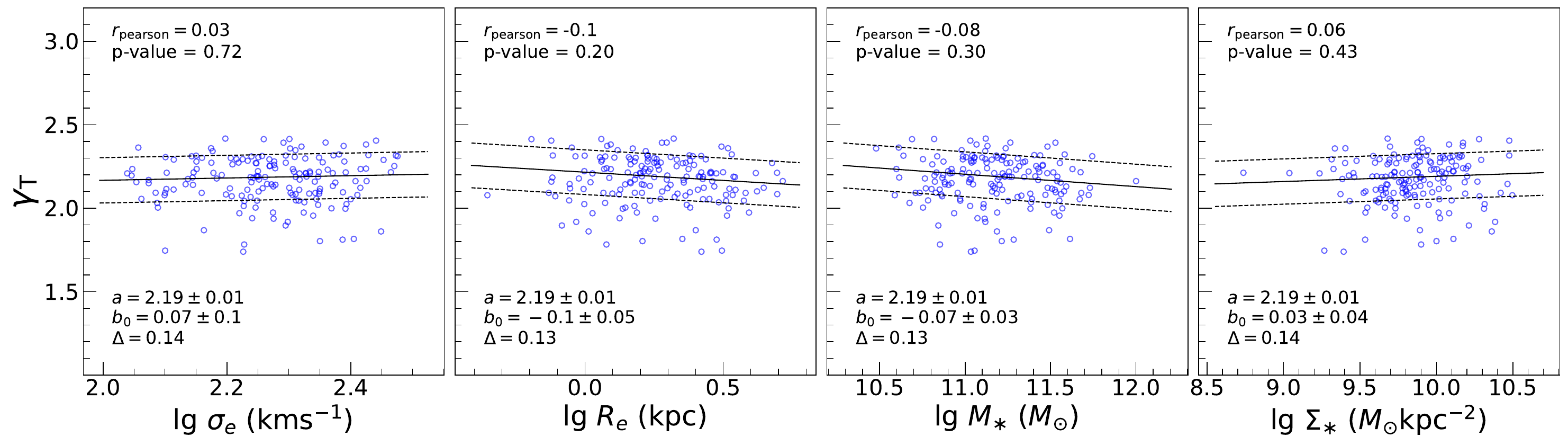}}
    \caption{
    Correlations between the total density slope ($\gamma_{\rm T}$) \newadd{from stars+gNFW with \jamsph\ model} and global galaxy properties for the MAGNUS sample. From left to right, the panels display the dependence of \tslope\ on central velocity dispersion ($\sigma_e$), effective radius ($R_e$), stellar mass ($M_{\ast}$), and surface mass density ($\Sigma_{\ast}$). The black solid line represents the robust best-fit linear relation $\gamma_{\rm T} = a + b(x-x_0)$, where $x$ denotes the logarithmic galaxy property corresponding to each panel. The dashed lines indicate the $1\sigma$ intrinsic scatter. The best-fit coefficients and intrinsic scatter are reported at the bottom of each panel, with the Pearson correlation coefficient ($r_{\rm {pearson}}$) and corresponding $p$-value displayed at the top.}
    \label{fig:corr_slope}
\end{figure*}

Quantifying these dependencies is critical for evolutionary studies, as any apparent redshift evolution in \tslope\ could potentially arise from the selection of galaxy samples with different structural properties across cosmic time. We examined the MAGNUS sample for linear relations between \tslope\ and four key parameters: central velocity dispersion ($\sigma_e$), effective radius ($R_e$), stellar mass ($M_{\ast}$), and surface mass density ($\Sigma_{\ast}$), defined as $\Sigma_{\ast} = \frac{M_{\ast}}{2 \pi R_e^2}$. For this analysis, we utilized the \tslope\ measurements derived from the fiducial stars+gNFW with \jamsph\ model. We parameterized the scaling relations as linear functions of the form $\gamma_{\rm T}(x) = a + b_0(x - x_0)$, where $a$ is the intercept, $b_0 = \mathrm{d}\gamma_{\rm T}/\mathrm{d}x$ represents the slope (gradient), and $x_0$ is the sample median of the logarithmic galaxy property ($x$) under consideration. We performed robust linear fits using the \textsc{LtsFit} package, which provides the best-fit coefficients, intrinsic scatter, Pearson correlation coefficients ($r$), and associated $p$-values. The results are displayed in \autoref{fig:corr_slope}. 

\begin{figure}
    \centering
    {\includegraphics[width=\linewidth]{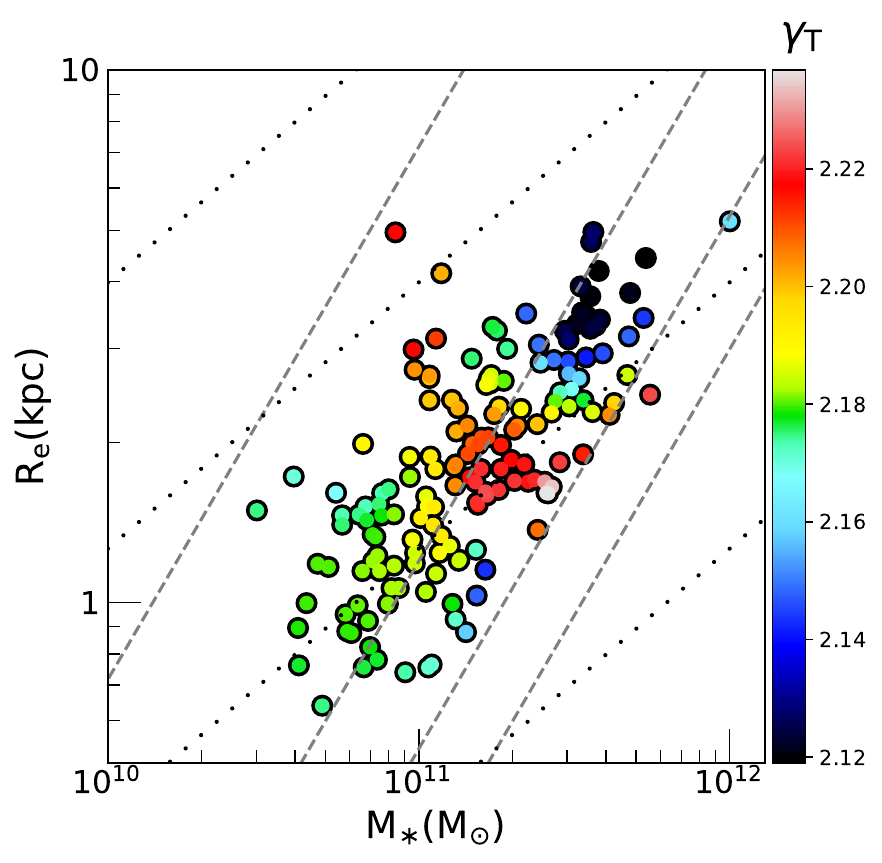}}
    \caption{\newadd{Distribution of the total density slope, $\gamma_{\rm T}$, in the stellar mass--size plane. The \tslope\ values are derived using the fiducial stars+gNFW with \jamsph\ model. The color map displays values smoothed via the Locally Weighted Regression method using the \textsc{LOESS} package \citep{ATLAS_XX_LOESS_Cappellari_2013b} to recover the underlying mean trends by suppressing intrinsic scatter. The gray dashed lines indicate contours of constant velocity dispersion ($\sigma_\mathrm{e} = 100, 200, 300,$ and $400$ \kmps, increasing from left to right), while the black dotted lines denote contours of constant surface mass density ($\Sigma_\mathrm{e} = 10^8, 10^9, 10^{10},$ and $10^{11}\,\mathrm{M}_{\odot}\,\mathrm{kpc}^{-2}$, increasing from top to bottom).}}
    \label{fig:mass_size_tslope}
\end{figure}

Overall, we find that the total density slope of MAGNUS sample exhibits negligible to weak correlations with these global properties. For velocity dispersion, we measured a slope of $\mathrm{d} \gamma_{\rm T}/\mathrm{d} \log_{10}\sigma_e = 0.07 \pm 0.10$ with an intrinsic scatter of 0.14. The Pearson coefficient ($r=0.03$, $p=0.72$) indicates no statistically significant correlation. Similarly, we found no significant trend with stellar mass, $\mathrm{d} \gamma_{\rm T}/\mathrm{d} \log_{10}M_{\ast} = -0.07 \pm 0.03$. The effective radius ($R_e$) shows a weak negative correlation ($\mathrm{d} \gamma_{\rm T}/\mathrm{d} \log_{10} R_e = -0.10 \pm 0.05$), suggesting that more compact galaxies tend to have slightly steeper density slopes, although the significance is marginal. Finally, the dependence on surface mass density is also statistically insignificant, with a gradient of $\mathrm{d} \gamma_{\rm T}/\mathrm{d} \log_{10} \Sigma_{\ast} = 0.03 \pm 0.04$ ($p=0.43$).

These observed trends remain consistent across other variations of our composite mass models. However, \tslope\ derived from power-law models exhibits a relatively stronger dependence on $\sigma_e$ and $\Sigma_{\ast}$, while trends with $R_e$ and $M_{\ast}$ remain at the same level as the composite models. For velocity dispersion, the measured slope from the power-law + \jamcyl\ model is $\mathrm{d} \gamma_{\rm T}/\mathrm{d} \log_{10}\sigma_e = 0.35 \pm 0.08$ with an intrinsic scatter of 0.12. Although the Pearson coefficient ($r=0.05$, $p=0.5$) indicates no statistically significant correlation within the MAGNUS sample, the gradient is noticeably steeper. For the power-law + \jamsph\ model, this gradient is $\mathrm{d} \gamma_{\rm T}/\mathrm{d} \log_{10}\sigma_e = 0.15 \pm 0.09$, though still statistically insignificant ($r=0.09$, $p=0.26$). For surface mass density, both power-law models yield a gradient of $\mathrm{d} \gamma_{\rm T}/\mathrm{d} \log_{10} \Sigma_{\ast} \approx 0.13 \pm 0.03$ ($p=0.1$).

We also performed an identical analysis on the full local MaNGA comparison sample. While the full sample showed similarly weak trends, the MaNGA primary sample (see \autoref{sec:combined_analysis}) exhibits a relatively stronger trend with $\sigma_e$, comparable to the gradient observed in the MAGNUS power-law models. This trend is also visible in Fig. \ref{fig:slp_in_fix_veldis}. Aside from velocity dispersion, the MaNGA samples show weak to no correlation with other galaxy properties. Given the lack of statistically significant correlations with structural parameters in either dataset, we proceed with determining the redshift evolution of \tslope\ without applying corrections for these galaxy properties. \\

\newadd{In Fig. \ref{fig:mass_size_tslope}, we present the MAGNUS sample in the stellar mass--size plane, color-coded by the locally weighted mean total density slope. We utilize the 2D LOESS algorithm \citep{LOESS_algorithm_Cleveland1988}, implemented via the \textsc{LOESS} package \citep{ATLAS_XX_LOESS_Cappellari_2013b} with a smoothing fraction of $\mathtt{frac}=0.5$, to recover the underlying average trend by suppressing intrinsic scatter. This projection facilitates the simultaneous inspection of correlations between $\gamma_{\rm T}$ and four key structural parameters: $M_*$, $R_{\rm e}$, $\sigma_{\rm e}$, and $\Sigma_*$. Consistent with the individual correlations, this plot reveals no significant dependence of $\gamma_{\rm tot}$ on any of these properties within the MAGNUS sample. This lack of correlation reinforces our primary finding that the observed redshift evolution is likely intrinsic rather than driven by structural selection effects, justifying our decision to not apply corrections for these galaxy properties.}

\subsection{Redshift Evolution of the Total Density Slope} \label{sec:z_evolution_result}
The evolution of the total density slope at intermediate redshifts remains a subject of debate, with cosmological simulations and lensing studies often yielding contradictory predictions \citep{Remus_2017, Sahu_AGEL_2024}. To resolve this discrepancy, we quantified the evolutionary trend, $\mathrm{d} \gamma_{\rm T}/\mathrm{d} z$, first within the MAGNUS sample alone and subsequently in combination with the local MaNGA ETG sample.

\begin{figure*}
    \centering
    {\includegraphics[width=\textwidth]{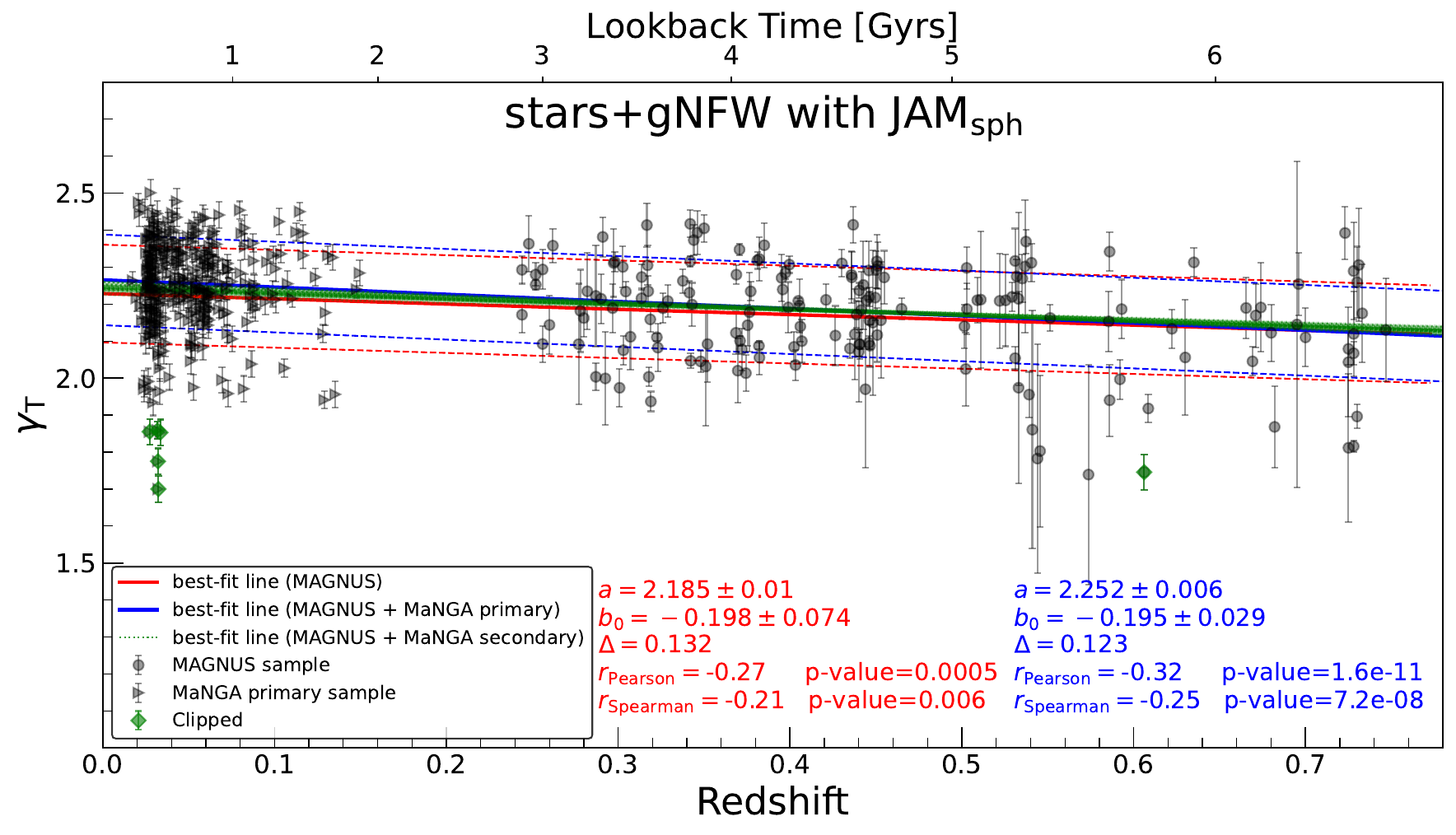}}
    \caption{Evolution of the total mass density slope ($\gamma_{\rm T}$) as a function of redshift. The \tslope\ for both sample are measured using the fiducial stars+gNFW with \jamsph\ dynamical model. The linear trends were determined using the \textsc{LtsFit} package according to the parameterization $\gamma_{\rm T} = a + b_0(z - z_0)$, where $z_0$ represents the median redshift of the sample, $a$ is the intercept, and $b_0$ quantifies the redshift gradient ($\mathrm{d}\gamma_{\rm T}/\mathrm{d}z$). The red solid line delineates the best-fit relation for the MAGNUS sample alone, while the blue solid line represents the fit to the combined dataset (MAGNUS + primary MaNGA subsample). The surrounding dashed lines of similar color mark the $1\sigma$ intrinsic scatter for the respective relations. The ensemble of green lines displays the fits derived from combining MAGNUS with each of the 100 importance-sampled secondary MaNGA sub-samples. The derived best-fit coefficients, intrinsic scatter ($\Delta$), and correlation statistics (Pearson and Spearman coefficients with associated $p$-values) are reported within the panel, color-coded in red and blue to correspond with the MAGNUS-only and combined MAGNUS+ MaNGA primary sample, respectively.}
    \label{fig:redshift_slope}
\end{figure*}

\subsubsection{Evolution within the MAGNUS Sample}\label{sec:z_evolution_magnus}
Empirical evidence indicates that massive ETGs rarely exhibit density slopes flatter than $\gamma_{\rm T} = 1.5$ or steeper than $\gamma_{\rm T} = 2.5$ \citep[e.g.][Fig.~11 left]{Cappellari_review_2026}. We adopted this range as a prior for the power-law models, whereas for the composite models (stars+NFW and stars+gNFW), we applied this constraint posteriori. After excluding galaxies falling outside the range $1.5 \le \gamma_{\rm T} \le 2.5$ and removing low-quality fits (Qual $\leq 0$), the final MAGNUS sample varies slightly by model choice where we retain 168 galaxies for the power-law models and 163 galaxies for the composite models. We found that the median total density slopes for the sample are $\gamma_{\rm T} \approx 2.220 \pm 0.013$ for the power-law models and $\gamma_{\rm T} \approx 2.187 \pm 0.013$ for the composite models, with uncertainties derived via bootstrap resampling. \newadd{Although the statistical errors are small, these two medians are consistent within the estimated median systematic uncertainties ($\sigma_{\rm syst} \sim 0.05$).}

To characterize the redshift evolution, we performed a robust linear fit to the \tslope\ -- $z$ relation using the \textsc{LtsFit} package. The relation is parameterized as $\gamma_{\rm T}(z) = a + b_0(z - z_0)$, where $z_0$ is the median redshift of the sample, $a$ is the normalization, and $b_0 \equiv \mathrm{d} \gamma_{\rm T}/\mathrm{d}z$ represents the evolutionary gradient. We adopted the symmetrized $1\sigma$ measurement uncertainty for individual galaxies (mean of the upper and lower percentiles) and clipped outliers at the $3\sigma$ level. The resulting redshift gradients are summarized in \autoref{table:redshift_slope}. We found that the NFW and gNFW models yield comparable evolutionary trends, with a gradient of $\mathrm{d} \gamma_{\rm T}/\mathrm{d} z \approx -0.21 \pm 0.07$ and an intrinsic scatter of $\sim 0.14$. The power-law models suggest a somewhat shallower evolution, $\mathrm{d} \gamma_{\rm T}/\mathrm{d} z \approx -0.10 \pm 0.07$, with similar intrinsic scatter.\newadd{ The redshift gradients from all six dynamical models are consistent within $1\sigma$ uncertainties and present a coherent qualitative picture:} the total density slopes of massive ETGs are becoming steeper with decreasing redshift (significance ranging from $1.5\sigma$ to $3\sigma$). 
This analysis represents the first dynamical study of such a large ETG sample over this redshift baseline to demonstrate that this \newadd{observed} trend is robust against the choice of mass model and velocity anisotropy. \newadd{The steepening is mild, with the slope changing by only a few hundredths over several billion years.} We visualize this evolution for the fiducial stars+gNFW with \jamsph\ model in \autoref{fig:redshift_slope}. Note that, the excluded galaxies have a lower median slope of $\gamma_{\rm T} \approx 2.0$ across all six models, which is naturally explained by their higher median redshift ($\langle z \rangle = 0.54$). While their inclusion would not qualitatively alter the conclusion of a steepening total density slope with decreasing redshift, we exclude them from the quantitative analysis to avoid relying on fits that are demonstrably unreliable.


\begin{figure*}
    \centering
    {\includegraphics[width=\textwidth]{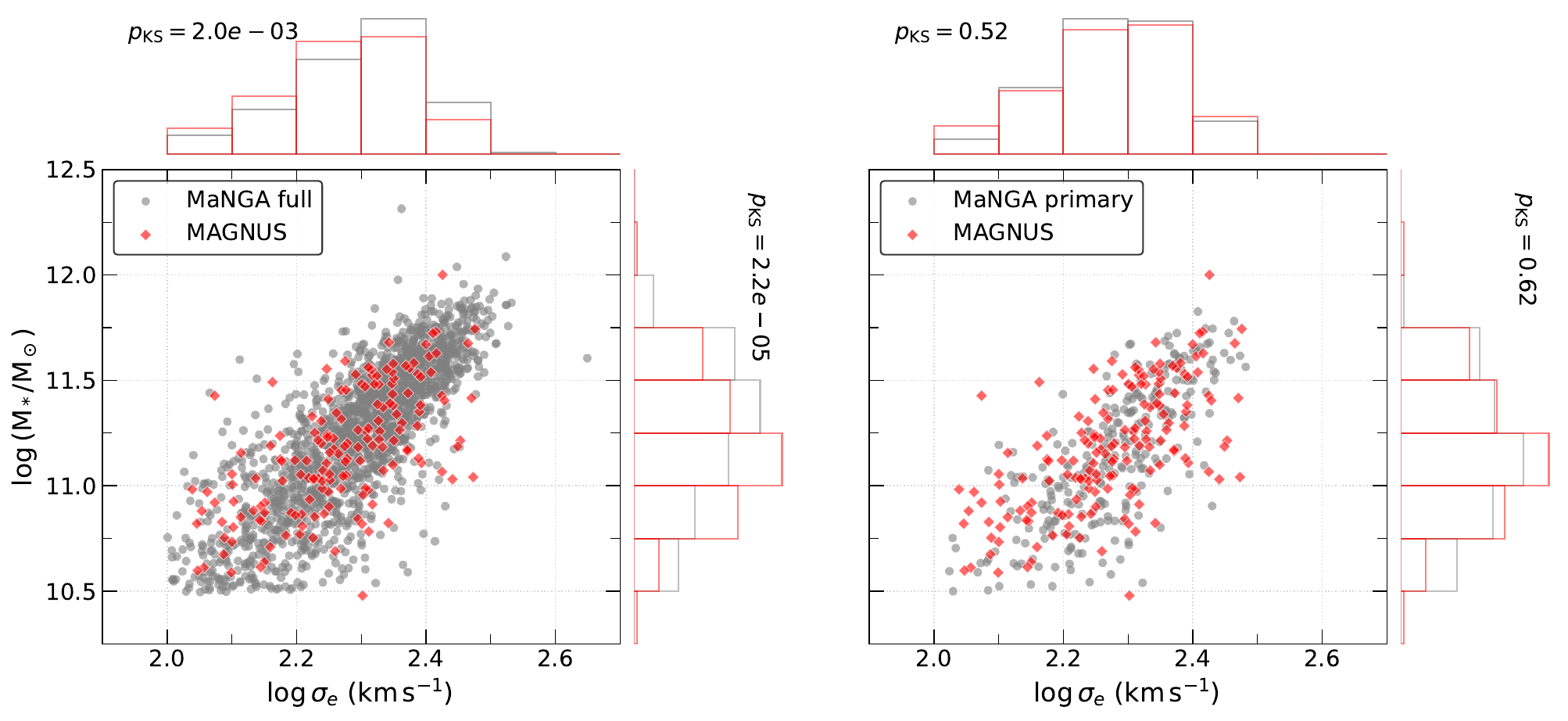}}
    \caption{Distribution of the MAGNUS and MaNGA ETG sample in ($\sigma_e, M_{\ast}$) plane. The left panel displays the full MaNGA ETG parent sample (gray circles) compared with the MAGNUS sample (red diamonds). The right panel shows the MAGNUS sample against the statistically matched primary MaNGA subsample. Normalized histograms for each parameter, color-coded to match the corresponding sample markers, are attached to the axes illustrate the respective distributions. The $p$-values derived from two-sample Kolmogorov–Smirnov (K–S) tests are reported alongside the histograms, quantifying the statistical consistency of the distributions.}
    \label{fig:matched_MaNGA}
\end{figure*}

\begin{figure}
    \centering
    {\includegraphics[width=1.05\linewidth]{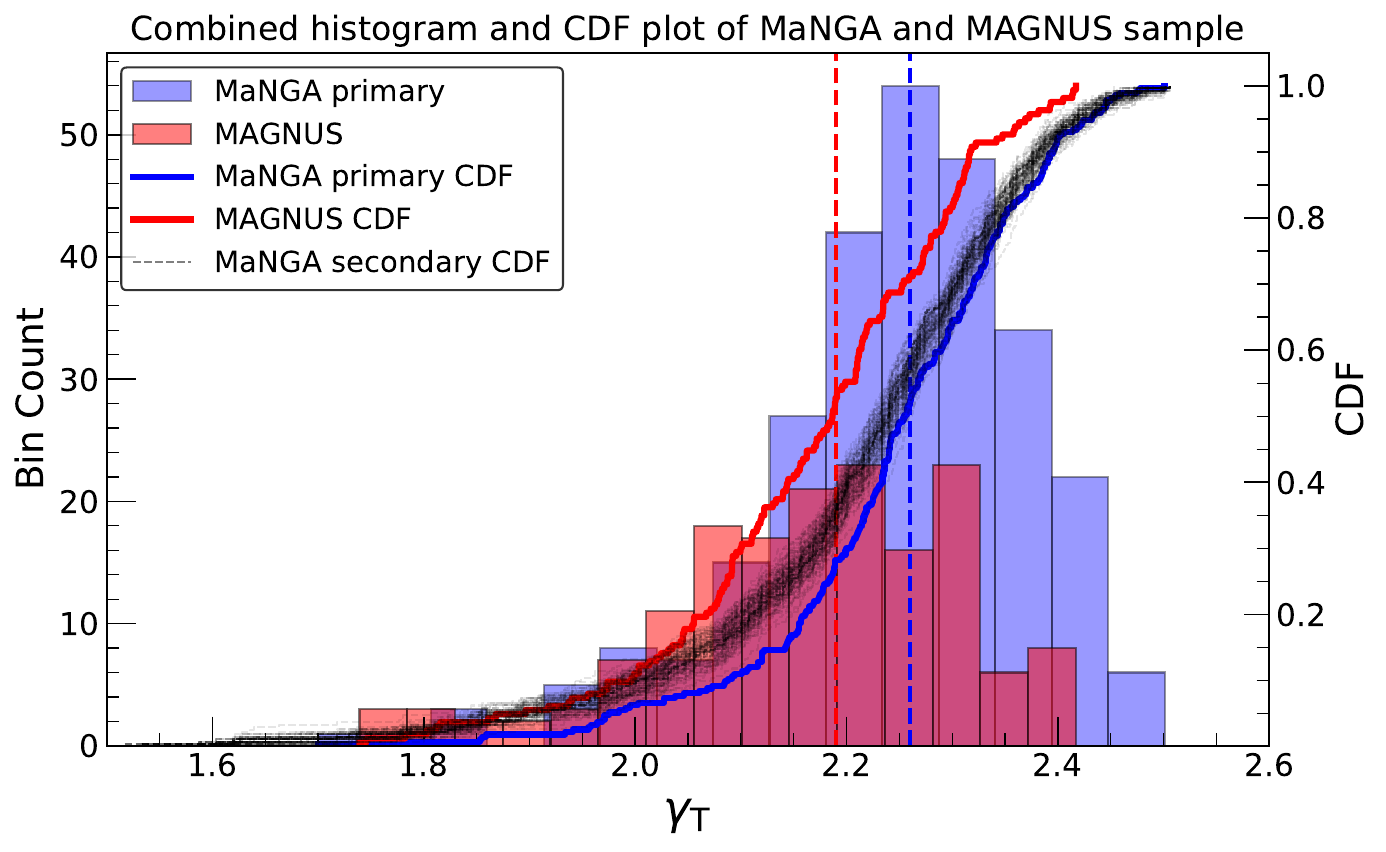}}
    \caption{Distributions of the total density slope ($\gamma_{\rm T}$) derived using the fiducial stars+gNFW with \jamsph\ model. The \tslope\ distributions are presented as histograms for the intermediate-redshift MAGNUS sample (red) and the local primary MaNGA subsample (blue). Vertical dashed lines of the corresponding colors mark the median slope for each sample. The solid red and blue curves denote the Cumulative Distribution Functions (CDFs) for the MAGNUS sample and the primary MaNGA subsample, respectively. The set of black dashed curves displays the CDFs for the 100 importance-sampled secondary MaNGA sub-samples.}
    \label{fig:hist_cdf}
\end{figure}

\subsubsection{Combined Analysis with the MaNGA Sample} \label{sec:combined_analysis}
Having quantified the evolutionary trend within the MAGNUS dataset, we now extend our analysis by incorporating the MaNGA ETG sample. Combining these datasets provides a long-baseline, enabling a significantly more precise constraint on the evolution of \tslope\ since $z < 1$.

Consistent with our treatment of the MAGNUS sample, we excluded MaNGA galaxies falling outside the physically motivated slope range ($1.5 \le \gamma_{\rm T} \le 2.5$). This constraint removed fewer than $\sim 2\%$ of the original 1,794 galaxies (the exact number varies slightly by mass model). To ensure a robust comparison, we employed a two-tiered strategy to select specific comparison sample from this filtered MaNGA catalog.

\textbf{Primary Comparison Sample:} First, we restricted the MaNGA dataset to include only those galaxies with high-quality dynamical fits (Qual $\geq 2$). This selection criterion naturally yields a subset of 277 ETGs (with a median redshift of $z \sim 0.044$) that closely matches the structural properties of the MAGNUS sample. Two-sample Kolmogorov–Smirnov (K–S) tests confirm that the stellar mass ($M_{\ast}$) and velocity dispersion ($\sigma_e$) distributions of this MaNGA subsample are statistically indistinguishable from those of the MAGNUS sample, yielding $p$-values of 0.62 and 0.52, respectively. \autoref{fig:matched_MaNGA} illustrates this structural agreement. We adopt this high-quality, statistically matched subset as our primary comparison sample.

\textbf{Secondary Comparison Sample:} To verify that our results are not driven solely by the restriction to high-quality fits, we constructed a secondary comparison set drawn from the full MaNGA ETG parent sample (Qual $\geq 1$). Since the full sample distribution differs from MAGNUS significantly (see Fig. \ref{fig:matched_MaNGA}), we utilized an importance sampling technique (detailed in Appendix \ref{sec:importance_sampling}) to generate 100 independent realizations, each consisting of 275 ETGs drawn without replacement. Each realization was constructed to be statistically matched to the MAGNUS sample in $M_{\ast}$ and $\sigma_e$ (verified via K-S tests with p-value $<$ 0.01).

In the following analysis, we treat the results derived from the primary sample as our main findings. The analysis of the secondary samples serves as a robustness check, for which, we report the ensemble average of the results derived from the 100 independent realizations.

The total density slopes for the MaNGA sample were adopted from DynPop I. Since DynPop I did not employ a single power-law mass model, we restrict this comparative analysis to the four composite models (stars+NFW and stars+gNFW with cylindrical and spherical alignments). \newadd{Although DynPop I does not provide individual measurement uncertainties for \tslope, it estimates systematic errors based on the RMS scatter between \tslope\ values from these four model assumptions (see Table 3 from DynPop I). We adopt these quality-dependent error values as our uncertainties and assign errors of $\sigma_{\gamma_{\rm T}} \approx 0.035$, $0.024$, and $0.056$ to galaxies with quality flags (\texttt{Qual}) of 3, 2, and 1, respectively.}\\


For the primary MaNGA subsample, we measured median slopes of $\gamma_{\rm T} \approx 2.260 \pm 0.01$ across all four composite models (ranging from 2.25 to 2.27). Similarly, the ensemble average of the median slopes for the 100 secondary subsamples is $\gamma_{\rm T} = 2.240 \pm 0.009$ (for the fiducial stars+gNFW with \jamsph\ model). Comparing these local baselines to the MAGNUS medians derived in \autoref{sec:z_evolution_magnus} ($\gamma_{\rm T} \approx 2.19$), we observe a systematic offset, with the local galaxies exhibiting steeper density profiles.

This evolutionary shift is small but statistically significant. We compared the cumulative distribution functions (CDFs) of the MAGNUS and MaNGA datasets using both the two-sample Kolmogorov–Smirnov (K–S) test and the Anderson-Darling (A–D) test, the latter being particularly sensitive to deviations in the distributional tails. For all four dynamical models and both MaNGA comparison sets (primary and secondary), the tests strongly reject the null hypothesis that the samples are drawn from the same underlying distribution ($p \leq 0.01$). \autoref{fig:hist_cdf} displays the combined histograms and CDFs for the stars+gNFW with \jamsph\ model, illustrating the clear distributional shift toward steeper slopes in the local Universe.

To quantify the cosmic evolution of the total density slope, we performed robust linear fits to the \tslope\--$z$ relation for the combined datasets. For the primary analysis (MAGNUS + Primary MaNGA), we fitted all four composite dynamical models. For the secondary analysis, we performed 100 independent fits, each combining MAGNUS with one of the importance-sampled MaNGA realizations, focusing on the fiducial stars+gNFW with \jamsph\ model. We utilized the \textsc{LtsFit} package to determine the best-fit parameters, employing the same linear parameterization defined in \autoref{sec:z_evolution_magnus} and applying a $3\sigma$ outlier clipping threshold. The results from the combined primary dataset are consistent across all NFW and gNFW model variants, yielding a global redshift gradient of \newadd{$\mathrm{d} \gamma_{\rm T}/\mathrm{d} z \approx -0.20 \pm 0.03$} with an intrinsic scatter of $\sigma_{\rm int} \approx 0.12$. This result implies a statistically significant \newadd{($\sim 6\sigma$)} steepening of the density profile over cosmic time. \newadd{The steeping is mild in amplitude and it is only detected thanks to our large sample of high quality data.} \\

The secondary analysis reinforces this finding. Across the 100 importance-sampled realizations, we recovered a mean gradient of \newadd{$\langle \mathrm{d} \gamma_{\rm T}/\mathrm{d} z \rangle = -0.150$ with a standard deviation of $0.017$}. While slightly shallower than the primary result, it robustly confirms the trend of the evolution. \autoref{fig:redshift_slope} illustrates these findings, where the solid red line denotes the fit to the primary combined sample, while the faint green lines show fits for the secondary realizations. The fit parameters for the combined MAGNUS + MaNGA primary sample are detailed in \autoref{table:redshift_slope}. In summary, the inclusion of the local baseline significantly reduces the statistical uncertainty on the evolutionary gradient compared to the MAGNUS-only measurement. Both the primary and secondary combined analyses provide compelling evidence that the total density slopes of massive ETGs have \newadd{mildly} steepened over the last 6-7 billion years ($z < 1$).\\

\begin{figure}
    \centering
    {\includegraphics[width=\linewidth]{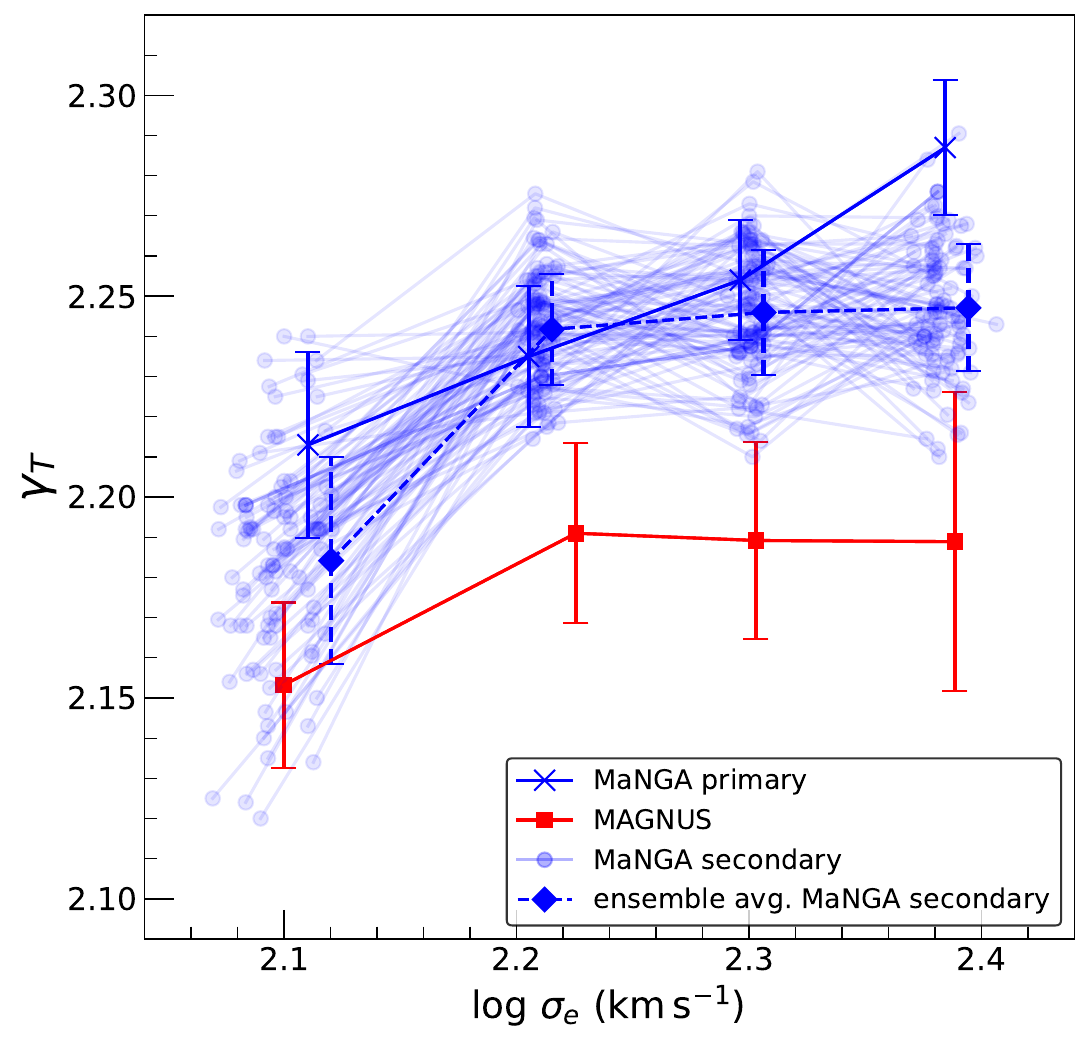}}
    \caption{Comparison of the total mass density slope ($\gamma_{\rm T}$) as a function of central velocity dispersion ($\sigma_e$). The galaxies were grouped into fixed velocity dispersion bins of width 40 \kmps, with \tslope\ values derived from the fiducial stars+gNFW with \jamsph\ model. The red squares (connected by a solid red line) and blue crosses (connected by a solid blue line) display the median values for the MAGNUS sample and the primary MaNGA subsample, respectively. To illustrate the variation within the local baseline, the light blue circles represent the binned medians for the individual importance-sampled secondary MaNGA sub-samples. The blue diamonds connected by a blue dashed line trace the ensemble average of these secondary sets. Error bars indicating bootstrap uncertainties are plotted for the MAGNUS, primary MaNGA, and secondary ensemble avaerage.}
    \label{fig:slp_in_fix_veldis}
\end{figure}

Finally, we tested whether the observed redshift evolution is a global trend or if it is driven by specific kinematic regimes. To verify this, we binned both the MAGNUS and MaNGA datasets by velocity dispersion ($\sigma_e$) using a fixed bin width of 40 \kmps. For each bin, we computed the median \tslope\ and associated bootstrap uncertainties using the fiducial stars+gNFW with \jamsph\ model. \autoref{fig:slp_in_fix_veldis} presents these binned trends as a function of $\log \sigma_e$. Comparing the MAGNUS sample with the primary MaNGA subsample, we observe that the high-redshift galaxies display systematically lower median \tslope\ values across nearly the entire dynamic range of velocity dispersion. On average, the MAGNUS bins are offset by approximately $3\%$ compared to their local counterparts, a difference consistent with the global ensemble offset mentioned above. This confirms that the steepening of the density slope is a pervasive feature of the population and persists after controlling for $\sigma_e$. 

We repeated this test using the secondary MaNGA subsamples. In this case, we observe a partial overlap between the high- and low-redshift populations in the lowest velocity dispersion bin, however, a clear separation emerges for $\log (\sigma_e/\text{\kmps}) > 2.15$. Notably, the MAGNUS sample exhibits a flattening of the $\gamma_{\rm T}-\sigma_e$ relation above a threshold of $\log \sigma_e \approx 2.2$ ($\sigma_e \approx 150$ \kmps), where the median slope remains roughly constant. While this feature is less distinct in the primary MaNGA subsample, it is reproduced at the ensemble level in the secondary MaNGA realizations. This behavior mirrors findings in local ETG studies, which have identified a break in the structural scaling relations at $\sigma_e \approx 100-150$ \kmps\ \citep[e.g.,][]{Poci_2017, Li_2019, Dynpop_III_Zhu_2024, DynPop_VI_Li_2024}. Importantly, varying the bin width or the specific binning scheme does not alter our qualitative conclusion that at fixed $\sigma_e$, massive ETGs at $z \sim 0.5$ have shallower density profiles than those at $z \sim 0$.

Collectively, these results provide a robust evidence that massive ETGs have experienced a gradual though mild evolution since $z \sim 1$, characterized by a steepening of their total density slopes with decreasing redshift. A critical strength of this analysis is the homogeneity of the methodology, the galaxy properties for both the MAGNUS and MaNGA samples were derived using identical kinematic extraction techniques, surface brightness parameterizations, and dynamical modeling assumptions. Consequently, potential systematics associated with the JAM formalism or tracer definitions are minimized, or expected to impact both epochs equally. This differential robustness strongly supports the interpretation that the observed shift in \tslope\ is a genuine physical signature of galaxy assembly over the last 6-7 billion years.

\begin{deluxetable*}{c c c c c c c}
    \tablecaption{\label{table:redshift_slope}List of best-fit parameters characterizing the redshift evolution of the total density slope (\tslope) for various dynamical models. The evolutionary trends were quantified using the linear relation \tslope\ $= a + b_0(z - z_0 )$, where $z_0$ represents the median redshift of the sample and $b_0$ denotes the redshift gradient ($\mathrm{d}\gamma_{\rm T}/\mathrm{d}z$). We report the best-fit coefficients, and the intrinsic scatter ($\Delta$) derived from the fit. Results are presented for both the MAGNUS sample alone and the combined MAGNUS + MaNGA primary dataset.
    }
    \tablehead{\colhead{Models} & \multicolumn{3}{c}{ MAGNUS only } & \multicolumn{3}{c}{MAGNUS and MaNGA primary} \\
              \colhead{} & \colhead{$\text{a}$} & \colhead{$\text{b}_0 = \frac{d\gamma_{\rm T}}{dz}$} & \colhead{$\Delta$}  & \colhead{$\text{a}$} &
              \colhead{$\text{b}_0 = \frac{d\gamma_{\rm T}}{dz}$} & \colhead{$\Delta$}}
\startdata
\jamcyl: power-law & $2.237 \pm 0.008$ & $-0.076 \pm 0.062$ & $0.122$ & & & \\
\jamsph: power-law & $2.232 \pm 0.009$ & $-0.115 \pm 0.069$ & $0.124$ & & & \\
\jamcyl: Stars + NFW & $2.182 \pm 0.010$ & $-0.235 \pm 0.076$ & $0.138$ & $2.262 \pm 0.006$ & $-0.216 \pm 0.029$ & $0.121$ \\
\jamsph: Stars + NFW & $2.182 \pm 0.010$ & $-0.192 \pm 0.074$ & $0.136$ & $2.249 \pm 0.007$ & $-0.192 \pm 0.030$ & $0.125$ \\
\jamcyl: Stars + gNFW & $2.185 \pm 0.010$ & $-0.209 \pm 0.075$ & $0.132$ & $2.264 \pm 0.006$ & $-0.213 \pm 0.028$ & $0.117$ \\
\jamsph: Stars + gNFW & $2.185 \pm 0.010$ & $-0.198 \pm 0.074$ & $0.134$ & $2.252 \pm 0.006$ & $-0.195 \pm 0.029$ & $0.123$ \\
\enddata
\end{deluxetable*}


\section{Discussion}
\label{sec:discussion}

\subsection{Robustness against Velocity Dispersion Selection} \label{sec:high_vd_cut}
As discussed in \autoref{sec:combined_analysis}, the total density slope flattens above a characteristic velocity dispersion threshold. While our sample selection process adopted $\sigma_e \geq 100$ \kmps, recent work by \citet{Dynpop_III_Zhu_2024} and \citet{DynPop_VI_Li_2024} suggests this structural break may occur at $\sigma_e \approx 150$ \kmps. To ensure our evolutionary trends are not driven by galaxies below this threshold, we repeated our analysis using this stricter selection criterion.

Before performing the analysis, we can anticipate the outcome based on the binned trends shown in \autoref{fig:slp_in_fix_veldis}. For the MAGNUS sample, the median \tslope\ remains effectively constant at $\log (\sigma_e/\text{\kmps}) > 2.2$ ($\sigma_e > 150$ \kmps). In contrast, the median slope for the primary MaNGA subsample increases slightly in this regime, while the secondary MaNGA subsamples remain flat at the ensemble level. Crucially, even above this higher threshold, the distributions of the high- and low-redshift samples remain distinct. Consequently, we expect that the CDFs will retain their separation and the qualitative trend of redshift evolution will persist.\\

Applying the stricter cut of $\sigma_e \geq 150$ \kmps\ reduces the MAGNUS and full MaNGA ETG samples to 130 and 1,537 galaxies (decreases of $20\%$ and $14\%$, respectively). The primary MaNGA subsample decreases by $14\%$ to 235 ETGs. Despite these reductions, the median \tslope\ values remain robust. For the stars+gNFW with \jamsph\ model, the MAGNUS median is $2.188 \pm 0.015$, identical to the baseline result. Similarly, the primary MaNGA median ($2.260 \pm 0.015$) and the secondary ensemble mean ($2.241 \pm 0.009$) are unchanged. K-S and A-D tests also confirm that the populations remain statistically distinguishable ($p \ll 0.01$).

Consistent with our anticipation, the inferred redshift evolution not only persists but strengthens in this high-mass region. For the MAGNUS-only sample, the gradient for the composite models steepens to $\mathrm{d} \gamma_{\rm T}/\mathrm{d} z \approx -0.27 \pm 0.08$. A similar steepening is observed in the combined analysis, where the gradient for the MAGNUS + primary MaNGA sub-sample increases to \newadd{$\mathrm{d} \gamma_{\rm T}/\mathrm{d} z \approx -0.225 \pm 0.030$,} and the mean gradient from the secondary MaNGA realizations shifts to \newadd{$\langle \mathrm{d} \gamma_{\rm T}/\mathrm{d} z \rangle = -0.160$ (standard deviation $0.015$)}. Across all cases, the magnitude of the evolutionary gradient increases by approximately \newadd{$10\%$} compared to the baseline analysis, while maintaining similar statistical precision. These results confirm that the observed steepening of the total density slope is not a selection artifact of lower-dispersion galaxies.

\subsection{Comparison with previous dynamical studies} \label{sec:comparison_prev_dyn}
In this section, we place our results in the context of previous dynamical studies of redshift evolution of the total slope at $z < 1$. \citet{Derkenne_2021} analyzed 90 ETGs in the Frontier Fields ($z \sim 0.35$) using archival MUSE data. Although the MAGNUS sample incorporates data from six of the seven fields utilized in that study, only $\sim 20$ galaxies overlap due to differences in selection criteria. While the Frontier Fields sample shares a similar velocity dispersion distribution ($\text{median } \sigma_e \approx 175$ \kmps) with MAGNUS, it is statistically distinct in terms of total dynamical mass (median $\log_{10} M_{\rm dyn}/M_{\odot} = 10.82$). Using a stars+gNFW with \jamcyl\ model and defining the total density slope as the mean logarithmic gradient ($\Delta \log \rho/ \Delta \log r$) over $\sim 3 R_e$, they reported a median slope of $\gamma_{\rm T} = 2.11 \pm 0.03$ (standard deviation $0.23$). Comparing this result to the full ATLAS\textsuperscript{3D} sample ($\gamma_{\rm T} \approx 2.14$), they concluded that there was no significant evolution over the last $4-6$ Gyr.

Subsequently, \citet{Derkenne_2023} modeled 28 galaxies (including 20 ETGs) from the MAGPI survey ($z \sim 0.3$), which match the mass and $\sigma_e$ range of the MAGNUS sample. Employing a stars+gNFW with \jamcyl\ model, they measured the total density slope as the mean logarithmic slope over the radial range $[0.1R_e, 2R_e]$. To evaluate the $\gamma-z$ evolution, they homogenized the comparison by re-measuring the slopes of the Frontier Fields and ATLAS\textsuperscript{3D} samples (using results from \citealt{Derkenne_2021} and \citealt{Poci_2017}, respectively) within the same radial aperture, while excluding galaxies with $\sigma_e < 100$ \kmps. They found that the median slope of the MAGPI sample is $\gamma_{\rm T} = 2.22 \pm 0.05$, consistent with the homogenized measurement of the ATLAS\textsuperscript{3D} sample (median $\gamma_{\rm T} \approx 2.25$), and suggested that the total density slope has remained constant over the past $3-4$ Gyr. However, they noted that the derived slope is sensitive to the radial aperture: when measured within $1 R_e$, the median MAGPI slope becomes $2.19$. This is identical to our measurement for the MAGNUS sample ($\gamma_{\rm T} = 2.19 \pm 0.01$). Notably, they found that the re-measured Frontier Fields sample exhibits a significantly shallower median slope of $\gamma_{\rm T} = 2.01 \pm 0.04$. They attributed the discrepancy between the steeper MAGPI/ATLAS\textsuperscript{3D} slopes and the shallower Frontier Fields slopes to environmental effects, as the Frontier Fields galaxies reside in dense cluster cores while MAGPI targets diverse environments. While investigating environmental dependencies is beyond the scope of this work, evaluating this aspect using the combined MAGNUS and MaNGA datasets remains a compelling avenue for future research.

In the local Universe ($z \sim 0$), \citet{Poci_2017} analyzed 260 ETGs from the ATLAS\textsuperscript{3D} survey using three different mass models, including the stars+gNFW with \jamcyl\ parameterization (`Model I'). The ATLAS\textsuperscript{3D} sample typically has lower mass (median $\log_{10} M_{\rm dyn}/M_{\odot} = 10.58$) and velocity dispersion (median 130 \kmps) than MAGNUS. For the gNFW+\jamcyl\ model, they found a median slope within $1 R_e$ of $2.19 \pm 0.02$ (intrinsic scatter $0.17$) for the 142 ETGs with $\sigma_e > 125$ \kmps. If we further restrict this sample to ETGs with reliable quality flags (Qual $\ge 1$) and slopes within the physical range $1.5 \le \gamma_{\rm T} \le 2.5$, the subsample (115 ETGs) yields a median $\gamma_{\rm T} = 2.20 \pm 0.01$. While this value is flatter than our primary MaNGA baseline ($2.26\pm 0.01$), it is steeper than the full ATLAS\textsuperscript{3D} population ($2.14$), consistent with the known positive correlation between $\gamma_{\rm T}$ and $\sigma_e$. \citet{Li_2019} analyzed an earlier release of MaNGA ($\sim 2000$ galaxies) using JAM, finding a mean slope of $\gamma_{\rm T} = 2.24 \pm 0.22$ for galaxies with $\sigma_e > 100$ \kmps. This is fully consistent with the results from our primary and secondary MaNGA subsamples. Similar results, all consistent with a local $\gamma_{\rm T}\approx2.2$ above the break in $\sigma_e$ have been reported in other local dynamical studies \citep[e.g.,][]{Cappellari_2015, Yildirm_2017, Bellstedt_2018}, as reviewed in our \autoref{sec:intro}. Collectively, these comparisons indicate that while differences in sample selection and slope definitions exist, our findings fit consistently within the broader landscape of dynamical results, bridging the gap between specific intermediate-redshift surveys and large local catalogs.

\subsection{Comparison with Gravitational Lensing Studies} \label{sec:comparison_lensing}

Studies utilizing strong gravitational lensing provide an independent probe of the total density slope over cosmic time. Generally, these investigations use a joint ``Lensing and Dynamics" (L\&D) analyses and report a non-negligible redshift gradient ($\mathrm{d} \gamma_{\rm T}/\mathrm{d} z < 0$), indicating that ETGs become structurally steeper at lower redshifts. This method combines the robust projected mass constraint from lensing with the stellar velocity dispersion by solving the Jeans equations (typically assuming a spherical symmetry). In this way, the  L\&D method effectively constrains the global mass-weighted density slope within the effective radius \citep{Koopmans_2006, Dutton_2014} which is physically comparable to the dynamical slopes derived in our work.

Early L\&D analyses consistently pointed toward a steepening mass profile. \citet{Koopmans_2006} analyzed 15 lens galaxies from the SLACS survey ($0.06 < z < 0.33$) combined with 5 higher-redshift systems \citep{Treu_Koopmans_2004}, measuring an evolutionary gradient of $\mathrm{d} \gamma_{\rm T}/\mathrm{d} z \approx -0.23 \pm 0.16$ and an average slope of $\gamma_{\rm T} \approx 2.01$. Expanding the sample to 73 SLACS lenses, \citet{Auger_2010} reported a slightly steeper average slope of $2.08 \pm 0.03$. Subsequent studies incorporating higher-redshift samples reinforced this evolutionary trend. By combining SLACS with the SL2S and LSD surveys (extending to $z \sim 1$), \citet{Ruff_2011} measured a gradient of $\mathrm{d} \gamma_{\rm T}/\mathrm{d} z \approx -0.25 \pm 0.11$. Similarly, \citet{Bolton_2012}, adding the intermediate-redshift BELLS sample, reported a significantly stronger evolution of $\mathrm{d} \gamma_{\rm T}/\mathrm{d} z \approx -0.60 \pm 0.15$ over the interval $0.1 < z < 0.6$. 

Consistent results were also reported by \citet{Li_2018} using a combined sample of SL2S, BELLS, and GALLERY lenses. More refined analyses distinguishing between population-level evolution and individual galaxy evolution have also been conducted. \citet{Sonnenfeld_2013} measured a population gradient of $\partial \gamma_{\rm T}/\partial z = -0.31 \pm 0.10$ for the combined SLACS and SL2S sample. After accounting for the  size-mass growth of ETGs, they inferred an intrinsic evolutionary rate for individual galaxies of $\mathrm{d} \gamma_{\rm T}/\mathrm{d} z = -0.10 \pm 0.12$.  

While recent studies comparing methodology suggest discrepancies when using lensing-only constraints \citep{Etherington_2023, Sahu_AGEL_2024}, the L\&D results from the same works remain broadly consistent with the dynamical picture. Lensing-only approaches fit a mass profile directly to the distorted image configuration; while precise, they are sensitive to the local density slope at the Einstein radius and can be affected by the mass-sheet degeneracy (MSD) if the profile deviates from a simple power law \citep[e.g.,][]{Tan_2024_project_dinos_I}.

Overall, the evolutionary trends derived from lensing ETG samples are remarkably consistent with the results presented in this work. Our measured gradients for the combined MAGNUS+MaNGA dataset \newadd{($\mathrm{d} \gamma_{\rm T}/\mathrm{d} z \approx -0.19$ to $-0.22$)} fall squarely within the range reported by L\&D studies ($-0.10$ to $-0.31$). This agreement is particularly significant given that lens ETG galaxies mostly overlap with respect to stellar mass and velocity dispersion with our sample. The convergence of these two independent methods—pure dynamics and joint lensing+dynamics—provides compelling evidence that the steepening of the total density slope is a fundamental feature of massive galaxy evolution since $z \sim 1$. To rigorously quantify any residual systematic discrepancies, a combined SLACS and SL2S sample augmented with spatially resolved kinematics offers an ideal testbed. By applying an identical mass and anisotropy parameterizations to model the resolved velocity fields of these lens galaxies, we could decisively test the consistency between lensing, dynamical, and joint L\&D analyses.

\subsection{Comparison with Cosmological Simulations} \label{sec:comparison_sims}


Comparing our observational constraints with predictions from cosmological hydrodynamical simulations reveals significant tensions regarding the evolution of the total density slope at $z < 1$. \newadd{However, it is important to note that the comparison is limited by differences in sample selection, as the simulated catalogs may not fully reproduce the specific combination of stellar mass and velocity dispersion characterizing our sample}. The \textbf{Magneticum} Pathfinder simulations \citep{Remus_2017} predict a monotonic shallowing of the density profile, with $\gamma_{\rm T}$ decreasing from $\sim 2.3$ at $z=1$ to an isothermal value ($\sim 2.0$) at $z=0$. Similarly, the \textbf{IllustrisTNG} suite \citep{Wang_2019} predicts a multi-phase evolutionary history where the density slope stabilizes below $z < 1$, maintaining a nearly constant isothermal profile ($\gamma_{\rm T} \approx 2.0$) during this epoch. Both of these predictions conflicts with the statistically significant gradient measured in our combined MAGNUS+MaNGA sample.

Qualitatively, the \textbf{Horizon-AGN} simulations \citep{Peirani_2019} offer a closer match to our findings, predicting a continuous steepening of the total density slope from $z=2$ to $z=0$. However, a quantitative discrepancy remains. While Horizon-AGN predicts slopes converging to $\gamma_{\rm T} \approx 2.0$ at the present day, our dynamical measurements indicate that massive ETGs are super-isothermal ($\gamma_{\rm T} \approx 2.26$) at $z \sim 0$.

The diversity in simulated trends—ranging from shallowing to steepening—highlights the sensitivity of the total density slope to sub-grid physics implementation, particularly AGN feedback and its coupling to the dark matter halo \citep{Peirani_2019, Wang_2020, Mukherjee_2021}. The fact that most current simulations converge toward $\gamma_{\rm T} \approx 2.0$, whereas observational studies (both ours and previous lensing-based results) consistently find $\gamma_{\rm T} \geq 2.1$, suggests that current models may overestimate the efficiency of feedback in expanding the central mass distribution or underestimate the baryon concentration in the cores of massive ellipticals.

\subsection{Physical Implications for Galaxy Assembly} \label{sec:physical_implications}

The prevailing ``two-phase'' formation scenario posits that massive ETGs assemble the bulk of their central mass at high redshift ($z > 2$) via dissipative processes, followed by a protracted phase of accretion dominated by dissipationless (dry) mergers \citep{Oser_2010, van_Dokkum_2015, Naab_Ostriker_2017}. Dry mergers deposit material in the galaxy outskirts, driving significant size growth and causing the total density profile to become shallower over time \citep{Hilz_Naab_Ostriker_2013, van_Dokkum_2010}.

Our finding that the total density slope has \textit{steepened} (or at least remained super-isothermal) since $z \sim 1$ stands in tension with a purely dissipationless merger history. The observed steepening ($\mathrm{d}\gamma_{\rm T}/\mathrm{d}z < 0$) therefore implies that dissipative processes must play a non-negligible role even in the late-stage assembly of massive ETGs. This interpretation is consistent with theoretical constraints suggesting that dry mergers alone cannot fully account for the structural scaling relations of present-day ETGs \citep{Nipoti_2009a, Nipoti_2009b}. Mechanisms such as the accretion of gas-rich satellites or adiabatic contraction can channel baryons to the galaxy center. The dominant old stellar populations of these galaxies \citep{DynPop_II_Lu_2023, Mozumdar_MAGNUS_I} rule out major starbursts at $z < 1$, but a ``drizzle'' of gas-rich minor mergers could provide sufficient dissipation to counteract the shallowing effect of dry accretion without significantly altering the global stellar ages. Furthermore, cosmological simulations often produce shallower, near-isothermal profiles due to strong AGN feedback expanding the central mass distribution \citep{Peirani_2019, Wang_2020, Mukherjee_2021}. Our results suggest that feedback coupling in the most massive ETGs might be weaker or less efficient at rearranging the central potential than currently predicted by these models. 

Alternatively, the mild increase in \tslope\ may be intrinsic to the dry merger scenario itself, driven by the non-homologous growth of massive galaxies. Observations indicate that the size evolution of ETGs outpaces their mass accumulation below $z \sim 2$, following a scaling relation of $\Delta \log R_e \sim 2 \Delta \log M_*$ \citep{van_Dokkum_2010, Patel_2013, van_Dokkum_2015}.  It is plausible that while mass is being accreted onto the outskirts, it is insufficient to ``keep up" with the rapid expansion of the physical scale of the effective radius. Consequently, as the measurement aperture ($R_e$) encompasses these diffuse outer regions, the mass-weighted average slope may appear to steepen (or fail to shallow as predicted by homologous growth models), reflecting the preservation of the dense core relative to the growing envelope.


To summarize, the observed evolution, while statistically significant, is mild in magnitude. This implies that the fundamental mass structure of massive ETGs was already largely established by $z \sim 1$, with the near-isothermal density profile being in place well before this epoch. Our findings suggest a scenario where dry mergers dominate the size growth, but dissipative processes—or inefficient feedback mechanisms—cannot be neglected, as they appear necessary to maintain the steep total density slopes observed over the last 7 billion years.




\section{Summary and Conclusion} \label{sec:conclusion}

In this work, the third installment of the MAGNUS series, we investigated the total density slope evolution of massive ETGs over the last 6 billion years ( $z < 1$). We performed a detailed dynamical analysis of the full MAGNUS sample (212 ETGs; $0.25 < z < 0.75$), constraining their mass distributions by applying the Jeans Anisotropic Modeling (JAM) technique to spatially resolved stellar kinematics derived from high-S/N MUSE-DEEP data cubes. This analysis relied on Multi-Gaussian Expansion (MGE) models of the surface brightness constructed from high-resolution HST imaging. After applying quality control and physically motivated constraints, we obtained a final sample of $\sim 165$ galaxies. To constrain the evolution of $\gamma_{\rm T}$ across cosmic time, we combined this intermediate-redshift dataset with a local Universe sample ($z \sim 0.05$) of 1794 ETGs from the MaNGA survey. We adopted the dynamical properties for these local ETGs from the DynPop catalogue of \citet{Dynpop_I_Zhu_2023}, specifically because that study utilized an identical JAM and MGE framework, thereby ensuring a strictly homogeneous comparison between the two samples.

To ensure the robustness of our results against model assumptions, we explored six distinct dynamical parameterizations for the MAGNUS sample, varying the mass profile (power-Law, stars+NFW, stars+gNFW) and the alignment of the velocity ellipsoid (cylindrical vs. spherical). However, for the comparative analysis with MaNGA, we focused on the four composite models common to both datasets. Furthermore, to account for selection effects in the local baseline, we employed a two-tiered comparison strategy. We selected a primary MaNGA sample (277 ETGs) of high-quality fits (Qual $\ge 2$) that naturally matches the MAGNUS sample, and a secondary set of 100 importance-sampled realizations drawn from the full ETG sample. Our main conclusions are summarized as follows:

\begin{itemize}
    \item  We find that the total density slopes, \tslope\ of intermediate-redshift ETGs are systematically shallower than their local counterparts. The median slope for the MAGNUS sample is \newadd{$\gamma_{\rm T} \approx 2.190 \pm 0.013$}, whereas for the local MaNGA baseline it is  \newadd{$\gamma_{\rm T} \approx 2.260 \pm 0.013$}. The \tslope\ distributions of the two epochs are also statistically distinct.

    \item  Combining the MAGNUS sample with the primary MaNGA baseline, we measure a redshift gradient of \newadd{$\mathrm{d} \gamma_{\rm T}/\mathrm{d} z \approx -0.20 \pm 0.03$ (-0.19 to -0.22 depending on the model)}. Similarly, MAGNUS with 100 secondary importance-sampled realizations also yield a mean gradient of \newadd{$\langle \mathrm{d} \gamma_{\rm T}/\mathrm{d} z \rangle = -0.150 \pm 0.017$} for the stars+gNFW with \jamsph\ model. Together, these results provide a \newadd{$\sim$ 6$\sigma$} evidence that the total density profiles of massive ETGs have steepened at level of $\sim 5\%$ over the last 6-7 billion years.

    \item  Within the MAGNUS sample, the total density slope exhibits negligible correlations with stellar mass, effective radius, and surface mass density. This suggests that the observed redshift evolution is not driven by simple scaling relations with these parameters.

    \item  We observed that the correlation between \tslope\ and velocity dispersion flattens above $\sigma_e \approx 150$ \kmps. Crucially, restricting our analysis to this high-dispersion regime ($\sigma_e > 150$ \kmps) yields evolutionary trends that are qualitatively identical to—and even slightly stronger than—those derived from the baseline $\sigma_e > 100$ \kmps\ selection. This confirms that the observed steepening of \tslope\ is not artifact of including lower-dispersion systems.

    \item  The derived evolutionary trends are statistically consistent across \newadd{power-law and composite models}, regardless of the velocity ellipsoid alignment (cylindrical vs. spherical). 
\end{itemize}

Galaxy evolution is a complex, multi-faceted process. As a result, a study regarding total density slope evolution is inevitably subject to challenges, such as sample selection, limited sample sizes, methodological systematics, progenitor bias, and the diversity of host environments etc,. Addressing all these factors simultaneously remains difficult. However, within the limitations of this study, this work provides the most precise estimate of \tslope\ evolution in massive ETGs since $z \sim 1$, to date.

Combining these results with the previous studies in the MAGNUS series offers a comprehensive view of the assembly history of massive ETGs since $z \sim 1$. In \textbf{MAGNUS I} \citep{Mozumdar_MAGNUS_I}, we established that the global scaling relations, morphologies, and stellar population properties were largely in place by $z \sim 1$. In \textbf{MAGNUS II} \citep{Mozumdar_2025_MAGNUS_II}, we identified a kinematic evolution, characterized by a decrease in rotational support over cosmic time, consistent with a merger-driven history. In this work (\textbf{MAGNUS III}), we observed that this merger history may have also induced a subtle but significant steepening of the total mass profile. Together, these findings depict massive ETGs not as complete static relics, but as dynamically evolving systems well into the latter half of the Universe's history.

\begin{acknowledgments}
P.M. and T.T. acknowledge support by NSF through grant AST-2407277. C.D.F. acknowledges support for this work from the National Science Foundation under grant No. AST- 2407278.
\end{acknowledgments}

%



\software{
MgeFit \citep{MGE_Cappellari_2002},
JamPy \citep{Jampy_Cappellari_2008,Jampy_spherical_Cappellari_2020,Cappellari2026jam},
pPXF \citep{Cappellari_2017_ppxf,LEGAC_ppxf_Cappellari_2023},
loess \citep{ATLAS_XX_LOESS_Cappellari_2013b},
emcee \citep{emcee_Foreman_2013},
NumPy \citep{Numpy2020},
SciPy \citep{Virtanen2020},
Astropy \citep{AstropyCollaboration13,AstropyCollaboration18}
}

\textbf{Data availability}: The total density slopes derived from six different dynamical models and associated uncertainties for the MAGNUS sample used in this work, are available in machine-readable format in the online journal.



\appendix

\begin{figure}
    \centering
    {\includegraphics[width=1.05\linewidth]{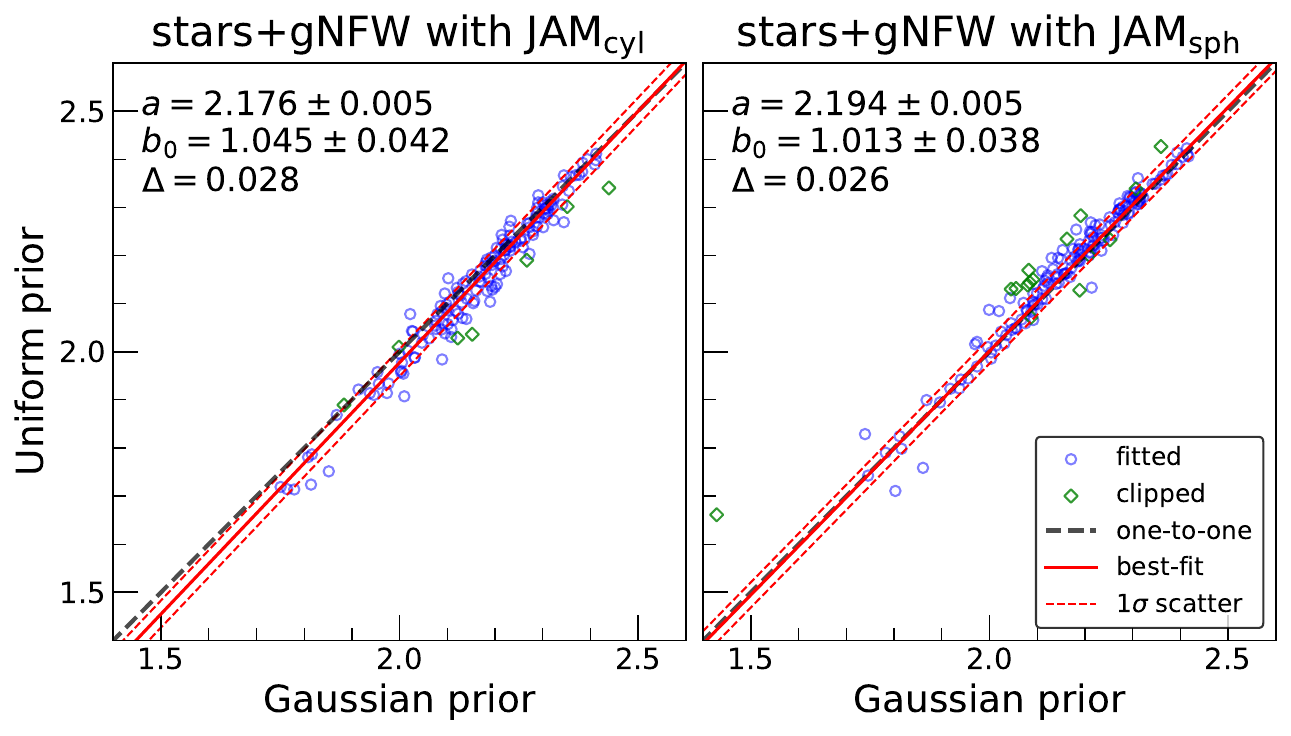}}
    \caption{Effect of the adopted anisotropy prior on the measured \tslope\ from stars+gNFW model with \jamcyl\ (left) and \jamsph\ (right) velocity ellipsoid orientations. For both case, the \tslope\ values are excluded at 3$\sigma$ level and denoted by green diamond shape markers while the fitted values are marked as blue circle. The one-to-one lines (black dashed) coincide closely with the best-fit lines (red solid). Coefficients of the best-fit line and rms scatter are also shown for both model.}    
    \label{fig:gamma_anisotropy_prior}
\end{figure}

\section{Robustness of Total Density Slope to Anisotropy Priors}
\label{sec:anisotropy_prior_slp}

In our dynamical modeling, the orbital anisotropy is parameterized by the ratio of the velocity dispersion components, $\sigma_z/\sigma_R$ for the \jamcyl\ case and $\sigma_{\theta}/\sigma_r$ for the \jamsph\ case. In our primary analysis, we imposed a Gaussian prior on these ratios, centered at isotropy, $\mathcal{N}(1.0, 0.07)$. For the \jamcyl\ configuration, we additionally enforced an upper bound of unity ($\sigma_z/\sigma_R \leq 1$), effectively adopting a half-Gaussian prior.

The motivations behind this choice of informed prior over an uniform or flat prior is described in Sect. \ref{sec:constant_anisotropy}. In contrast, the MaNGA sample analyzed in DynPop I where modeled with an uniform prior on $\sigma_z/\sigma_R$ and $\sigma_{\theta}/\sigma_r$ such that  $\mathcal{R}(q) < \sigma_z / \sigma_R < 1$ and $\mathcal{R}(q) < \sigma_\theta / \sigma_r < 2$, where the lower bound is defined by the intrinsic axial ratio $q$ as $\mathcal{R}(q)=\sqrt{0.3+0.7 q}$. 

To check the effect of these anisotropy priors on the measured \tslope\ and evaluate whether the observed evolutionary trends in \tslope\ are influenced by this difference in priors, we re-modeled the MAGNUS galaxies with `Qual' $\geq$ 1, using the same uniform prior adopted in the DynPop I. Everything else such as the priors on other model parameters and fitting procedure described in Sect. \ref{sec:fitting} etc., remain same. 

Fig. \ref{fig:gamma_anisotropy_prior} presents the comparison of \tslope\ with the Gaussian prior vs the flat prior for both stars+gNFW with \jamcyl\ and \jamsph\ models . The \tslope\ for each galaxy is the median of their respective posterior distributions. We quantified the relationship between the two sets of measurements using the \textsc{LtsFit} package, performing a robust linear regression with outliers clipped at the $3\sigma$ level. We found that while the choice of prior influences the specific best-fit anisotropy and mass parameters, it has a negligible impact on the derived total density slope. For both case, the slope of the best-fit line is close to unity, and almost overlap with the one-to-one line with an small intrinsic scatter. For the NFW models, we also found similar results. However, for the power-law models, we observed more deviation of the best-fit line from the one-to-one relationship and the intrinsic scatter is slightly higher than the composite models. These results confirm that our \tslope\ measurements are robust to the specific choice of anisotropy prior. 

We also measured the redshift gradient of the total density slope both for MAGNUS only and MAGNUS+MaNGA sample using $\gamma_{\rm T}$ values derived with a flat anisotropy prior for the  MAGNUS sample, applying the same quality and slope cuts as in the main analysis. This allows a direct comparison of the evolutionary trend under the two prior assumptions, using consistent posterior median $\gamma_{\rm T}$ values in both cases. We find that the choice of anisotropy prior does not qualitatively affect the conclusion of a steepening total density slope with decreasing redshift. For most model configurations, the flat prior yields gradients that are consistent with or slightly steeper than those derived under the Gaussian prior, confirming that the Gaussian prior provides a conservative rather than inflated estimate of the evolutionary trend.

\section{Importance sampling method}
\label{sec:importance_sampling}
To ensure a robust comparison between our target sample (MAGNUS) and the underlying parent population (MaNGA), we construct a control sample matched in stellar mass and velocity dispersion. We employ a 2D Sampling Importance Resampling (SIR) technique to select a subset of galaxies from the parent sample that reproduces the joint distribution of the target sample. We first discretize the 2D parameter space defined by the above two properties into a grid of 10×10 bins. For each galaxy k in the parent sample falling into the bin (i,j), we assign a selection weight $w_{k}$ proportional to the ratio of the number densities of the two samples in that bin:

\begin{equation}
\label{eq:weights}
    w_k = \frac{N_{\mathrm{target}}^{ij}}{N_{\mathrm{parent}}^{ij}}
\end{equation}
\\
where $N_{\mathrm{target}}^{ij}$ and $N_{\mathrm{parent}}^{ij}$ represent the number of galaxies in the (i,j)-th bin for the target and parent samples, respectively. In bins where $N_{\mathrm{parent}}^{ij} = $0, the weight is set to zero and in bins where $N_{\mathrm{target}}^{ij} = $0, by construction, the weight is zero. This means all the galaxies in a particular bin, share the same selection weight factor. We then construct the matched control sample by drawing galaxies from the parent sample without replacement, where the probability $P_k$ of selecting galaxy k is given by its normalized weight:

\begin{equation}
P_k = \frac{w_k}{\sum_{m \in S_{\mathrm{parent}}} w_m}
\end{equation}

where $S_{\mathrm{parent}}$ is the parent sample and m denotes each galaxy in the parent sample. This procedure ensures that the resulting control sample statistically shares the same 2D covariance structure as the target sample. We verify the quality of the match using a two-sample Kolmogorov-Smirnov (KS) test on the marginalized 1D distributions, requiring $\rm p>0.1$ to confirm that the control and target samples are drawn from effectively the same distribution.

\section{Modeling Results for Individual Galaxies}
\label{sec:all_galaxy_model}
In this appendix, we present the comprehensive modeling results for each galaxy in MAGNUS sample (198 in total). In Table \ref{table:slope_table}, we 
present the total density slopes derived from six different dynamical models for a representative subset of 10 galaxies. This table is provided to illustrate the format and content of the results. The comprehensive catalog, containing the slope measurements for the entire MAGNUS sample, will be available in a machine-readable format in the online journal. 

The figures below display the high-resolution HST imaging alongside the MGE surface brightness models, illustrating the accuracy of our photometric inputs. We further compare the spatially resolved observed $\rm{V_{rms}}$ maps with the predictions from our best-fitting stars+gNFW \jamsph\ models to demonstrate the quality of the kinematic fits. Finally, we include the resulting posterior probability distributions for the total density slope from also the stars+gNFW \jamsph\ model to highlight the statistical constraints derived for individual systems.

\figsetstart
\figsetnum{1}
\figsettitle{Photometric and Dynamical Modeling Results for the MAGNUS Sample}

\begin{figure*}
    \centering
    \includegraphics[width=0.95\textwidth]{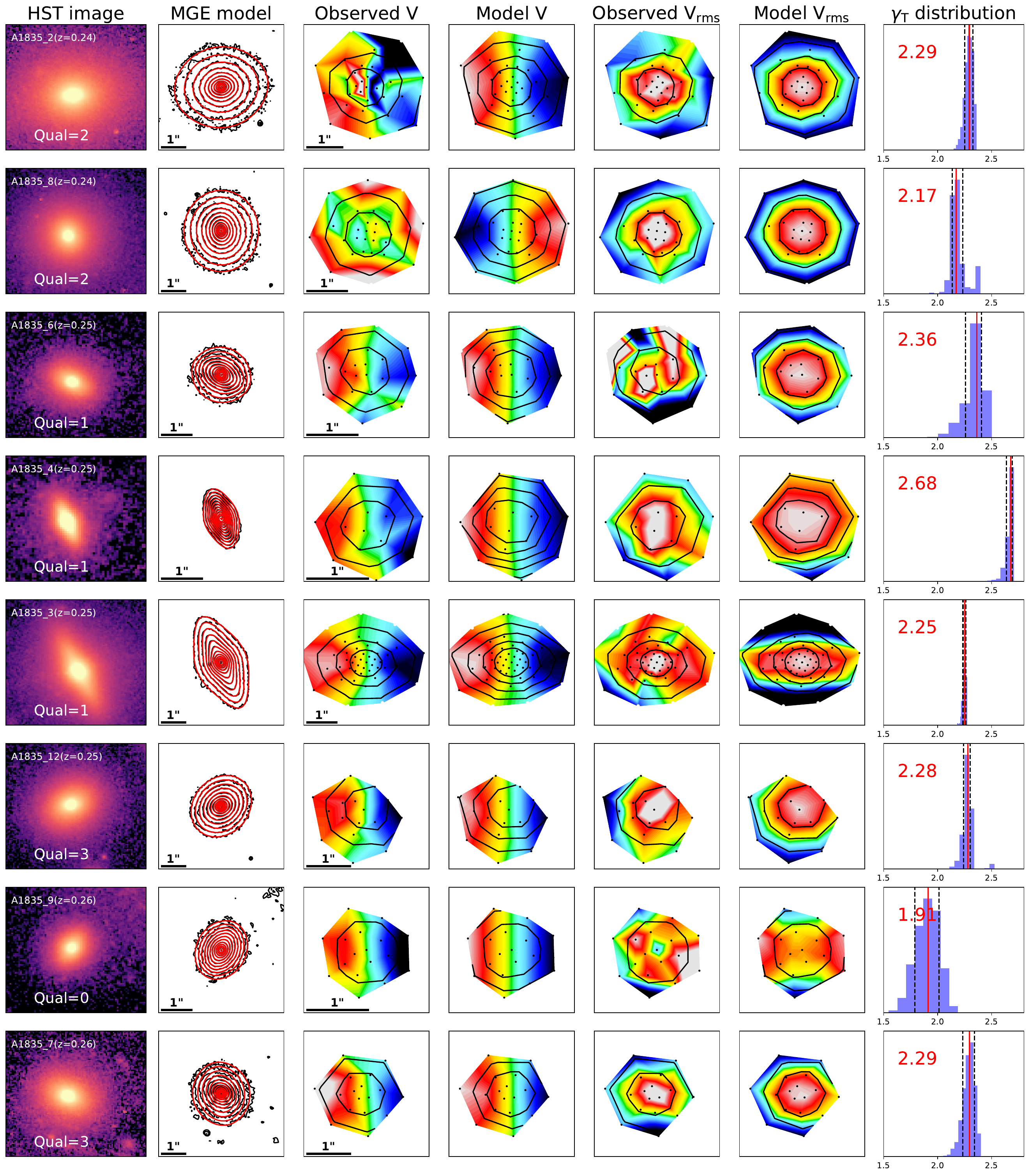}
    \caption{Spatially resolved modeling results for the MAGNUS sample using the stars+gNFW with \jamsph\ model. Each row corresponds to one galaxy and displays, from left to right: (1) HST imaging, with the galaxy ID and redshift ($z$) indicated at the top and the visual quality assessment of the dynamical model at the bottom; (2) Multi-Gaussian Expansion (MGE) fit, where black and red contours represent the observed and modeled surface brightness, respectively; (3--6) Observed and modeled stellar kinematics, showing the mean velocity ($V$) and root-mean-square velocity ($V_{\rm rms}$) fields. All kinematic maps are rotated to align the major axis with the $x$-direction. Black contours in the kinematic panels trace the surface brightness isophotes of the corresponding data (observed or modeled), while white circles denote clipped bins excluded from the \textsc{JAMPY} fit. A 1\arcsec\ scale bar is provided in the MGE panel (for imaging) and the observed velocity panel (for kinematics); (7)Posterior probability distribution of the total density slope, \tslope. The solid red line and numerical value indicate the median slope, while vertical black dashed lines mark the 16th and 84th percentiles. The full figure set is available in the online journal, with each set identified by its constituent galaxy IDs and redshift range.}
\label{fig:galaxy_atlas}
\end{figure*}

\figsetgrpstart
\figsetgrpnum{1.1}
\figsetgrptitle{Photometric and dynamical modeling results: Set 1}
\figsetplot{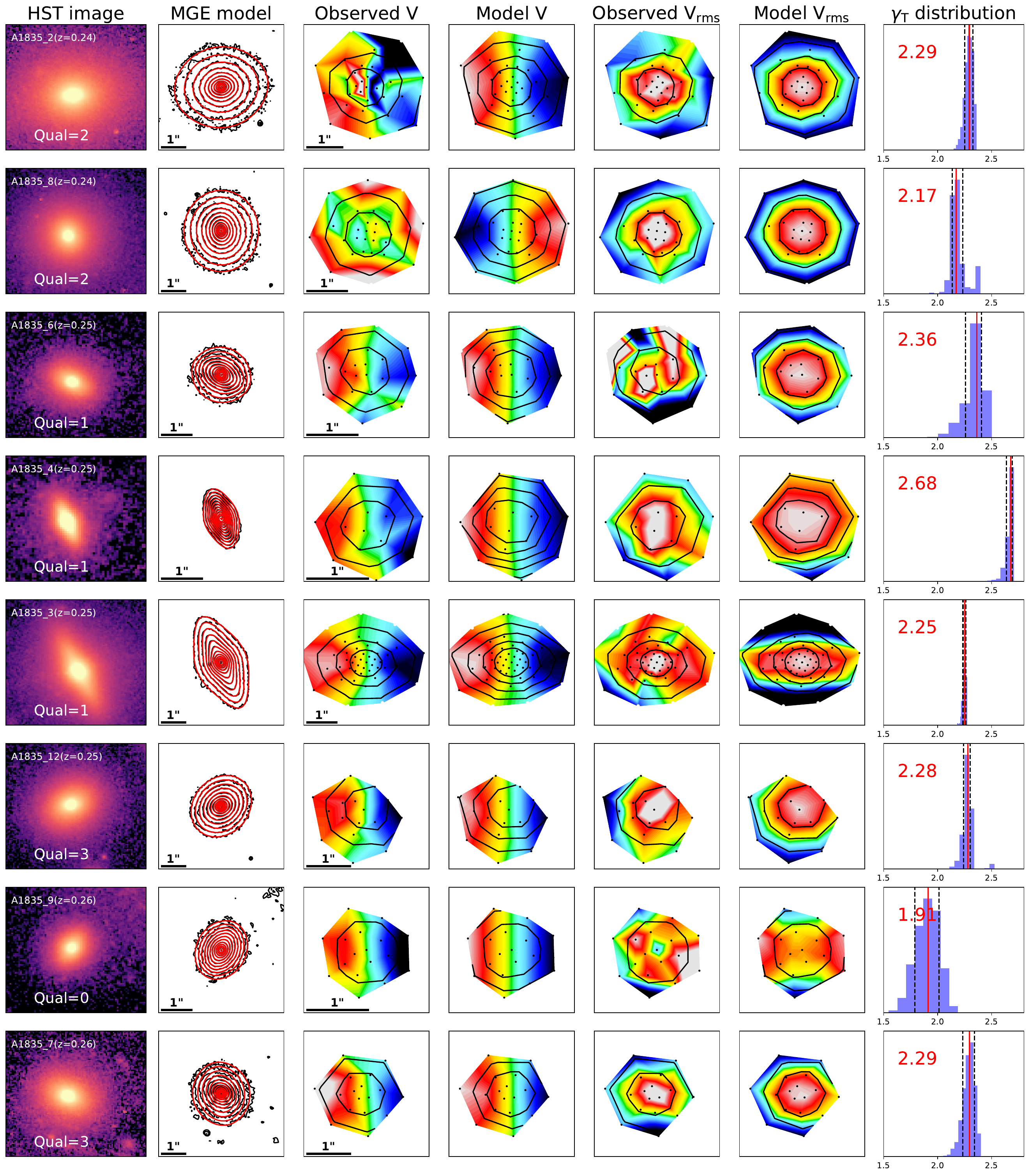}
\figsetgrpnote{Photometric and dynamical modeling results for galaxies A1835\_2, A1835\_8, A1835\_6, A1835\_4, A1835\_3, A1835\_12, A1835\_9, and A1835\_7, spanning the redshift range $0.24 < z < 0.26$. See Figure~\ref{fig:galaxy_atlas} for a full description of the figure layout and panel contents.}
\figsetgrpend

\begin{deluxetable*}{c c c c c c c c}
    \tablecaption{\label{table:slope_table} Mass-weighted total density slope, \tslope\ from six dynamical models and visual quality classification of the models.}

    \tablehead{\colhead{Galaxy} & \colhead{\tslope} & \colhead{\tslope} & \colhead{\tslope} & \colhead{\tslope} & \colhead{\tslope} & \colhead{\tslope} & \colhead{Quality} \\
     \colhead{ID} & \colhead{\tiny (power-law with \jamcyl)} & \colhead{\tiny (power-law with \jamsph)} & \colhead{\tiny (stars+NFW with \jamcyl)} & \colhead{\tiny (stars+NFW with \jamsph)} & \colhead{\tiny (stars+gNFW with \jamcyl)} & \colhead{\tiny (stars+gNFW with \jamsph)} & \colhead{} }
     
\startdata
A1063\_90 & $2.11^{+0.22}_{-0.21}$ & $2.09^{+0.23}_{-0.21}$ & $2.11^{+0.13}_{-0.07}$ & $2.11^{+0.10}_{-0.07}$ & $2.10^{+0.12}_{-0.07}$ & $2.09^{+0.13}_{-0.08}$ & 1 \\
A2744\_11950 & $2.21^{+0.06}_{-0.08}$ & $2.20^{+0.04}_{-0.06}$ & $2.29^{+0.02}_{-0.02}$ & $2.30^{+0.03}_{-0.02}$ & $2.34^{+0.06}_{-0.25}$ & $2.31^{+0.06}_{-0.04}$ & 3 \\
A370\_0409 & $2.10^{+0.06}_{-0.05}$ & $2.12^{+0.07}_{-0.06}$ & $2.10^{+0.06}_{-0.06}$ & $2.11^{+0.05}_{-0.06}$ & $2.11^{+0.08}_{-0.06}$ & $2.12^{+0.06}_{-0.06}$ & 1 \\
A370\_7278 & $2.25^{+0.07}_{-0.07}$ & $2.25^{+0.06}_{-0.07}$ & $2.25^{+0.03}_{-0.12}$ & $2.23^{+0.03}_{-0.02}$ & $2.24^{+0.03}_{-0.03}$ & $2.23^{+0.03}_{-0.03}$ & 3 \\
A370\_9731 & $2.39^{+0.07}_{-0.03}$ & $2.40^{+0.08}_{-0.09}$ & $2.00^{+0.05}_{-0.02}$ & $2.01^{+0.00}_{-0.08}$ & $2.00^{+0.05}_{-0.02}$ & $2.01^{+0.00}_{-0.09}$ & 3 \\
HE0238\_1 & $2.34^{+0.09}_{-0.06}$ & $2.30^{+0.13}_{-0.09}$ & $2.31^{+0.04}_{-0.03}$ & $2.32^{+0.05}_{-0.02}$ & $2.31^{+0.05}_{-0.03}$ & $2.31^{+0.04}_{-0.03}$ & 3 \\
LAB1\_1 & $2.30^{+0.14}_{-0.12}$ & $2.31^{+0.15}_{-0.10}$ & $2.20^{+0.08}_{-0.05}$ & $2.21^{+0.08}_{-0.04}$ & $2.21^{+0.07}_{-0.04}$ & $2.21^{+0.08}_{-0.05}$ & 1 \\
MACS1206\_3863 & $2.33^{+0.08}_{-0.06}$ & $2.36^{+0.07}_{-0.04}$ & $2.23^{+0.03}_{-0.02}$ & $2.23^{+0.02}_{-0.02}$ & $2.23^{+0.02}_{-0.02}$ & $2.24^{+0.03}_{-0.01}$ & 1 \\
RXS0437\_11 & $2.30^{+0.08}_{-0.08}$ & $2.30^{+0.09}_{-0.08}$ & $2.38^{+0.05}_{-0.04}$ & $2.37^{+0.04}_{-0.04}$ & $2.38^{+0.06}_{-0.04}$ & $2.38^{+0.06}_{-0.04}$ & 1 \\
SMACS2131\_217 & $2.18^{+0.07}_{-0.08}$ & $2.17^{+0.06}_{-0.07}$ & $2.29^{+0.05}_{-0.05}$ & $2.29^{+0.05}_{-0.05}$ & $2.29^{+0.07}_{-0.05}$ & $2.29^{+0.05}_{-0.05}$ & 3 \\
\enddata
\end{deluxetable*}

\bibliographystyle{aasjournalv7}
\bibliography{pritom}


\end{document}